\documentclass[12pt, notitlepage]{article}

\usepackage{preamble}

\title{Automated Market Making for Energy Sharing\protect\thanks{
We are grateful to Ulysse Griffon for his research assistance. We also thank the audiences at the following events for their feedback: The guest lecture in the \textit{CS 6501: Economics of Distributed Systems} course at the University of Virginia (Fall 2024), seminars at École Polytechnique (January 2025), the University of Virginia Computer Science Theory Seminar (February 2025), the UCSB DeFi Seminar (April 2025), the TLDR Conference at Columbia Business School (May 2025), the Second Tech4Finance Conference at Université Paris Dauphine (March 2025), Télécom Paris (June 2025), Princeton DeCenter (June and November 2025), ANR BlockFi Workshop Grenoble (December 2025). We also thank Julien Prat, Alfred Lehar, Ciamac Moallemi, Alkis Georgiadis-Harris for their valuable comments. Financial support from the Fellowship “The Latest in Decentralized Finance” (TLDR, 2024/2025 cohort) is gratefully acknowledged.
}}

\author{Michele Fabi\protect\footnote{Telecom Paris, CREST, IP Paris. Mail: michele.fabi@telecom-paris.fr} \ \ \,  \ \ Viraj Nadkarni\protect\footnote{Princeton University. Mail: viraj@princeton.edu} \ \ \, \\ Leonardo Leone\protect\footnote{Telecom Paris. Mail: leonardo.leone@telecom-paris.fr} \ \ \, \ \ Matheus X.V. Ferreira\protect\footnote{University of Virginia. Mail: matheus@virginia.edu} }

\date{\bigskip December 2025}

\begin{document}
	
\maketitle
 \begin{abstract}
We develop an axiomatic theory for Automated Market Makers (AMMs) in local energy sharing markets and analyze the Markov Perfect Equilibrium of the resulting economy with a Mean-Field Game. In this game, heterogeneous prosumers solve a Bellman equation to optimize energy consumption, storage, and exchanges. Our axioms identify a class of mechanisms with linear, Lipschitz continuous payment functions, where prices decrease with the aggregate supply-to-demand ratio of energy. We prove that implementing batch execution and concentrated liquidity allows standard design conditions from decentralized finance—quasi-concavity, monotonicity, and homotheticity—to construct AMMs that satisfy our axioms. The resulting AMMs are budget-balanced and achieve ex-ante efficiency, contrasting with the strategy-proof, ex-post optimal VCG mechanism. Since the AMM implements a Potential Game, we solve its equilibrium by first computing the social planner's optimum and then decentralizing the allocation. Numerical experiments using data from the Paris administrative region suggest that the prosumer community can achieve gains from trade up to 40\% relative to the grid-only benchmark.
\end{abstract}

\section{Introduction}
The traditional architecture of the electric power grid is increasingly strained by the proliferation of Distributed Energy Resources (DERs), such as rooftop solar, electric vehicles, and residential batteries.
Designed primarily for uni-directional power flows from centralized generators, the existing infrastructure struggles to accommodate the volatility and bi-directional flows inherent to markets with significant DER penetration.
Central to this landscape is the figure of the power producer-consumer, or \textit{prosumer}.
As the legacy grid model falters, these actors offer the potential to reduce reliance on centralized generation and provide essential distributed flexibility.
This necessity is underscored by recent instability events, such as the April 2025 blackout in the Iberian Peninsula \citep{entsoe2025blackout} triggered by disturbances at large-scale solar farms.
This event exposed the fragility inherit in centralized power architectures, highlighting the need for decentralized markets mechanisms that coordinate prosumer behavior to absorb such systemic shocks.
In response to these challenges, this paper investigates the potential of Automated Market Makers (AMMs) to facilitate decentralized market clearing within prosumer communities.
We propose a design where the AMM functions as an algorithmic intermediary that coordinates prosumers by quoting dynamic price curves based on aggregate supply and demand.
Our contribution is to provide an axiomatic characterization of this mechanism and to analyze the properties of the resulting market equilibrium within a mean-field game.

The AMM mechanism introduced in this paper is fundamentally different from AMMs in decentralized finance (DeFi).
First and foremost, the proposed energy AMM does not hold an inventory, and thus does not rely on liquidity providers. 
It rather utilizes the main grid as a buyer and seller of last resort.
This configuration allows one to designate a coordinator node within the local power grid to offer effectively “infinite liquidity’’ for residual imbalances.
The economic logic of the mechanism is to profit from operating within the spread between the grid's retail price---purchase price---and the feed-in tariff (FIT)---sale price.
This margin is often substantial; indeed, in many jurisdictions (such as Brazil) the FIT is effectively zero \citep{epe2022nota}.\footnote{In Brazil, the selling price is effectively zero for cash flows under the 2022 regulation. This policy reflects a broader issue of renewable curtailment---exceeding 10\% in some regions \citep{ONS_PEN2024}---caused by transmission congestion and insufficient local demand during peak generation hours. By signaling abundance through lower prices, the AMM incentivizes local consumption and storage, mitigating the need for such curtailment.}
\footnote{In the European Union, 2023 data show that substantial spreads at peak-hours, e.g.: Spain (Buy €0.32 vs. Sell €0.06--0.10/kWh), France (Buy €0.25 vs. Sell €0.13--0.18/kWh), and the Netherlands (Buy €0.32 vs. Sell €0.08--0.12/kWh). Similarly, in the United States, the shift from Net Metering to Net Billing (e.g., California's NEM 3.0) seems to have widened this gap, with export rates dropping to cost-avoidance levels (\$0.05/kWh) against retail rates exceeding \$0.30--0.40/kWh.}

\subsection{Contribution and Paper Structure}

This paper makes both conceptual and quantitative contributions to the design and analysis of local energy sharing markets.
In the first part of the paper, \cref{sec:market-design,AMM-construction}, we demonstrate that design principles from blockchain-based Constant Function Market Makers (CFMMs) apply in this context.
In particular, price curves derived from trading functions (i.e., potentials) that are suited for blockchain applications—namely those quasi-concave, homothetic, and monotone—are compliant with market design axioms that guarantee an efficient operation of local electricity markets.
These axioms impose anonymous execution of local trade, which distinguishes the AMM from order books and bilateral P2P matching where the order of bids and the identity of the bidders matters. Furthermore, a coalition-proofness requirement implies that the AMM payment function is \textit{linear} in the prosumer's power flows and quotes marginal buying and selling prices that are bounded and decreasing in the aggregate supply-to-demand ratio of power.
The resulting AMM operates through three distinct phases: (1) dynamic re-anchoring the trading function to align it in real-time with the energy price (buying price) and the feed-in-tariff (selling price); (2) determination of the internal clearing price via batch execution and concentrated liquidity; and (3) distribution of power export revenues and import costs proportionally to prosumers’ contribution to the community power balance. 

In the second part of the paper, Sections  \ref{sec:prosumer_model} and \ref{sec:equilibrium}, we move from the definition of the market mechanism to the analysis of strategic behavior \textit{within} the market. We model the prosumer community interaction as a dynamic stochastic game involving a large population of price-taking prosumers. Within this environment, each prosumer solves a dynamic optimization problem to determine their net power profile---jointly optimizing consumption, battery usage, and trading decisions---to maximize expected discounted profits over discrete trading sessions.

As prosumers interact, market prices (and rational expectations on them) arise endogenously from the aggregate net flows induced by the mixed-strategy Markov Perfect Equilibrium (MPE) of the game. Solving for such equilibrium can, in principle, be challenging due to the curse of dimensionality. However, a central finding of our paper establishes an equivalence between the decentralized MPE and a centralized Social Planner problem. We prove that the AMM frames the strategic environment as a \textit{potential game}: the mechanism acts as a coordination device where the gradient of prosumers' profit functions aligns with the gradient of a global potential function, representing the prosumer community's expected gains from trading with the power grid (importing and exporting electricity). Under our axioms, the ``invisible hand'' of the market is at work: a sub-optimal clearing of internal supply and demand leads to arbitrage opportunities over quoted prices, which prosumers are incentivized to exploit to restore efficiency.

The capability of coordinating prosumers purely by market forces, combined with its simplicity,  position the AMM as a practical alternative to market designs that have struggled to gain traction in recent years. For instance, pilot programs based on double auctions often struggle to scale because they require residential users to actively bid and commit to consumption schedules \textit{before} the actual power flows occur---a feature which may discourage non tech-savvy residential households. Conversely, while the Vickrey-Clarke-Groves (VCG) mechanism achieves a stronger notion of efficiency (ex-post efficiency) and satisfies strategy-proofness, it again requires  full type disclosure ahead of power flows, and generally runs a budget deficit. The AMM instead is budget-balanced and charges prosumers right \textit{after} power exchanges occur based solely on on real-time metering data---making it more suitable for possible integrations with automated smart-home energy management systems. Moreover, by decentralizing the power flow optimization on the user side, the AMM fosters the development of an open marketplace where energy startups can develop and offer power profile optimization tools. 

Besides conceptual relevance, our main equivalence result has practical implications. In fact, we use it  to develop a procedure that computes the equilibrium and simulates the behavior of a representative prosumer community. A key property for doing so is the scale-invariance (homogeneity of degree one) of the AMM payment function. This property allows us to circumvent numerical integration over a high dimensional type space by making the Planner's Problem asymptotically equivalent to its deterministic mean-field limit---the certainty-equivalent surplus of a representative prosumer.  Based on this, we compute the MPE by approximating the equilibrium mixed strategy of the infinite-dimensional population via state-dependent representative types and ``Monte-Carlo decentralization''---a procedure that solves the equilibrium mixing weights for representative prosumers, then simulates the  distribution of induced actions it via Monte-Carlo sampling and Euclidean projection onto the actual prosumers' action spaces to enforce feasibility. Furthermore, we approximate the infinite-horizon Bellman equation of the Planner by employing Model Predictive Control (MPC) with a finite lookahead window, a technique established in battery optimization theory and practice.

Finally, in the last part of our paper (\cref{quantitative}), we bring the model to data to quantify the benefits for both individual prosumers and their community. Using real-world weather and consumption data from the Île-de-France (Paris) region, we simulate a heterogeneous community of 1{,}000 agents over a full year (2023), allowing them to optimize their strategies over discrete 15-minute intervals (96 decision time-steps per day). The population is composed of 40\% pure consumers, 30\% solar prosumers (equipped with 1.6 $m^2$ panels at 15\% efficiency), and 30\% wind prosumers (using 1.9m-blade turbines), all managing 20 kWh residential batteries and a demand profile with 30\% flexible load. 

Our experiments reveal that the AMM generates substantial value by arbitrating the spread between electricity retail tariffs and the community's flexible needs. While the main grid charges a peak rate of 21.46 c€/kWh and offers a meager feed-in tariff (FIT) of 8.86 c€/kWh, the AMM allows prosumers to trade internally at dynamic clearing prices that frequently settle near 13--15 c€/kWh. This allows prosumers to satisfy demand at roughly half the cost of peak grid power, while simultaneously selling surplus generation at a premium of ~50–60\% over the standard FIT. By enabling agents to shift 30\% of their flexible load and synchronize 20 kWh batteries to these internal prices, the mechanism reduces the average daily energy cost per agent from €1.24 to €0.72—a 60\% reduction in individual spend—and achieving community's gains from internal trade of 42\% relative to uncoordinated trading with the power grid.

\subsection*{Related Work}

To the best of our knowledge, this work represents the first effort to bridge literatures on the economics and optimization of power systems, computational game theory, and decentralized finance (DeFi).

\paragraph{Distributed Optimization in Power Systems.} A substantial body of literature applies the Alternating Direction Method of Multipliers (ADMM) and related decomposition techniques to Optimal Power Flow (OPF) and grid coordination problems \citep{boyd2011admm,erseghe2014opf,Moret2024}. These approaches typically decompose the global optimization problem into coupled subproblems solved iteratively by network agents. While effective for computation, standard ADMM-based schemes do not generally ensure incentive compatibility during the iterative process, nor do they offer a clear economic interpretation of the intermediate ``prices'' (dual variables) prior to convergence. In contrast, the market mechanism proposed herein relies on axiomatic pricing rules which guarantee that  prices remain economically meaningful even when the community has not yet reached its equilibrium. 

There is also an extensive literature on battery optimization techniques. We draw upon recent advances in finite-horizon approximations for storage scheduling; in particular, \citep{prat2024finitehorizon} show that (under mild regularity conditions) the quality of approximation to the solution of an infinite-horizon Bellman equation using a rolling-horizon Model Predictive Control (MPC) improves exponentially with the size of the MPC lookahead window.

\paragraph{Equilibrium Analysis of Power Systems.} Strategic bidding and market clearing in power systems populated by DERs have traditionally been modeled as bilevel optimization problems or Mathematical Programs with Equilibrium Constraints (MPECs) \citep{hobbs2000strategic,Hasan2008,gabriel2012complementarity}. Our equilibrium analysis, however, diverges from that paradigm and shares closer  similarities with routing and congestion games. In these frameworks, agents optimize flows over a network subject to costs that depend on aggregate usage, often admitting a potential function that guarantees the existence and uniqueness of an equilibrium \citep{rosenthal1973congestion,monderer1996potential}. 

\paragraph{Automated Market Makers and  Decentralized Finance.} Constant Function Market Makers (CFMMs) have emerged as the dominant design paradigm for on-chain decentralized exchanges. The axiomatic foundations of these mechanisms have been extensively studied, demonstrating that trading functions which are quasi-concave, monotonic and homothetic induce markets with desirable properties \citep{angeris2020improved,angeris2023geometry,schlegel2023axioms,fabi2025economics}. In this paper we show that such trading functions can be implemented in this context to generate prices that effectively coordinate a community of prosumers.  AMMs with batch order execution—a feature of our proposed design—has been analyzed as a possible way to implement execution fairness \citep{canidio2023batching, he2024optimaldesignautomatedmarket}. We adapt these principles to the energy domain, replacing token inventories with grid-backed liquidity. 

The implementation of an AMM for energy grids falls into the scope of Decentralized Physical Infrastructure Networks (DePIN). While recent work has addressed incentive-compatible signal recovery in DePIN systems—such as verifying bandwidth claims \citep{milionis2025depin}—our work assumes reliable smart meter data and focuses on the economic behavior of agents.

\paragraph{Peer-to-Peer Markets and Prosumer Economics.} The rise of the "prosumer" has necessitated new economic models for distributed energy interaction \citep{gautier2018prosumersgrid,gautier2024economicsofprosumers}. Extensive surveys on peer-to-peer (P2P) energy trading and local flexibility markets highlight the diversity of proposed pricing mechanisms \citep{SOUSA2019367,CROWLEY2025125154,pinsonP2P,alfaverh2023dynamic}. Our analysis demonstrates that many existing P2P pricing rules can be recovered as special cases of the generalized trading functions presented in this paper \citep{8279516,8572734}, providing a unified theoretical framework for local energy exchange.

\paragraph{Mean-Field Games.} Our equilibrium results derived under large population approximations by The theory of Mean Field Games (MFG)  \citep{lasry2007meanfield,GueantLasryLions2011}, which deals with limits of games when the number of agents goes to infinity and payoff externalities occur only through aggregate statistics of population actions, such as the ratio of supply to demand in our case. While the core literature is focused on continuous-time stochastic games, our approach builds on recent discrete-time formulations \citep[e.g.,][]{Doncel_2019,guo2024mf}. Our algorithm combines MPC and ``Monte Carlo decentralization'' to solve numerically the Master Equation of the game, given by the Social Planner's value function, together with the evolving distribution of battery States of Charge (SoC) across the population.

\section{Axiomatic Design of an Energy Sharing Market}
\label{sec:market-design}

This section develops a formal model for a peer-to-peer (P2P) energy sharing market and introduces a set of axioms that characterize a desirable pricing mechanism. For the purpose of this axiomatic analysis, we treat individual energy flows as exogenously given; these flows will be endogenized as solutions to prosumer optimization programs and as Markov Perfect Nash Equilibrium (MPE) outcomes in Sections \ref{sec:prosumer_model} and \ref{sec:equilibrium}.

\subsection{The Microgrid}
We consider a local energy community (i.e., a microgrid) composed of a set $\mathcal{N} = \{1, 2, \dots, N\}$ of \textbf{prosumers}. These prosumers interact through a central \textbf{aggregator}, which facilitates local energy trades to import or export power. The aggregator is also connected to the main electrical grid.\footnote{In principle, we can allow every prosumer to be also connected to the main grid, forming  a double-star network. However, the axoms defined later on in this section impose a participation (individual-rationality) constraint such that prosumers always prefer to trade with the aggregator rather than directly with the external gird.}

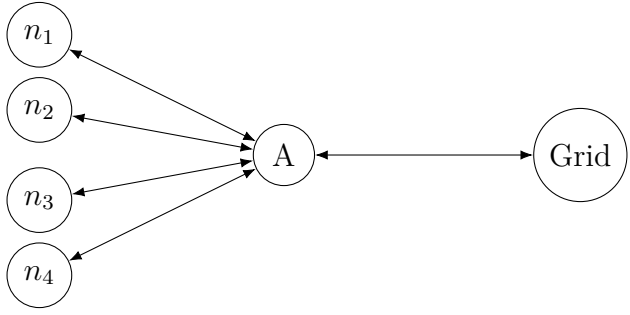
\begin{figure}
    \centering
\begin{tikzpicture}[>=Latex, node distance=8mm, scale=1, transform shape]
\node[draw, circle, minimum size=7mm] (A) {A};
.
\node[draw, circle, minimum size=7mm, right=2.9cm of A] (G) {Grid};
  
\draw[<->] (A) -- (G);

 \foreach \i/\y in {1/1.6,2/0.6,3/-0.6,4/-1.6}{
\node[draw, circle, minimum size=6mm, left=2.4cm of A, yshift=\y cm] (P\i) {$n_{\i}$};
    
 \draw[<->] (P\i) -- (A);
 }
\end{tikzpicture}
    \caption{Microgrid Network Representation}
    \label{fig:placeholder}
\end{figure}

Within this framework, the aggregator operates as the community's local \textbf{market maker}. Its function is to pool all energy supply and demand from the prosumers and to set a single, uniform price for buying and selling energy within the community, based on a deterministic function of these aggregates. This pooling mechanism provides ``infinite''  liquidity and abstracts away the need for direct, bilateral negotiations between prosumers.

By complementing the local market maker, the main grid functions as the \textbf{market of last resort}. It is an infinitely liquid external venue that guarantees the local community can always balance its \textit{net} energy position after all internal trading has been resolved by the aggregator. The main grid will buy any community-wide surplus and supply any community-wide deficit. For most of our analysis, we assume the grid's capacity to do so is not binding.

The main grid, operated by a retailer (or a competitive retail market), sets two distinct prices at each time $t$ for trading electricity. It buys energy from the community at an exogenous \textbf{Feed-in Tariff (FiT)}, denoted by $\underline{\lambda}_t$, and sells energy to the community at a higher \textbf{retail price}, $\overline{\lambda}_t > \underline{\lambda}_t$. These two prices form the external benchmark and the price boundaries for the local market.

\subsection{Market Definition}

We now formally define the local energy market and its properties. 
Market interactions involve each prosumer $n \in \mathcal{N}$ submitting a net energy flow, or \textit{netput} $x_n \in \mathbb{R}$, to the aggregator. Since we will describe pricing at a given time step, for notational simplicity, we omit the time index $t$. A positive flow ($x_n > 0$) represents a net supply to the local market, while a negative flow ($x_n < 0$) represents a net demand. The collection of these individual flows constitutes the market's \textbf{allocation vector}, $\bm{x} = (x_1, \dots, x_N) \in \mathbb{R}^N$. 

Although our full model is dynamic, with trades $x_{nt}(e)$ occurring at discrete trading sessions $t$ within epochs $e$ (e.g., days), the definition of market and payment rules applies to any single trading session. For this reason, for now we adopt a static view and omit the time and epoch index for simplicity.

The mechanism's function is to map this allocation to a vector of payments:

\begin{definition}[Market Mechanism]
A \textit{market mechanism} for energy sharing is a pair $(\bm{x}, \bm{P}(\bm{x}))$, where:
\begin{itemize}
    \item $\bm{x} = (x_1, \dots, x_N) \in \mathbb{R}^N$ is the energy \textbf{allocation vector}.
    \item $\bm{P}: \mathbb{R}^N \to \mathbb{R}^N$ is the \textbf{payment function}, which maps an allocation $\bm{x}$ to a payment vector $\bm{P}(\bm{x}) = (P_1(\bm{x}), \dots, P_N(\bm{x}))$.
\end{itemize}
A positive payment $P_n(\bm{x}) > 0$ represents a revenue for prosumer $n$, while a negative payment $P_n(\bm{x}) < 0$ represents a cost.
\end{definition}

For our analysis, it is useful to decompose the net flows and payments into more intuitive components. Specifically, An individual prosumer's net flow $x_n$ can be split into non-negative \textbf{supply ($s_n$)} and \textbf{demand ($d_n$)}:
\begin{equation}
    x_n = s_n - d_n, \quad \text{where} \quad s_n = \max\{x_n, 0\} \quad \text{and} \quad d_n = \max\{-x_n, 0\}.
\end{equation}
Summing these across all prosumers gives the \textbf{aggregate local supply ($s$)} and \textbf{aggregate local demand ($d$)}:
\begin{equation}
    s = \sum\nolimits_{n \in \mathcal{N}} s_n, \quad \text{and} \quad d = \sum\nolimits_{n \in \mathcal{N}} d_n.
\end{equation}
The community's net position relative to the main grid is $x_A = s - d$. This can be decomposed into the total energy \textbf{exported to ($s_A$)} or \textbf{imported from ($d_A$)} the main grid:
\begin{equation}
    s_A = \max\{s-d, 0\}, \quad \text{and} \quad d_A = \max\{d-s, 0\}.
\end{equation}
Similarly, each prosumer's net payment $P_n(\bm{x})$ is the difference between a non-negative \textbf{revenue function $R_n(\bm{x})$} and a non-negative \textbf{cost function $C_n(\bm{x})$}:
\begin{equation}
    P_n(\bm{x}) = R_n(\bm{x}) - C_n(\bm{x}).
\end{equation}

\subsection{Axiomatic Characterization}

Our goal is now to define design a mechanism based on a set of desirable features, detailed by the axioms that will follow. 

\subsubsection{Anonymity}

In many P2P markets, such as those based on limit order books, the identity and arrival time of an order can influence its execution price. This creates opportunities for preferential treatment or strategic exploitation like front-running. To avoid this, a local energy sharing market should be \textbf{anonymous}. Informally, this means that a prosumer's payment should depend only on their own energy contribution and the collective contributions of others, not on anyone's identity.

\begin{axiom}[Anonymity]
\label{def:anonymity}
A market mechanism $(\bm{x}, \bm{P}(\bm{x}))$ is \textbf{anonymous} if for any prosumer $n \in \mathcal{N}$ and any permutation $\pi$ of the other prosumers, the payment to $n$ remains unchanged:
\[
P_n(x_n, \bm{x}_{-n}) = P_n(x_n, \pi(\bm{x}_{-n})),
\]
where $\bm{x}_{-n}$ is the vector of allocations for all prosumers other than $n$, and $\pi(\bm{x}_{-n})$ is the vector with those allocations permuted.
\end{axiom}

\paragraph{Anonymity and Batched Pricing.}
The axiom of anonymity forces the payment function to disregard the identities of the prosumers, meaning all trades can be processed as a single batch. Formally, a prosumer's payment can only depend on their own allocation, $x_n$, and the \textit{multiset} of all other allocations, denoted $\Sigma(\bm{x}_{-n})$. 

\begin{proposition} \label{prop:anonimity}
A market is anonymous if and only if there exists a function $\Phi: \mathbb{R} \times \mathrm{Multiset}(\mathbb{R}^{N-1}) \to \mathbb{R}$ such that for any prosumer $n$:
\[
P_n(x_n, \bm{x}_{-n}) = \Phi(x_n, \Sigma(\bm{x}_{-n})).
\]
\end{proposition}

This functional form is the essence of an Automated Market Maker (AMM) in this context. Because the rules are anonymous, the mechanism is stripped of any discretion; it cannot rely on subjective information about participants (e.g., identity, reputation, or past behavior). Pricing must be determined solely from the submitted quantitative data via a pre-defined, deterministic rule. This rule is, in effect, an algorithm, which is precisely what makes the market maker "automated." Consequently, mechanisms that rely on order arrival or identity, like limit order books and bilateral matching, are precluded. 

Notice also that Anonymity is equivalent to the principle of \textbf{Fairness}, which requires that prosumers who contribute identically are paid identically. A market is fair if $x_n = x_m \implies P_n(\bm{x}) = P_m(\bm{x})$ for any two prosumers $n$ and $m$.

\paragraph{Equivalence to Fairness and Aggregation.}

While the function $\Phi$ could depend on the full distribution of trades, a powerful simplification is to make payments depend only on the total aggregate supply ($s$) and demand ($d$). This leads to a more tractable pricing rule, $\Psi$:
\[
P_n(\bm{x}) = \Psi(x_n, s, d).
\]
The mechanisms constructed in this paper will adhere to this aggregation-based pricing principle.

\subsubsection{Coalition-Proofness}

A robust market mechanism should not incentivize strategic manipulation by groups of participants. The axiom of \textbf{coalition-proofness} ensures that a group of prosumers (e.g., all sellers or all buyers) cannot benefit by merging their individual trades into a single, larger trade, or by splitting a large trade into several smaller ones. This guarantees that the mechanism treats individual contributions additively.

\begin{axiom}[Coalition-Proofness]
A market mechanism is \textbf{coalition-proof} if for any coalition of pure sellers $\mathcal{I} \subseteq \{n \in \mathcal{N} \mid x_n > 0\}$ and any coalition of pure buyers $\mathcal{J} \subseteq \{n \in \mathcal{N} \mid x_n < 0\}$, the following additivity conditions hold:
\begin{equation}
\sum_{n \in \mathcal{I}} \Psi(s_n, s, d) = \Psi\left(\sum_{n \in \mathcal{I}} s_n, s, d\right),  \qquad
\sum_{n \in \mathcal{J}} \Psi(-d_n, s, d) = \Psi\left(-\sum_{n \in \mathcal{J}} d_n, s, d\right) \label{eq:cproof}
\end{equation}
\end{axiom}
\noindent{}This additivity requirement implies a linearity restriction on the functional form of the payment function:

\begin{proposition}[Linear Payment Function]
\label{prop:affine-payment}
A market mechanism satisfying aggregation-based pricing is coalition-proof if and only if the payment function $P_n$ is linear in the prosumer’s own supply $s_n$ and demand $d_n$. Specifically, there exist price functions $r(s,d)$ and $c(s,d)$ such that:
\[
P_n(\bm{x}) = s_n \cdot r(s,d) - d_n \cdot c(s,d),
\]
where $s = \sum_{n \in \mathcal{N}} s_n$ and $d = \sum_{n \in \mathcal{N}} d_n$ are the aggregate supply and demand.
\end{proposition}

This proposition is a crucial step in our design. It proves that any anonymous, coalition-proof mechanism must operate by establishing a uniform \textbf{marginal selling price}, $r(s,d)$, for all sellers and a uniform \textbf{marginal buying price}, $c(s,d)$, for all buyers. This result dramatically simplifies our analysis: instead of reasoning about a complex, high-dimensional payment function $\bm{P}(\bm{x})$, we can now focus on the properties of these two intuitive, one-dimensional price functions. All remaining axioms will be defined as constraints on the behavior of $r(s,d)$ and $c(s,d)$.

\subsubsection{Core Economic Axioms}

Having established that a rational market mechanism must use linear pricing, we now introduce three fundamental axioms that govern the behavior of the marginal price functions, $r(s,d)$ and $c(s,d)$. These axioms ensure the mechanism is internally consistent, financially viable, and economically beneficial to its participants.

\paragraph{No-Arbitrage.}
First, the mechanism must be free from internal arbitrage. A participant must not be able to generate a risk-free profit by simultaneously buying and selling energy from the aggregator. This requires that the marginal selling price can never exceed the marginal buying price.

\begin{axiom}[No-Arbitrage] \label{no-arb}
A market mechanism satisfies no-arbitrage if for any pair $(s,d)$:
\begin{equation}
    r(s,d) \le c(s,d).
\end{equation}
\end{axiom}

\paragraph{Budget-Balance.}
A core requirement for a self-sustaining market is that it must be financially viable without external subsidies. The axiom of budget-balance ensures this by requiring that the total payments collected from buyers are at least as great as the total revenues paid out to sellers.

\begin{axiom}[Budget-Balance] \label{BB} 
A market mechanism is budget-balanced if for any state $(s,d)$:
\begin{equation}
    d \cdot c(s,d) \ge s \cdot r(s,d).
\end{equation}
If the condition holds with equality, the mechanism is \textbf{exactly budget-balanced}.
\end{axiom}

\paragraph{Individual Rationality (IR).}
Finally, the mechanism must provide an incentive for prosumers to participate over their outside option of trading directly with the grid. The prices offered by the local market must be strictly better than the grid's prices when local supply and demand can be matched, and they should converge to the grid's prices when the community has a net surplus or deficit that must be cleared externally.

\begin{axiom}[Individual Rationality] \label{ax:IR} 
A market mechanism satisfies Individual Rationality (IR) if the price functions $r(s,d)$ and $c(s,d)$ adhere to the following boundary conditions:
\begin{equation}
\label{eq:IR}
\begin{cases}
r(s,d) \geq \underline{\lambda}, & \text{if } s < d, \\
r(s,d) = \underline{\lambda}, & \text{if } s \ge d,
\end{cases}
\quad \text{and} \quad
\begin{cases}
c(s,d) \leq \overline{\lambda}, & \text{if } s > d, \\
c(s,d) = \overline{\lambda}, & \text{if } s \le d.
\end{cases}
\end{equation}
\end{axiom}

This IR condition is the economic justification for the local market's existence and is achieved in AMMs via \textbf{concentrated liquidity}.

\paragraph{Remark on Continuity.}
A direct consequence of Individual Rationality is that the total cost and revenue functions, $C_n(\bm{x})$ and $R_n(\bm{x})$, must be Lipschitz continuous with constants $\overline{\lambda}$ and $\underline{\lambda}$ respectively, implying that for any two allocations $\bm{x}$ and $\bm{x}'$ that differ only for prosumer $n$:
\begin{align*}
    |C_n(\bm{x}) - C_n(\bm{x}')| &\le \overline{\lambda} \cdot |d_n - d'_n| \\
    |R_n(\bm{x}) - R_n(\bm{x}')| &\ge \underline{\lambda} \cdot |s_n - s'_n|
\end{align*}

\subsubsection{Axioms on Qualitative Price Properties}
The final set of axioms governs the dynamic behavior of the price functions, ensuring they respond to market conditions in an economically intuitive way.

\paragraph{Monotonicity (Positive Prices).}
To ensure that trades are meaningful, selling energy must generate revenue and buying energy must incur a cost. This is guaranteed if the marginal prices are strictly positive.

\begin{axiom}[Monotonicity]
A market mechanism is \textbf{monotonic} if its price functions are strictly positive:
\begin{equation}
    r(s,d) > 0 \quad \text{and} \quad c(s,d) > 0,
\end{equation}
for all $s,d > 0$.
\end{axiom}

\paragraph{Responsiveness.}
A well-behaved market should naturally counteract imbalances through price signals. When aggregate supply increases, prices should strictly decrease to encourage demand, and when aggregate demand increases, prices should strictly increase to encourage supply. These conditions ensure that prosumers' actions are \textbf{strategic substitutes}, preventing extreme coordination failures. This property also provides incentives for \textbf{peak shaving}.

\begin{axiom}[Responsiveness] \label{ax:resp}
A mechanism is \textbf{responsive} if its price functions' partial derivatives satisfy:
\begin{equation}
\frac{\partial c}{\partial s} < 0, \quad
\frac{\partial c}{\partial d} > 0, \quad
\frac{\partial r}{\partial s} < 0, \quad \text{and} \quad
\frac{\partial r}{\partial d} > 0.
\end{equation}
\end{axiom}

\noindent{}Notice that Responsiveness combined with Individual Rationality (\cref{eq:IR}) imply that internal prices are strictly more favorable than the grid bounds whenever the community is in a net imbalance:

\begin{equation}
\begin{cases}
\label{eq:prices_IR+RESP}
r(s,d) > \underline{\lambda}, & \text{if } s < d, \\
r(s,d) = \underline{\lambda}, & \text{if } s \ge d,
\end{cases}
\quad \text{and} \quad
\begin{cases}
c(s,d) < \overline{\lambda}, & \text{if } s > d, \\
c(s,d) = \overline{\lambda}, & \text{if } s \le d.
\end{cases}
\end{equation}

\paragraph{Remark on Concavity.}
The Responsiveness axiom does not, by itself, imply the convexity or concavity of the prosumers' total cost and revenue functions, However, the axiom is closely related to the geometric properties of the underlying market mechanism. For example, in the AMM constructions introduced later in \cref{AMM-construction} it is a direct consequence of implementing a strictly \textbf{quasi-concave bonding curve}. 

\paragraph{Homogeneity.}
For many theoretical models, it is useful to assume that prices depend only on the \textit{ratio} of supply to demand, not on their absolute scale. A market with 100 kWh of supply and 50 kWh of demand should have the same price as one with 10 kWh of supply and 5 kWh of demand. This is the property of homogeneity of degree zero.

\begin{axiom}[Homogeneity] \label{ax:homo}
A mechanism is \textbf{homogeneous} if its price functions are homogeneous of degree zero, meaning for any scaling factor $\alpha > 0$:
\begin{equation}
    r(\alpha s, \alpha d) = r(s,d) \quad \text{and} \quad c(\alpha s, \alpha d) = c(s,d).
\end{equation}
This implies the prices can be written as a function of the supply-to-demand ratio, $s/d$.
\end{axiom}

The Homogeneity axiom plays a key role in \cref{sec:equilibrium} (Equilibrium Analysis) since scale invariance of prices implies proportional scaling of the total prosumer community welfare in the population size $N$. This property is thus a necessary condition to approximate the stochastic finite-player game implemented by the AMM with its deterministic Mean-Field limit as $N \to \infty$.

\section{AMM Construction and Applications}
\label{AMM-construction}

We now present a procedure to construct \emph{Automated Market Makers (AMMs)} for energy trading that satisfies the design principles outlined in Section \ref{sec:market-design}. While other market mechanisms exist, an AMM offers a streamlined solution for local energy markets. Prices are quoted automatically after energy transfers occur, making the market user-friendly as prosumers do not need to commit to a trade schedule ahead of time. Furthermore, AMMs mitigate the lack of liquidity common in small residential communities by pooling supply and demand, which can lead to tighter bid-ask spreads and faster matching than bilateral exchange mechanisms.

\paragraph{CFMM-style construction.} Our approach is to derive prices from a \textit{trading function}, $\psi(E,M)$, which defines a potential over the money ($M$) and energy ($E$) in the AMM's pool. This logic is akin to the Constant-Function Market Makers (CFMMs) popular in decentralized finance \citep[e.g.,][]{AngerisChitra2020AFT,AngerisEtAl2022Handbook}. The core of a CFMM is an invariant relationship, $\psi(E,M) = K$, where $K$ is constant for a trading session. Standard CFMM theory indicates that the marginal price of energy, $P$, is the slope of the function's level sets:
\[
\rho(E,M) \;=\; \abs{\frac{\partial_E \psi}{\partial_M \psi}}.
\]
A canonical starting point is the constant-product function $\psi_{CPMM}(E,M) = E \cdot M \equiv K$, used by the original Uniswap protocol, which yields a price of $\rho=M/E$. While this specific function is not directly compatible with our axioms, its underlying principle forms the foundation for our adapted models.

\paragraph{Infrastructural and Operational Adaptations.}
Adapting a financial CFMM to a physical commodity market requires two key modifications. First, an energy AMM must be embedded in physical infrastructure. Unlike purely digital assets, energy is delivered through a distribution grid, and prosumers' contributions are mediated by trusted devices like \textbf{smart meters} that record net energy flows. The AMM must also remain consistent with the external market, importing deficits and exporting surpluses at the grid's prices.

Second, the operational logic is fundamentally different. An energy market lacks the traditional liquidity providers (LPs) of DeFi, as prosumers themselves supply the energy that initializes the pool for each session. This inspires our mechanism guided by a dynamic bonding curve. Through a process of \textbf{“re-anchoring,”} the level set of the trading function is shifted at the start of each session based on the available energy and market conditions. The market clearing process then proceeds in two steps: (1) internal matching of $\Delta E = \min\{d_t,s_t\}$ units of energy along the newly anchored curve, and (2) external balancing of any residual with the grid.

\paragraph{Remark on Implementation.}
While we often consider a blockchain setting where a smart contract can automate this process, the construction does not require it. The AMM can be constructed by any trusted operator able to measure aggregate flows and apply the pricing rules. These physical requirements—and the need to prevent sensor manipulation—are key challenges in the broader field of Decentralized Physical Infrastructure (DePIN) projects \citep{milionis2025depin}.

\subsection{Energy Provision and Trade}

We will now construct the AMM starting from the constant-product invariant and adjusting it progressively. As mentioned previously, the bonding curve is now dynamic and adjusted at each trading session. We let $\{\psi_t(\cdot, \cdot)\}_{t\in \mathcal T}$ denote a family of trading functions parameterized by time and $K_t$ the value achieved at the target level set for time $t$.

\begin{figure}[h!]
    \centering
    \includegraphics[width=.6\linewidth]{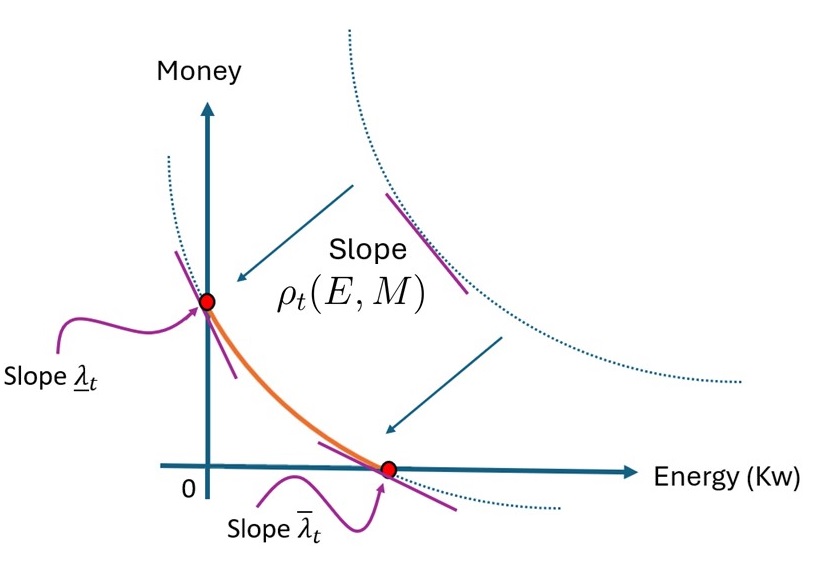}
    \caption{Constraining the AMM with Upper/Lower Bound Prices}
    \label{fig:translation}
\end{figure}

\paragraph{Concentrated Liquidity} To prevent arbitrage with the main grid’s feed-in tariff \(\underline{\lambda}_t\) and retail price \(\overline{\lambda}_t\), the AMM must \emph{concentrate liquidity} and quote a local price \(\rho_t(E,M) \in [\underline{\lambda}_t, \,\overline{\lambda}_t]\). Figure~\ref{fig:translation} shows how we shift the bonding curve to ensure the (internal) energy price cannot exceed or drop below these boundary rates. 

\paragraph{Batching and re-anchoring} At the beginning of the trading session $t$, prosumers charge the pool with the total energy supplied $s_t$. The pool starts with no monetary reserves, so its initial state is $(E_0,M_0)= (s_t,0)$. We use this initial state along with the main grid's prices $(\underline{\lambda}_t,\overline{\lambda}_t)$ to calculate a session-specific level set 
\[
K_t \;\equiv\; K\bigl(s_t,\underline{\lambda}_t,\overline{\lambda}_t\bigr).
\]
thus fixing a bonding curve for the trading session: $$\left\{(E,M) \in \mathbb{R}^2_+: \psi_t(E,M) = K\bigl(s_t,\underline{\lambda}_t,\overline{\lambda}_t )\right\}.$$

\begin{figure}[h!]
    \centering
    \includegraphics[width=\linewidth]{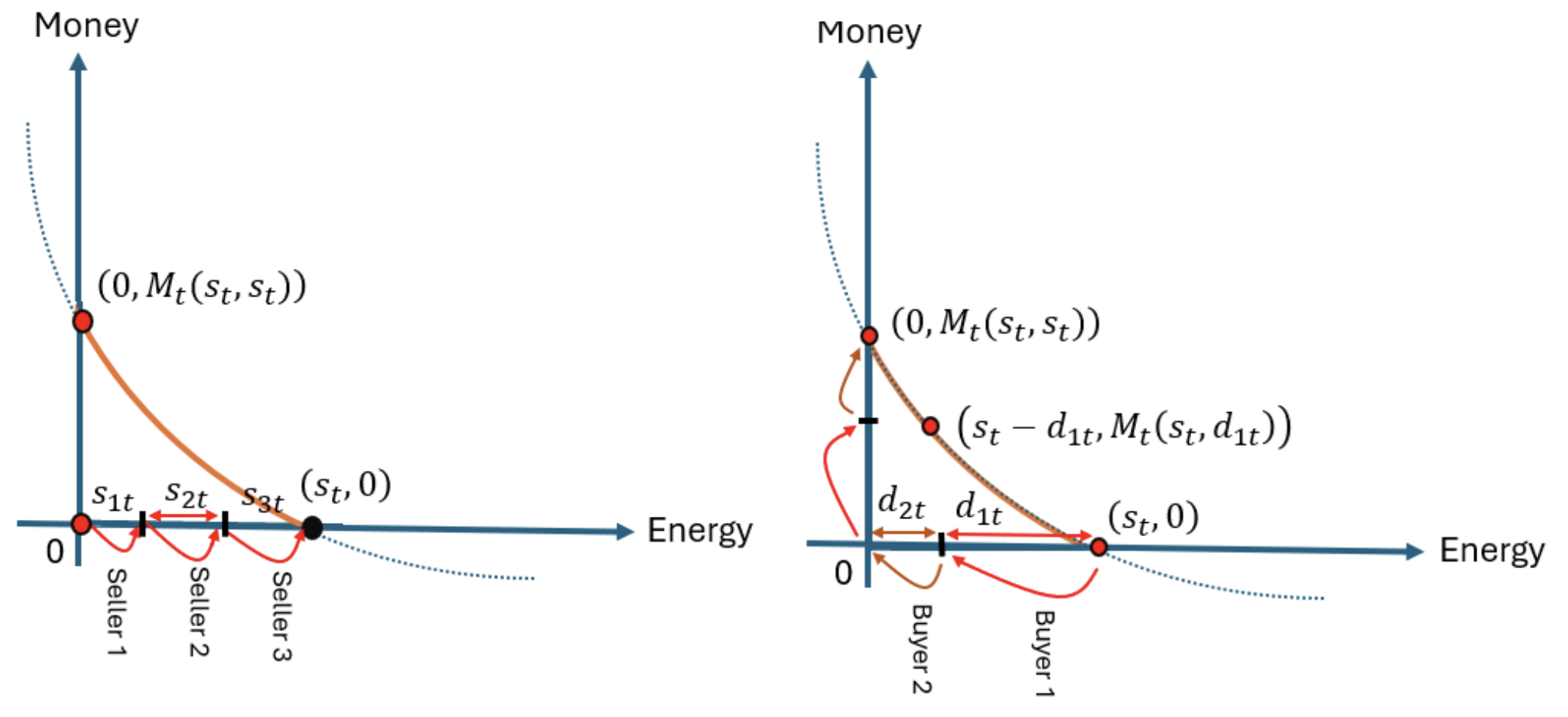}
    \caption{(Left) Pool Initialization with Supply; (Right) Depletion of Capacity in Exchange for Money}
    \label{fig:sessions}
\end{figure}

The bonding curve determines the amount of money the pool collects as a function of the energy reserve (given the energy available at the start of the trading session). Once the session's bonding curve is fixed, we can derive  how much money will the AMM collect from prosumers:  
\begin{equation}
M(E_0,\Delta E) \;=\;\psi^{-1}_{E= E_0 - \Delta E}\left(K(s_t,\underline{\lambda}_t,\overline{\lambda}_t)\right),
\label{eq:value}    
\end{equation}
this can be computed  given the initial supply $E_0 = s_t$ can be calculated from \cref{eq:value} setting $\Delta E = \min \{s_t ,d_t\}$. Figure~\ref{fig:sessions} illustrates how capacity is deposited and depleted for the case of three sell orders and two buy orders, starting with an empty battery and no initial capital.

\begin{figure}[h!]
    \centering
    \includegraphics[width=\linewidth]{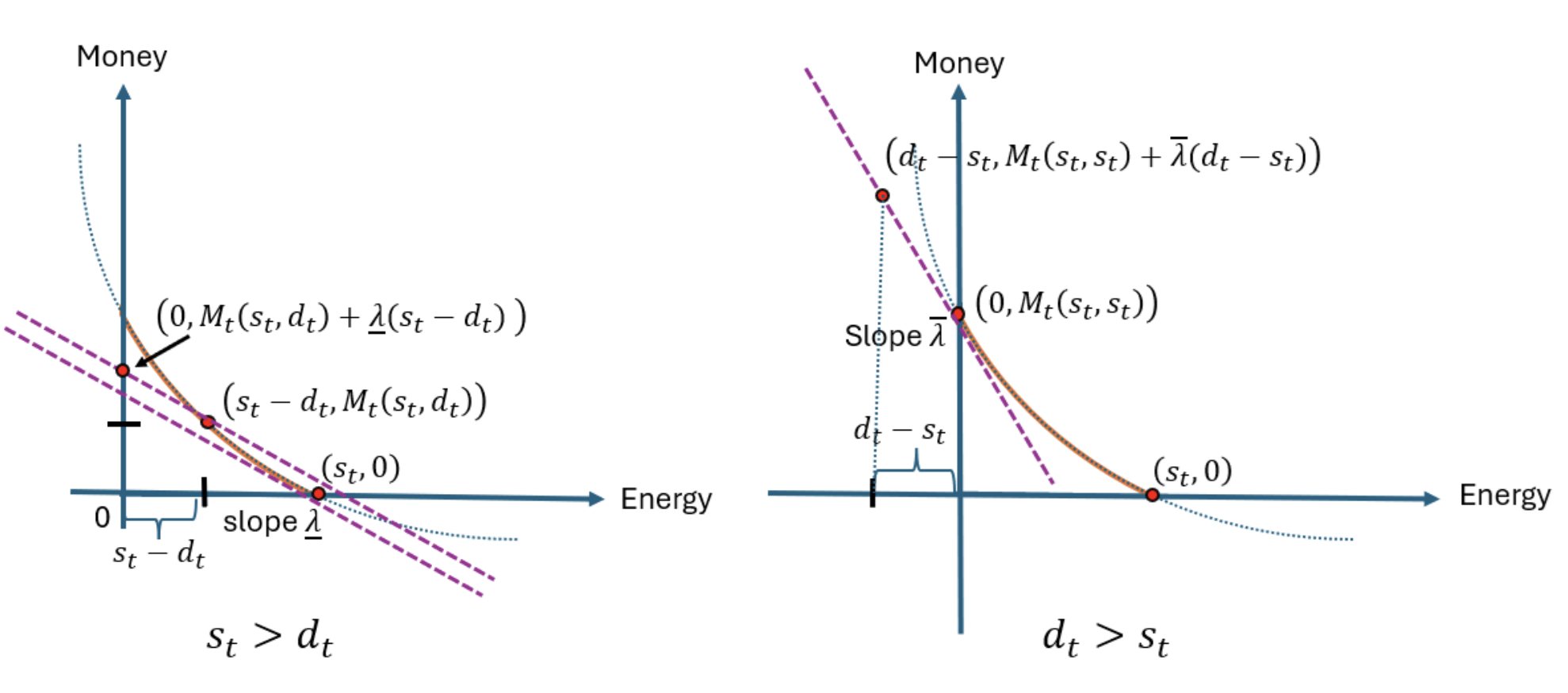}
    \caption{Managing Surpluses (left) and Deficits (right) via Tangent-Line Shifts}
    \label{fig:imbalance}
\end{figure}

\paragraph{Imbalance Management} After the internal P2P exchange is complete, the AMM must settle any net energy imbalance by trading with the main grid. This process handles two distinct scenarios, both of which can be understood geometrically using tangent lines to the bonding curve, as illustrated in Figure~\ref{fig:imbalance}.

When local supply exceeds demand, an energy surplus of $(s_t - d_t)$ remains after all internal demand has been met. At this point, the pool's state is $(s_t - d_t, {M}_t(s_t,d_t))$. This surplus is then sold to the main grid at the feed-in tariff $\underline{\lambda}_t$. The additional revenue generated from this external sale is $\underline{\lambda}_t (s_t - d_t)$. Geometrically, this corresponds to the monetary value gained by moving along the tangent line from the final internal trading state, as shown in the left panel of Figure~\ref{fig:imbalance}.

When local demand exceeds supply, the entire internal supply $s_t$ is consumed. The internal trading phase concludes with the pool in a state of full depletion: $(0, M_t(s_t, s_t))$. To satisfy the remaining demand, the AMM must import the energy shortfall of $(d_t - s_t)$ from the main grid at the retail price $\overline{\lambda}_t$. The additional cost incurred is $\overline{\lambda}_t (d_t - s_t)$. Geometrically, this cost is represented by the tangent line at the point of full depletion, as illustrated in the right panel of Figure~\ref{fig:imbalance}.

\paragraph{Payments.}
Finally, the AMM distributes costs and revenues among prosumers. The smart contract first categorizes prosumers as net buyers or net sellers based on their energy contribution. It then calculates the total costs for buyers and total revenues for sellers, distributing them proportionally based on each prosumer's share of the total demand ($d_{nt}/d_t$) or supply ($s_{nt}/s_t$), respectively. This distribution process depends on the net energy balance of the community, leading to three distinct cases:

\noindent\textbf{Case I: $d_t = s_t$ (Balanced Market).}
The total cost and revenue are equal to the pool's final monetary value, $C_{total} = R_{total} = M_t(s_t,s_t)$. The resulting per-unit prices are:
\begin{equation}
    c_t(s_t, d_t) = \frac{M_t(s_t,s_t)}{d_t}, \phantom{+ \overline{\lambda}_t(d_t - s_t)} \qquad 
    r_t(s_t, d_t) = \frac{M_t(s_t,s_t)}{s_t}
\end{equation}

\noindent\textbf{Case II: $d_t < s_t$ (Excess Supply).}
The total cost for buyers is determined by the internal trade, $C_{total} = M_t(s_t, d_t)$. The total revenue for sellers includes this amount plus the income from selling the surplus $(s_t - d_t)$ to the grid at price $\underline{\lambda}_t$. The prices are:
\begin{equation}
    c_t(s_t, d_t) = \frac{M_t(s_t, d_t)}{d_t}, \qquad 
    r_t(s_t, d_t) = \frac{M_t(s_t,d_t) + \underline{\lambda}_t(s_t - d_t)}{s_t}
\end{equation}

\noindent\textbf{Case III: $d_t > s_t$ (Excess Demand).}
The total revenue for sellers is determined by the full depletion of their supply, $R_{total} = M_t(s_t, s_t)$. The total cost for buyers includes this amount plus the cost of importing the deficit $(d_t - s_t)$ from the grid at price $\overline{\lambda}_t$. The prices are:
\begin{equation}
    c_t(s_t, d_t) = \frac{M_t(s_t,s_t) + \overline{\lambda}_t(d_t - s_t)}{d_t}, \qquad
    r_t(s_t, d_t) = \frac{M_t(s_t,s_t)}{s_t}
\end{equation}

While payments can be generalized to other repartition schemes, the proportional method is unique in ensuring the market is coalition-proof, as established in the previous section. Furthermore, if the AMM is not required to be strictly budget-balanced in every session, a dynamic fee can be introduced based on aggregate supply or demand.

\subsection{Bonding Curves and Other Pricing Functions}

\begin{figure}[H]
    \centering
    \includegraphics[width=\linewidth]{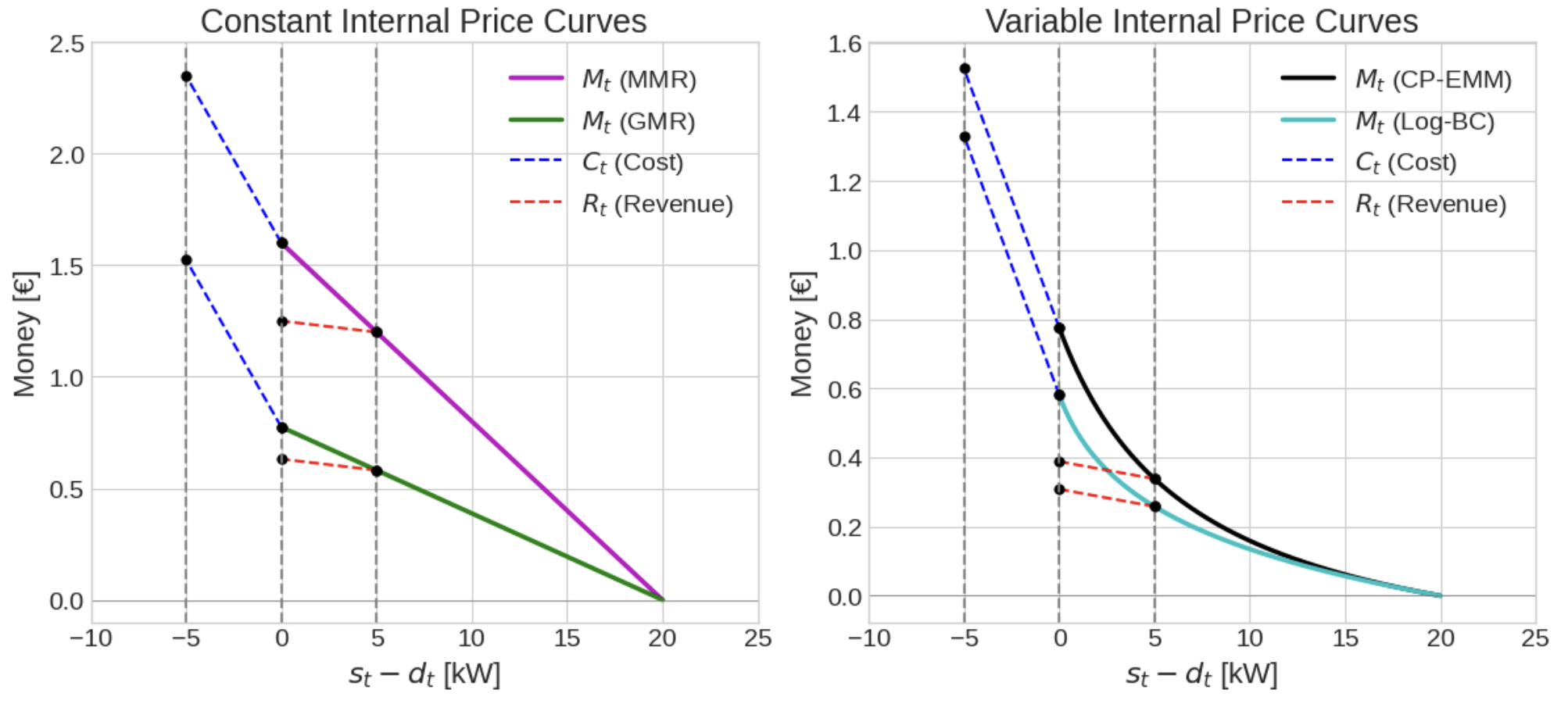}
    \caption{Comparison of Bonding Curves}
    \label{fig:price_curves}
\end{figure}

We now present three applications of the general construction from the previous section: linear, hyperbolic, and logarithmic bonding curves. Figure~\ref{fig:price_curves} compares the resulting buy ($c_t$) and sell ($r_t$) price curves for each model as a function of the supply-demand ratio $y_t =s_t/d_t$.

\begin{figure}[H]
    \centering
    \includegraphics[width=\linewidth]{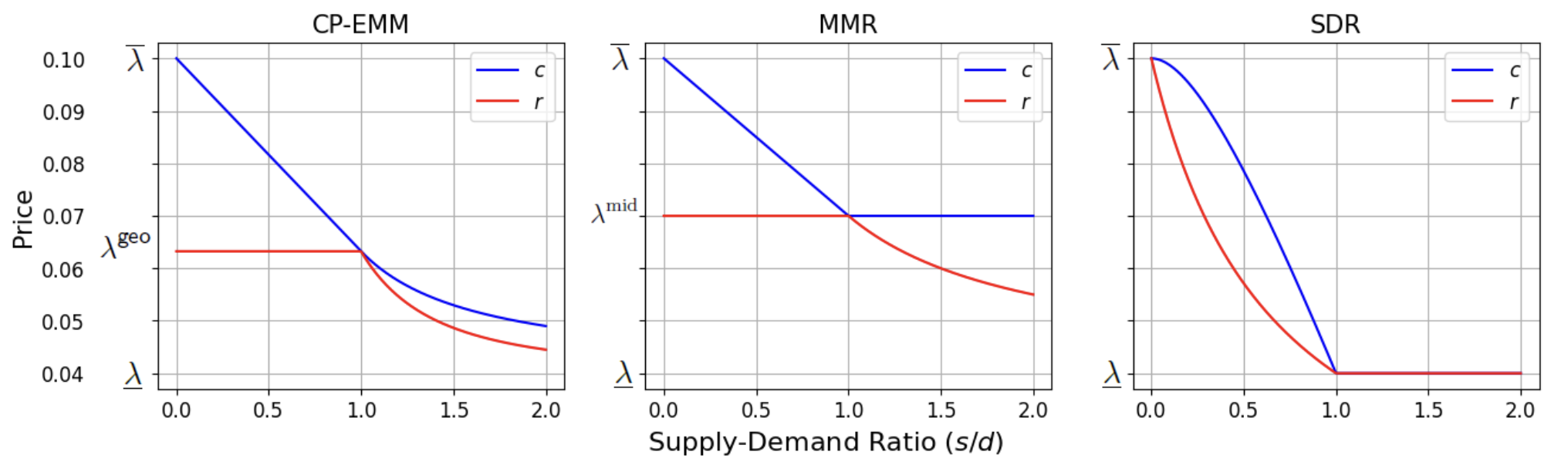}
    \caption{Comparison of buy ($c$) and sell ($r$) price curves for the Constant Price (MMR), Hyperbolic (CP-EMM), and  SDR price mechanism, plotted against the supply-demand ratio.}
    \label{fig:price_curves_comparison}
\end{figure}

\subsubsection*{Linear Bonding Curve (Constant Internal Price)}
To implement a constant internal price $ \rho_t(E,M) = \rho_t \in [\underline{\lambda_t}, \overline{\lambda_t}]  $ for all internally matched energy, we use a linear trading function. The trading function and corresponding session invariant are:
\begin{equation}
   \psi_t(M,E) = M + \rho_t E, \qquad K_t = \rho_t s_t.
\end{equation}
Solving for the monetary reserve $M$ yields a simple, linear Pool Value Function:
$$
M_t(s_t, \Delta E) = \rho_t \Delta E.
$$
This simple form gives rise to the following price functions after accounting for imbalances with the grid:
\begin{equation}
 c_t(y) = \rho_t + (\overline{\lambda}_t - \rho_t) \left(1- y\right)^+, \qquad 
 r_t(y) = \rho_t - (\rho_t - \underline{\lambda}_t) \left(1- 1/y\right)^+.
\end{equation}

\subsubsection*{Hyperbolic Bonding Curve (CP-EMM)}
This AMM is constructed from a hyperbolic trading function adapted from the standard constant-product formula. We first define a session-specific parameter \footnote{The symbol $\chi$ will be redefined later on to represent collections of Lagrange multipliers and representative prosumer types.}$$\chi_t \equiv \frac{\sqrt{\overline{\lambda}_t \underline{\lambda}_t}}{\sqrt{\overline{\lambda}_t} - \sqrt{\underline{\lambda}_t}}.$$The trading function and session invariant are then defined as:
\begin{equation}
   \psi_t(M,E) = \left(M +  \sqrt{ K_t  \underline{\lambda}_t}\right) \left(E + \sqrt{ \frac{K_t}{\overline{\lambda}_t} }\right), \qquad K_t = (s_t \chi_t)^2.
\end{equation}
This construction gives rise to pricing based on an adjusted geometric-market-rate. When the market is balanced ($y=1$), it implements an over-provision penalty to incentivize an efficient match between local generation and consumption. The final price functions are:
\begin{align}
r_t(y) &= \chi_t \left[ \frac{\chi_t}{(1 - 1/y)^+ + \frac{\chi_t}{\sqrt{\overline{\lambda}_t}}} - \sqrt{\underline{\lambda}_t} \right] + (1 - 1/y)^+ \underline{\lambda}_t, \\[12pt]
c_t(y) &= y \cdot \chi_t \left[ \frac{\chi_t}{(1 - 1/y)^+ + \frac{\chi_t}{\sqrt{\overline{\lambda}_t}}} - \sqrt{\underline{\lambda}_t} \right] + (1 - y)^+ \overline{\lambda}_t.
\end{align}

In CREF APPENDIX, we show that the hyperbolic trading function gives rise to pricing based on an \textit{adjusted} geometric-market-rate (GMR): when the market is perfectly balanced ($s_t = d_t$), the effective price for internally traded energy is the geometric mean of the grid prices, $\lambda_t^{\textsf{geo}} = \sqrt{\underline{\lambda}_t \cdot \bar{\lambda}_t }$. When supply exceeds demand ($s_t > d_t$), the AMM implements an over-provision penalty to incentivize an efficient match between local generation and consumption. 

\subsubsection*{Logarithmic Bonding Curve}

This model is derived from a price function that depends on the ratio of total supply to demand withdrawn.\footnote{The expression is derived from a price function of the form $P(E_0 - \Delta E, M(E_0,\Delta E)) = \left( a (1 - \Delta E/E_0) + b \right)^{-1}$, where the constants $a,b \in \mathbb{R}$ are determined by the boundary conditions $P(E_0,0) = \underline{\lambda}_t$ and $P(0, M(E_0,E_0)) = \overline{\lambda}_t$.} The marginal price for internal trades is:
\begin{equation*}
\rho_t(E_0, M(E_0,\Delta E)) = \frac{\overline{\lambda}_t \underline{\lambda}_t}{(\overline{\lambda}_t - \underline{\lambda}_t)(1 - \Delta E/E_0)^{+} + \underline{\lambda}_t}
\end{equation*}
Integrating this price function yields the Money Reserve function:
\begin{equation*}
M_t(E_0, \Delta E) =  E_0 \frac{\overline{\lambda}_t \underline{\lambda}_t}{\overline{\lambda}_t - \underline{\lambda}_t} \ln\left( \frac{\overline{\lambda}_t}{(\overline{\lambda}_t - \underline{\lambda}_t)(1 - \Delta E / E_0)^+ + \underline{\lambda}_t} \right)
\end{equation*}
The final per-unit prices for sellers ($r_t$) and buyers ($c_t$) are determined by substituting this $M_t$ into the general payment rules:
\begin{align}
r_t(y) &= \frac{\overline{\lambda}_t \underline{\lambda}_t}{\overline{\lambda}_t - \underline{\lambda}_t} \cdot \ln\left( \frac{\overline{\lambda}_t}{(\overline{\lambda}_t - \underline{\lambda}_t)(1 - 1/y )^+ + \underline{\lambda}_t} \right) + (1 - 1/y )^+ \underline{\lambda}_t \\[12pt]
c_t(y) &= y  \cdot \frac{\overline{\lambda}_t \underline{\lambda}_t}{\overline{\lambda}_t - \underline{\lambda}_t} \ln\left( \frac{\overline{\lambda}_t}{(\overline{\lambda}_t - \underline{\lambda}_t)(1 - 1/y)^+ + \underline{\lambda}_t} \right) + (1 - y)^+ \overline{\lambda}_t
\end{align}
The behavior of this pricing rule is similar to the Hyperbolic one. While a corresponding trading function $\psi_t$ exists, its form is convoluted. It is therefore more intuitive in this case to define the mechanism directly via its bonding curve.
\subsection{Other Pricing Rules from the Literature}

We now review several common pricing mechanisms from the energy sharing literature and evaluate them against our axiomatic framework. We will see that some of them can be generated by the above trading functions. 

\paragraph{Bill Sharing (BS).}
An intuitive mechanism, Bill Sharing (BS) assigns the community's total grid costs to buyers and total grid revenues to sellers \cite{XIONG2022300}. The effective prices are $c_t(y) = \overline{\lambda}_t (1 - y)^+$ and $r_t(y) =\underline{\lambda}_t (1 - 1/y)^+$. While simple, this mechanism is structurally flawed. It violates \textbf{Individual Rationality}, as a prosumer is not guaranteed a better price than their outside option. For example, a seller receives zero revenue if the community has a net deficit.

\paragraph{Mid-Market Rate (MMR).}
The Mid-Market Rate (MMR) rule sets the internal trading price to the average of the grid's bid and ask prices \cite{muntasir2023developing}. This mechanism is a special case of our \textbf{Linear Bonding Curve} model where the internal price is set to the arithmetic mean, $\rho_t = (\overline{\lambda}_t + \underline{\lambda}_t)/2$. When imbalances occur, the cost or revenue is shared proportionally. The linear model is a general framework that also allows for other constant-price rules, such as the Geometric-Market-Rate (GMR), by setting $\rho_t  = \sqrt{\overline{\lambda}_t \underline{\lambda}_t}$.

\paragraph{Supply-Demand Ratio (SDR).}
The SDR mechanism defines a responsive price that interpolates between the grid's prices based on the supply-demand ratio $y$ \cite{XIONG2022300}. The selling price is a decreasing function of market surplus, and the buying price is determined by the budget-balance condition:
\begin{align*}
    r_t(y) &= 
    \begin{cases} 
        \dfrac{\overline{\lambda}_t \cdot \underline{\lambda}_t}{ (\overline{\lambda}_t - \underline{\lambda}_t) y + \underline{\lambda}_t} & \text{if } 0 \leq y \leq 1, \\[12pt]
           \underline{\lambda}_t & \text{if } y > 1,
    \end{cases},    \qquad
    c_t(y) &= 
    \begin{cases} 
        y \cdot r(y) + (1 - y) \overline{\lambda}_t   & \text{if } 0 \leq y \leq 1, \\[6pt]
        r(y) & \text{if } y > 1.
    \end{cases}
\end{align*}
Table~\ref{tab:axiom_comparison} summarizes our axiomatic analysis. Both MMR and SDR are robust mechanisms that satisfy all axioms unconditionally. Our CP-EMM, representative of AMMs derived from strictly quasi-concave potentials (e.g., Log-EMM), also satisfies the axioms, with a single conditional exception for Monotonicity.

\begin{table}[H]
\centering
\renewcommand{\arraystretch}{1.3}
\setlength{\tabcolsep}{12pt} 

\begin{tabular}{lcccc}
\toprule
\textbf{Axiom} & \textbf{BS} & \textbf{MMR} & \textbf{SDR} & \textbf{CP-EMM} \\
\midrule
Anonymity        & \cmarkg & \cmarkg & \cmarkg & \cmarkg \\
Coalition-Proof  & \cmarkg & \cmarkg & \cmarkg & \cmarkg \\
No-Arbitrage     & \cmarkg & \cmarkg & \cmarkg & \cmarkg \\
Budget-Balance   & \cmarkg & \cmarkg & \cmarkg & \cmarkg \\
Individual Rationality & \xmarkr & \cmarkg & \cmarkg & \cmarkg \\
Monotonicity     & \xmarkr & \cmarkg & \cmarkg & \condpassb \\
Responsiveness   & \cmarkg & \cmarkg & \cmarkg & \condpassb \\
Homogeneity      & \cmarkg & \cmarkg & \cmarkg & \cmarkg \\
\bottomrule
\end{tabular}
\caption{Comparison of pricing mechanisms against key design axioms. The CP-EMM, the Log-bonding curve and the GMM satisfy the same axioms.}
\label{tab:axiom_comparison}
\flushleft
\condpassb \footnotesize{Axiom holds only if $\underline{\lambda}_t > 0$.}
\end{table}

\subsection{Market Design Guarantees}

We now formalize the link between the properties of the underlying trading function and the axiomatic compliance of the resulting AMM.

\begin{proposition}[Axiom Compliance via Geometric Construction]
\label{prop:general-compliance}
Let an AMM be constructed using batch execution, proportional payment rules, and with its liquidity concentrated \textbf{strictly within} the grid's prices, $\rho_t(E,M) \in (\underline{\lambda}_t, \overline{\lambda}_t)$. Then, a market mechanism satisfies the axioms of Anonymity, Coalition-Proofness, Budget-Balance, Individual Rationality, No-Arbitrage, Monotonicity, Homogeneity, and Responsiveness if and only if the underlying trading function $\psi(E,M)$ is \textbf{(i) strictly increasing in E and M, (ii) homothetic, and (iii) quasi-concave}. 
\end{proposition}

Both directions of the equivalence arise from the  interaction between the axioms of \cref{sec:market-design} and the geometry of the trading function. 

In the forward direction, once batch execution and proportional payments are fixed, the mechanism is entirely determined by the trajectory traced along a single bonding curve during internal matching. Increasingness of the curve then guarantees that both reserves represent positively valued assets, ensuring strictly positive prices. Homotheticity guarantees that only the supply--demand ratio matters, yielding homogeneous pricing. Quasi-concavity ensures convex level sets and thus a price schedule that reacts monotonically to imbalances, producing the responsiveness required for peak shaving and strategic substitutability. Finally, concentrating liquidity strictly within the grid's bid--ask interval ensures that all internal trades occur at strictly better prices than the outside option, enforcing individual rationality, preventing arbitrage with the main grid, and, together with proportional sharing, guaranteeing budget balance.

In the reverse direction, the axioms themselves impose strong shape restrictions that precisely recover the geometric structure of such a trading function. Anonymity eliminates dependence on order identity and forces the mechanism to aggregate trades into a single batch, while coalition-proofness requires that payments be linear in individual quantities, leaving proportional sharing as the unique feasible allocation rule. No-arbitrage, individual rationality, and budget-balance together imply the existence of a well-defined \emph{demand curve}, describing how much energy the mechanism is willing to hold at a given price. Under the axioms, this curve must be flat outside the interval $[\underline{\lambda}_t, \overline{\lambda}_t]$ and strictly decreasing within it; as shown in the Myersonian framework of \citet{milionis2023myersonian}, such curves are in one-to-one correspondence with strictly increasing, homothetic, quasi-concave trading potentials whose liquidity is concentrated inside the same interval. Thus the axioms force the mechanism to admit exactly the geometric representation used in our construction, completing the necessity direction.

\paragraph{Remark.} Since MMR and SDR also satisfy the axioms, we can simply derive the quasi-concave curves for them. It turns out that they are all a special type of a linear trading function. Specifically, they can all be represented as $M+\rho E = c$, where $\rho$ is $(\overline{\lambda}_t + \underline{\lambda}_t)/2$ for the arithmetic MMR, $\sqrt{\overline{\lambda}_t \underline{\lambda}_t}$ for the geometric MMR, and $\dfrac{\overline{\lambda}_t \cdot \underline{\lambda}_t}{ (\overline{\lambda}_t - \underline{\lambda}_t) \min(1,y) + \underline{\lambda}_t}$ for the SDR pricing rules. 


\section{Prosumer Optimization}
\label{sec:prosumer_model}

Having defined a market mechanism, we now model the behavior of individual prosumers operating within it. This section develops a dynamic optimization model for a single prosumer facing the AMM prices. We will use this formalism analyze a prosumer's optimal decision (their best-response) given a belief on the price path. This forms the basis for the equilibrium characterization of Section \ref{sec:equilibrium}. We also use the prosumer model to quantify the benefits of participating in the AMM compared to direct grid trading in \cref{quantitative}.

\subsection{Prosumer Model}

We model the prosumer's decision-making process using (epoch-based) discrete-time dynamic programming. Time unfolds over epochs $e=1, 2, \dots$, here representing days. Each epoch is divided into $T$ timesteps $t=1, \dots, T$, each of duration $\mathrm{\Delta}$ hours (e.g., $\Delta=0.25$ for 15-minute intervals, $T=96$). Table \ref{tab:prosumer_notation} summarizes the notation needed to define the prosumer's optimization problem.

\paragraph{Prosumer Characteristics.}
A prosumer $n$ is characterized by a set of static parameters and dynamic (state) variables. The static parameters include physical constraints—battery energy capacity $B_n$ (kWh), battery power limit $K_n$ (kW), and grid connection limit $X_n$ (kW)—and parameters governing energy generation and consumption profiles. Within each epoch $e$, the prosumer observes their initial battery state $b_{n0}(e)$ (kWh) and forecasts for their net load profile over the $T$ timesteps. This profile consists of local generation $\omega_{nt}$ (kW) and baseline consumption $\alpha^{\text{base}}_{nt}$ (kW). Additionally, the prosumer has a total flexible energy requirement $\alpha^{\text{flex}}_{n}$ (kWh) that must be met cumulatively over the epoch.

\paragraph{Decision Variables.}
In each timestep $t$ of epoch $e$, the prosumer chooses three power flows: consumption $p_{nt} \ge 0$, battery charge/discharge $k_{nt} \in [-K_n, K_n]$ (with $k_{nt}>0$ indicating charging), and net grid trade $x_{nt} \in [-X_n, X_n]$. We decompose $x_{nt}$ into non-negative power sold, $s_{nt} = \max\{x_{nt}, 0\}$, and power bought, $d_{nt} = \max\{-x_{nt}, 0\}$.

\paragraph{Power Balance.}
At each timestep $t$, the instantaneous power flows must balance. Formally,  local generation ($\omega_{nt}$) plus power imported from the grid ($d_{nt}$) must equal the power consumed locally ($p_{nt}$), power stored in the battery ($k_{nt}$, positive if charging), and power exported to the grid ($s_{nt}$): $\omega_{nt} + d_{nt} = p_{nt} + k_{nt} + s_{nt}$ (kW).

\paragraph{Consumption Constraints.}
Power consumption $p_{nt}$ must always cover the essential, non-shiftable baseline load, $p_{nt} \ge \alpha^{\text{base}}_{nt}$ (kW). Additionally, the total energy consumed \textit{above} this baseline over the entire epoch must exactly meet the flexible load target: $\sum_{t=1}^{T} (p_{nt} - \alpha^{\text{base}}_{nt}) = \alpha^{\text{flex}}_{n}$.\footnote{It is  equivalent to have $\sum_{t=1}^{T} (p_{nt} - \alpha^{\text{base}}_{nt}) \geq \alpha^{\text{flex}}_{n}$ since the constraint will bind at the optimum.} This cumulative constraint models deferrable energy needs, such as electric vehicle charging or running specific appliances, that must be completed within the epoch but whose timing can be optimized.\footnote{For notational simplicity we treat the timestep as the unit of time in the constraints. This makes energy quantities (kWh) numerically equivalent to average power (kW). To ensure the objective function is correctly calculated in Euros,  we multiply the price parameters by the timestep duration $\Delta$. This converts hourly tariffs (€/kWh) into effective prices per unit of power (€/kW).} 

\paragraph{Battery Constraints.}
For prosumers equipped with storage ($B_n > 0$), the battery's state of charge (SoC) $b_{nt}$ (kWh) at the end of timestep $t$ evolves according to the battery dynamics $b_{nt} = b_{n,t-1}(e) + k_{nt}$, where $k_{nt}$ is the charging ($>0$) or discharging ($<0$) power and $b_{n,0}(e)$ is the initial state for the epoch. The SoC is constrained by the physical battery capacity limits, $0 \le b_{nt} \le B_n$ (kWh). Furthermore, the rate of charge or discharge is limited by the battery power limits, $-K_n \le k_{nt} \le K_n$ (kW).

\paragraph{Grid Connection Limits.}
Power exchanged with the grid is constrained by the connection capacity $X_n$. Both selling $s_{nt}$ and buying $d_{nt}$ are non-negative power flows, individually bounded by this limit: $0 \le s_{nt} \le X_n$ (kW) and $0 \le d_{nt} \le X_n$ (kW). These bounds implicitly enforce the net trade constraint $|s_{nt} - d_{nt}| \le X_n$.

\paragraph{Inter-Epoch Link.}
The prosumer value function is connected across consecutive epochs through the battery state. The terminal state of charge at the end of epoch $e$, $b_{nT}(e)$, becomes the initial state of charge for the subsequent epoch $e+1$: $b_{n0}(e+1) = b_{nT}(e)$.

\paragraph{Objective Function.}
The prosumer's objective is to maximize the total expected discounted (net) profit over an infinite horizon. This dynamic optimization problem can be expressed recursively using the Bellman equation. For a single epoch $e$, given the initial battery state $b_{n0}(e)$, the value function $\Pi_n(b_{n0}(e))$ represents the maximum achievable expected future profit:\footnote{To be fully rigorous, the prosumer's payoff in a given epoch is quasi-linear and defined as $\sum_{t=1}^T P_n(\mathbf{x}_t) + \mathcal{I}_n(\mathbf{p})$, where $\mathcal{I}_n$ is the (convex) characteristic function of the feasible set:  $\mathcal{I}_n(\mathbf{p}) = 0$ if the consumption schedule satisfies the load requirements, i.e., $p_{nt} \ge \alpha^{\text{base}}_{nt}$ for all $t$ and $\sum_{t} (p_{nt} - \alpha^{\text{base}}_{nt}) = \alpha^{\text{flex}}_{n}$, and $\mathcal{I}_n(\mathbf{p}) = -\infty$ otherwise.}
\begin{equation}
\Pi_n(b_{n0}(e)) = \max_{\{\mathbf{p}, \mathbf{k}, \mathbf{s}, \mathbf{d}\}} \mathbb{E} \left\{ \sum_{t=1}^T P_n(\mathbf{x}_t(e)) + \gamma \Pi_n(b_{n0}(e+1)) \right\}
\label{eq:bellman_abstract}
\end{equation}
Here, the maximization is over the sequences of decisions within epoch $e$. The term $\sum_{t=1}^T P_n(\mathbf{x}_t(e))$ represents the total net profit (€) accumulated within the current epoch $e$, where $P_n(\mathbf{x}_t(e))$ is the net payment received (positive for revenue, negative for cost) for the power allocation $\mathbf{x}_t(e)$ at timestep $t$. The term $\gamma \Pi_n(b_{n0}(e+1))$ is the discounted continuation value, starting from the next epoch's initial state $b_{n0}(e+1)$, with $\gamma \in [0, 1)$ being the discount factor between epochs.

\subsection{The Best-Response Program}
\label{cref:BR-linear}

We now analyze the prosumer's problem of optimizing profits by best responding to a (perfectly) forecasted  path of prices $\bar{\mathbf{r}}$ and $\bar{\mathbf{c}}$ (€/kW):
\begin{equation}
\max_{\{\mathbf{s}, \mathbf{d}, \mathbf{k}, \mathbf{p}, \mathbf{b}\}} \quad \left[ \sum_{t=1}^{T} (\bar{r}_t s_{nt} - \bar{c}_t d_{nt}) + \gamma \bar{\Pi}_n(b_{nT}) \right]
\label{eq:prosumer_lp_objective}
\end{equation}
subject to the following constraints for $t=1, \dots, T$:
\bigskip

\begin{itemize}[itemsep=1pt, topsep=2pt, labelwidth=!, labelindent=0pt, leftmargin=*]
    \item Power Balance: $\omega_{nt} + d_{nt} = p_{nt} + k_{nt} + s_{nt}$ \hfill (Lagrange multiplier: $\theta_{nt}$)
    \item Baseline Consumption: $p_{nt} \ge \alpha^{\text{base}}_{nt}$ \hfill 
    \item Flexible Load Target: $\sum_{t=1}^{T} (p_{nt} - \alpha^{\text{base}}_{nt}) = \alpha^{\text{flex}}_{n}$ \hfill 
    \item Battery Dynamics: $b_{nt} = b_{n,t-1}(e) + k_{nt}$ \hfill
    \item Battery State Limits: $0 \le b_{nt} \le B_n$ \hfill
    \item Battery Power Limits: $-K_n \le k_{nt} \le K_n$ \hfill 
    \item Grid Transmission Limits: $0 \le s_{nt} \le X_n$, $0 \le d_{nt} \le X_n$ \hfill 
\end{itemize}

\bigskip 

\noindent{}Appendix \ref{sec:KKT-prosumer-matrix} shows how to express \cref{eq:prosumer_lp_objective} and its constraints in a more compact matrix form. Since the objective function is concave and the feasible set defined by the linear constraints is convex, this is a convex optimization problem. 

\paragraph{Remark.} The value function in \cref{eq:prosumer_lp_objective} can account for intra-epoch discounting by scaling prices $(\bar{c}_t,\bar{r}_t)$ by $\gamma^{t/T}$.

\paragraph{Trading Behavior.} The optimal action plan of the prosumer follows from the KKT and stationary condition of the Lagrangian associated to the above optimization problem. In particular, the trading strategy can be understood looking at the dual space of the optimization problem, in particular by analyzing how the Lagrange multiplier of the Power Balance constraint, $\theta^*_{nt}$ relates to the market prices $(\bar{r}_t, \bar{c}_t)$. 

Intuitively, the multiplier $\theta^*$ represents the \textit{shadow price} of power, or the price that the prosumer would attribute to its power budget. This leads the prosumer to sell power in periods where $\theta^*_{nt} \leq \bar{r}_t$, to buy power in periods where  $\theta^*_{nt} \geq \bar{r}_t$, and to not trade in the intermediate range, $\bar{r}_t < \theta_{nt}^* < \bar{c}_t $.
In the previous relationships for active trading, strict inequalities denote binding transmission limits ($X_n$), whereas equalities imply internal constraints (e.g., consumption constraints, battery capacity) determine the volume by equating marginal value to price. Formally, 
\begin{equation}
x_{nt}^* =
\begin{cases}
    X_n & \text{if } \theta_{nt}^* < \bar{r}_t \quad (\text{Transmission-Constrained Selling}) \\
    s_{nt}^* \in (0, X_n) & \text{if } \theta_{nt}^* = \bar{r}_t \quad (\text{Internally-Constrained Selling}) \\
    0 & \text{if } \bar{r}_t < \theta_{nt}^* < \bar{c}_t \quad (\text{No Trade}) \\
    -d_{nt}^* \in (-X_n, 0) & \text{if } \theta_{nt}^* = \bar{c}_t \quad (\text{Internally-Constrained Buying}) \\
    -X_n & \text{if } \theta_{nt}^* > \bar{c}_t \quad (\text{Transmission-Constrained Buying})
\end{cases}
\label{eq:trading_rule_detailed}
\end{equation}

\subsection{Computation via Model Predictive Control}

\begin{algorithm}[h!]
\caption{Rolling-Horizon MPC for Prosumer Optimization}
\label{alg:full_mpc}
\begin{spacing}{1.10} 
\begin{algorithmic}[1]
\Require Total epochs (i.e., horizon) $H$, prosumer type $\theta_n$, initial state $b_{n0}(1)$, price forecasts $\{\bar{\mathbf{r}}(e'), \bar{\mathbf{c}}(e')\}_{e'=1}^{H+L_{max}}$, tolerance $\epsilon$.
\Ensure Sequences of optimal plans $\{\boldsymbol{a}_n^*(e)\}$, profits $\{\Pi_n^*(e)\}$, first-epoch dual vars $\{\Lambda^*(e)\}_{e=1}^H$.

\State Initialize storage: $\boldsymbol{A}_n^*, \boldsymbol{\Pi}_n^*, \boldsymbol{\Lambda}^* \gets \emptyset$.
\State $b_{n0}^{\text{current}} \gets b_{n0}(1)$.
\For{$e = 1$ \textbf{to} $H$} \Comment{Outer loop: Rolling horizon}
    \State $L \gets 1$; $\textsf{converged} \gets \text{false}$; $b_{res}^{(0)} \gets -\infty$. \Comment{Inner loop: Find lookahead horizon $L$}
    \While{\textbf{not} \textsf{converged}}
        \State $(\mathbf{z}^{*(L)}, \Lambda^L) \gets \arg\max_{\mathbf{z}^{(L)} \in \mathcal{A}_n^L(b_{n0}^{\text{current}})} \left( \mathbf{c}_{\pi}^{(L)\top} \mathbf{z}^{(L)}  \right)$. \Comment{Solve L-epoch LP}
        \State Extract L-epoch battery plan $\mathbf{k}^{*(L)} \subset \mathbf{z}^{*(L)}$.
        \State $b_{res}^{(L)} \gets b_{n0}^{\text{current}} + \sum_{e'=e}^{e+L-1} \sum_{t=1}^T k_{nt}^*(e') $. \Comment{Terminal state after L epochs}
        \If{$L > 1$ \textbf{and} $|b_{res}^{(L)} - b_{res}^{(L-1)}| < \epsilon$}
            \State $\textsf{converged} \gets \text{true}$.
        \Else
            \State $b_{res}^{(L-1)} \gets b_{res}^{(L)}$; $L \gets L+1$.
            \If{$e+L-1 > H+L_{max}$} $\textsf{converged} \gets \text{true}$. \EndIf
        \EndIf
    \EndWhile

    \Statex \Comment{Extract results for the first epoch ($e$) from converged solution:}
    \State $\boldsymbol{a}_n^*(e) \gets \{s_{nt}^*(e), d_{nt}^*(e), k_{nt}^*(e), p_{nt}^*(e)\}_{t=1}^T \subset \mathbf{z}^{*(L)}$.
    \State $\Lambda^*(e) \gets \{\theta_{nt}^*(e), \mu_{nt}^*(e), \nu_n^*(e), \dots \}_{t=1}^T \subset \Lambda^L$.
    \State $\Pi_n^*(e) \gets \sum_{t=1}^T (\bar{r}_t(e)s_{nt}^*(e) - \bar{c}_t(e)d_{nt}^*(e)) $.

    \Statex \Comment{Store results and update state for next epoch:}
    \State $\boldsymbol{A}_n^* \gets \boldsymbol{A}_n^* \cup \{\boldsymbol{a}_n^*(e)\}$; $\boldsymbol{\Pi}_n^* \gets \boldsymbol{\Pi}_n^* \cup \{\Pi_n^*(e)\}$; $\boldsymbol{\Lambda}^* \gets \boldsymbol{\Lambda}^* \cup \{\Lambda^*(e)\}$.
    \State $b_{nT}(e) \gets b_{n0}^{\text{current}} + \sum_{t=1}^T k_{nt}^*(e) $.
    \State $b_{n0}^{\text{current}} \gets b_{nT}(e)$. \Comment{Update initial state for epoch $e+1$}
\EndFor
\State \Return $\boldsymbol{A}_n^*, \boldsymbol{\Pi}_n^*, \boldsymbol{\Lambda}^*$.
\end{algorithmic}
\end{spacing}
\end{algorithm}

\noindent{}Directly solving the infinite-horizon problem \eqref{eq:bellman_abstract} is computationally intractable. We therefore approximate the solution using \textbf{Model Predictive Control (MPC)} on rolling-horizon \citep{prat2024finitehorizon}. In each epoch $e$, the prosumer solves a deterministic Linear Program (LP) over a finite \textit{lookahead window} of $L$ epochs ($e, \dots, e+L-1$), based on available forecasts. 

Let $\mathbf{z}^{(L)}$ denote the stacked vector of decision variables over these $L$ epochs. The optimization problem is given by:
\begin{equation} \label{eq:mpc_lp_concise}
\max_{\mathbf{z}^{(L)} \in \mathcal{A}_n^L(b_{n0}(e))} \quad \mathbf{c}_{\pi}^{(L)\top} \mathbf{z}^{(L)}
\end{equation}
where $\mathcal{A}_n^L(b_{n0}(e))$ represents the feasible set over $L$ epochs starting from state $b_{n0}(e)$, and $\mathbf{c}_{\pi}^{(L)}$ is the vector of forecasted effective prices. After doing so, only the optimal plan for the \textit{first epoch}, $\boldsymbol{a}_n^*(e) \subset \mathbf{z}^{*(L)}$, is implemented. 

The lookahead horizon $L$ is adaptively increased until the terminal state stabilizes, ensuring the finite window approximation is sufficiently precise. The resulting terminal state $b_{nT}(e)$ then serves as the initial state $b_{n0}(e+1)$ for the subsequent epoch's optimization. 

Algorithm \ref{alg:full_mpc} shows the full implementation of this procedure.  

\section{Equilibrium Analysis}
\label{sec:equilibrium}


In the previous section, we analyzed a single prosumer's optimization problem by assuming perfect foresight of the price path. We now relax this assumption and study the strategic interactions among all prosumers in a multi-agent setting. 

As prosumers' decision-making unfolds over an infinite horizon, the appropriate framework to study their interaction is a dynamic stochastic game. We first characterize the \textbf{Bayes-Nash Equilibrium (BNE)} for a single epoch, taking continuation values as given. We then concatenate these stage games to analyze the resulting \textbf{Markov Perfect Equilibrium (MPE)} of the stochastic game.

\subsection{Equilibrium Concept}
\label{sec:welfare}

\paragraph{Strategic Environment and Prosumer Beliefs.}
At the beginning of each epoch $e$, the stage game is determined a public state $z_e$ (e.g., meteorological conditions) and by each prosumer $n$'s private state, consisting of their battery level $b_{n0}(e)$ and their epoch-specific type $\theta_n(e)$. 

The public state captures common sources of randomness as the joint distribution of types is conditional on it, $F(\boldsymbol{\theta} \mid z_e)$. Each prosumer observes their own type but only holds probabilistic beliefs about the types of others.

We focus on mixed-strategy equilibria. A prosumer with type $\theta$ has a feasible action set $\mathcal{A}(\theta)$. Because prosumers are atomistic, their individual optimization problem is a linear program (see \cref{cref:BR-linear}). Thus, any best response to an expected price path is an extreme point of this set, $\boldsymbol{a} \in \operatorname{ext}(\mathcal{A}(\theta))$. A mixed strategy is therefore a measurable mapping $\sigma : \Theta \;\to\; \varDelta\big(\operatorname{ext}(\mathcal{A}(\theta))\big),
$
which assigns to each \textit{type} a probability distribution over the extreme points available to that type. Therefore,  a \textbf{mixed-strategy profile} for the $N$-prosumer game is given by the following mapping:\footnote{Since $\mathcal{A}(\theta)$ depends only on a prosumer's type and not on their index, the strategy space is type-symmetric.}
\[
\boldsymbol{\sigma} : \times_{n=1}^N \Theta_n 
\;\to\; 
\prod_{n=1}^N \varDelta\big(\operatorname{ext}(\mathcal{A}(\theta_n))\big).
\]

 Given a mixed-strategy profile, and conditional on a realized type profile $\boldsymbol{\theta}$, a pure action profile $\mathbf{a}$ is played with probability $\prod_{n=1}^N \sigma_n(\mathbf{a}_n \mid \theta_n).$ 
 
 Under the common prior assumption, all prosumers hold the same belief about the expected price paths, which can be computed by taking expectations over types and induced actions:
\begin{align}
    \bar{r}_t(\boldsymbol{\sigma} \mid z_e) &= \int_{\boldsymbol{\Theta}} \sum_{\mathbf{a} \in \mathcal{A}_{\text{ext}}} \left[ r_t(y_t(\mathbf{a})) \cdot \prod_{n=1}^N \sigma_n(\mathbf{a}_n | \theta_n) \right] dF(\boldsymbol{\theta} \mid z_e) \label{eq:prices-r}\\
    \bar{c}_t(\boldsymbol{\sigma} \mid z_e) &= \int_{\boldsymbol{\Theta}} \sum_{\mathbf{a} \in \mathcal{A}_{\text{ext}}} \left[ c_t(y_t(\mathbf{a})) \cdot \prod_{n=1}^N \sigma_n(\mathbf{a}_n | \theta_n) \right] dF(\boldsymbol{\theta} \mid z_e) \label{eq:prices-c}    
\end{align}
where $y_t(\mathbf{a})$ is the aggregate supply-to-demand ratio resulting from the action profile $\mathbf{a}$.

\paragraph{Equilibrium Definitions.}
A strategy profile $\boldsymbol{\sigma}^*$ is a BNE if each prosumer's strategy is a best response to the strategies of all others, $\boldsymbol{\sigma}_{-n}^*$. Since prices are determined by the aggregate behavior induced by $\boldsymbol{\sigma}^*$, equilibrium solves a fixed-point problem in which beliefs about prices coincide with the expected prices arising in equilibrium. Let $\bar{\boldsymbol{\rho}}^* = (\bar{\boldsymbol{r}}^*, \bar{\boldsymbol{c}}^*)$ denote these equilibrium expected prices.\footnote{For notational convenience, we use variations of the symbol $\rho$ to define AMM prices in different contexts: in \cref{AMM-construction}, $\rho(E,M)$ denotes the AMM \textit{internal} marginal price, while in the current section, $\bar{\boldsymbol{\rho}}^*$ denotes the rational expectation of ``final'' AMM prices, accounting for the pro-rata repartition of the trading surplus or deficit.} Given $\bar{\boldsymbol{\rho}}^*$, the payoff from a pure (extreme-point) action plan $\mathbf{a}_n$ for prosumer $n$ is
$$    \pi_n(\mathbf{a}_n, \bar{\boldsymbol{\rho}}^*)
        = \sum_{t=1}^T 
            \big( \bar{r}_t^*\, s_{nt} - \bar{c}_t^*\, d_{nt} \big).$$

Since we consider a large population of prosumer, we model each prosumer as \emph{atomistic} and hence \emph{price-taker}: a unilateral deviation of a single prosumer from the strategy profile does not affect the net marginal prices quoted by the AMM. Formally,

\begin{assumption}[Atomicity]\label{as:mean_field}
For any prosumer $n$ and type $\theta_n$, a deviation form  $\sigma_n(\cdot \mid \theta_n)$ to $\sigma'_n(\cdot \mid \theta_n)$ leaves the prices quoted by the AMM unchanged:
\[
r(s(\boldsymbol{\sigma}), d(\boldsymbol{\sigma}))
=
r(s(\sigma_n', \boldsymbol{\sigma}_{-n}), d(\sigma_n', \boldsymbol{\sigma}_{-n})),
\qquad
c(s(\boldsymbol{\sigma}), d(\boldsymbol{\sigma}))
=
c(s(\sigma_n', \boldsymbol{\sigma}_{-n}), d(\sigma_n', \boldsymbol{\sigma}_{-n})).
\]
\end{assumption}

We can now define BNE in this context as follows:\footnote{An equivalence definition of incentive-compatibility can be stated requiring indifference among pure strategies in the mixing support: for each type $\theta_n$ and any $\mathbf{a}_n,\mathbf{a}_n' \in \operatorname{ext}(\mathcal{A}(\theta_n))$ with $\sigma_n^*(\mathbf{a}_n \mid \theta_n) > 0$ and $\sigma_n^*(\mathbf{a}_n' \mid \theta_n) > 0$, we have
$\pi_n(\mathbf{a}_n,\bar{\boldsymbol{\rho}}^*) = \pi_n(\mathbf{a}_n',\bar{\boldsymbol{\rho}}^*) 
= \max_{\mathbf{a} \in \operatorname{ext}(\mathcal{A}(\theta_n))} \pi_n(\mathbf{a},\bar{\boldsymbol{\rho}}^*)$.}

\begin{definition}[Mean-Field BNE] \label{cref:def-BNE}
A pair $(\boldsymbol{\sigma}^*, \bar{\boldsymbol{\rho}}^*)$ is a Mean-Field Bayes-Nash Equilibrium of the stage game if:
\begin{enumerate}[label=(\alph*)]
    \item \textbf{Incentive-Compatibility:} For every prosumer $n$ and type $\theta_n \in \Theta_n$, the mixed strategy $\sigma_n^*(\cdot | \theta_n)$ maximizes the prosumer's expected profit given the equilibrium prices $\bar{\boldsymbol{\rho}}^*$:
    \begin{equation} \label{eqm_optimality}
        \sigma_n^*(\cdot|\theta_n) \in \arg\max_{\sigma_n \in \Delta(\text{ext}(\mathcal{A}_n))} \sum_{\mathbf{a}_n \in \text{ext}(\mathcal{A}_n)} \sigma_n(\mathbf{a}_n) \cdot \pi_n(\mathbf{a}_n, \bar{\boldsymbol{\rho}}^*)
    \end{equation}

    \item \textbf{Rational Expectations:} The expected price vector $\bar{\boldsymbol{\rho}}^*$ is generated by the aggregate behavior of all prosumers playing according to the strategy profile $\boldsymbol{\sigma}^*$, conditional on the public state $z_e$, according to \cref{eq:prices-r,eq:prices-c}:
    $$  \bar{r}_t^* =  \bar{r}_t(\boldsymbol{\sigma} \mid z_e) \qquad  \bar{c}_t^* =  \bar{c}_t(\boldsymbol{\sigma} \mid z_e). $$

    \item \textbf{Atomicity}:  The prices quoted by the AMM $r(s(\boldsymbol{\sigma}),d(\boldsymbol{\sigma}))$ and $c(s(\boldsymbol{\sigma}),d(\boldsymbol{\sigma}))$ are not affected by a deviation of single prosumer $n$ from strategy $\sigma_n(.|\theta_n)$  to $\sigma'_n(.|\theta_n)$ (\cref{as:mean_field}).
\end{enumerate}
\end{definition}

\paragraph{Dynamic Extension: Markov Perfect Equilibrium.}
The definition of BNE describes equilibrium interaction for a single epoch. However, prosumers interact with the AMM repeatedly over epochs to maximize discounted future profits. The previous Equilibrium definition \ref{cref:def-BNE} can thus be extended to the dynamic setting by requiring strategies to be  Markovian in payoff-relevant state variables, such as the initial battery level and the public signal. Letting $\Pi_n$ denote prosumer $n$'s value function, we now define:

\begin{definition}[MPE]
A triple $(\boldsymbol{\sigma}^*, \bar{\boldsymbol{\rho}}^*, \Pi^*)$ is a Markov Perfect Equilibrium if, for every prosumer $n$, the value function satisfies
\[
\Pi_n(b_{n0}(e), z_e)
=
\max_{\sigma_n(\cdot \mid \theta_n)}
\Big\{
    \pi_n(\sigma_n, \boldsymbol{\sigma}_{-n}^*, \bar{\boldsymbol{\rho}}^*)
    + \gamma\, 
        \mathbb{E}\big[
            \Pi_n(b_{n0}(e+1), z_{e+1})
        \,\big|\,
            \sigma_n, \boldsymbol{\sigma}_{-n}^*
        \big]
\Big\},
\]
and the maximizer is the equilibrium strategy $\sigma_n^*(\cdot \mid \theta_n)$.  
The continuation value $\Pi^*$ is stationary under the equilibrium strategy profile.
\end{definition}

\subsection{Characterization of prosumer equilibria}

 So far, we have analyzed how prosumers would best-respond to a market making mechanism. Given their response, the question still remains whether the resulting power flows and prices maximize some notion of welfare.

To this end, we now show a powerful equivalence based on the Axioms of \cref{sec:market-design}. Under those axioms indeed the BNE of the prosumers' stage game leads to a (randomized) allocation of power flows that is the same as the one a Social Planner would choose to maximize the expected trade profits of the prosumer community with the power grid.

\begin{proposition}[Welfare Equivalence] \label{prop-welfare-equilvanence}
A strategy profile is a BNE of the stage-game if and only if it maximizes the expected value of trade profits with the grid:
\begin{equation} \label{eq:welfare-exp} W(\boldsymbol{\sigma}) =
\mathbb{E}_{\boldsymbol{\theta} \sim F(\cdot|z_e)} \left[ \sum_{t=1}^T \left( s_{At}(\boldsymbol{\sigma}(\boldsymbol{\theta})) \cdot \underline{\lambda}_t - d_{At}(\boldsymbol{\sigma}(\boldsymbol{\theta})) \cdot \overline{\lambda}_t \right) \right].
\end{equation}
\end{proposition}

This result is powerful because it establishes that, in this context, the First and Second Welfare Theorems hold \citep{mascolell1995} and the AMM essentially implements a \textit{competitive equilibrium} among prosumers: The AMM serves as a ``zero-sum redistribution engine" that \textit{complements} the main electrical market to coordinate prosumers into solving collectively a (soft) market clearing problem. 

To better see this, we can express the objective of the Planner (and hence of the prosumer collective) in terms of netput vectors:\footnote{The transformation follows using the properties of the absolute value operator: $x^+ = (|x| + x)/2$, $x^- = (|x| - x)/2$.}     
\begin{equation} \label{eq:welfare-netput}
W(\boldsymbol{\sigma})= \mathbb{E}_{\boldsymbol{\theta} \sim F(\cdot|z_e)} \bigg[ \bm{\lambda}_{\textbf{mid}}^\top \mathbf{X}(\boldsymbol{\sigma}(\boldsymbol{\theta})) - \bm{\lambda}_{\mathbf{spread}}^\top || \ \mathbf{X} (\boldsymbol{\sigma}(\boldsymbol{\theta})) \  ||_1  \bigg]    
\end{equation}
where $\mathbf{X} = (x_1, \dots, x_T)$, and $\bm{\lambda}_{\textbf{mid}}$, $\bm{\lambda}_{\textbf{spread}}$ are vectors of mid-prices  and mid-spreads. with respective elements $(\boldsymbol{\lambda}_{\textbf{mid}}^\top)_t = (\overline{\lambda}_t + \underline{\lambda}_t)/2$, and  $  (\boldsymbol{\lambda}_{\textbf{spread}})_t = (\overline{\lambda}_t - \underline{\lambda}_t)/2$. 

According to \cref{eq:welfare-netput}, the Planner achieves optimality by matching internal demand with supply to avoid the bid-ask spread (with penalty coming from the $L_1$-norm), while shifting any residual grid trades to periods where the mid-price is favorable. Since the community's total net energy balance over an epoch is mostly fixed, the planner objective is usually dominated by the minimization of the weighted $L_1$-minimization of a weighted netput vector:
\[
\min_{\boldsymbol{\sigma}} \mathbb{E}_{\boldsymbol{\theta} \sim F(\cdot|z_e)} \left[ \big\| \boldsymbol{\lambda}_{\text{spread}} \cdot \mathbf{X}(\boldsymbol{\sigma}(\boldsymbol{\theta})) \big\|_1 \right].
\]

The sufficiency part of  \cref{prop-welfare-equilvanence} ($\text{BNE} \implies \text{Max Welfare}$) can be established solely based on Exact Budget Balance (Axiom \ref{BB}) and Atomocity (Assumption \ref{as:mean_field}). In particular, from these two assumptions it emerges that the AMM creates a Potential Game \citep{monderer1996} where the potential function is exactly the community welfare. 

\begin{proposition}[Potential Game] \label{lemma-potential-game} The prosumer game is a potential game with potential function equal to the expected global welfare, $ W(\boldsymbol{\sigma})$.
\end{proposition}

This is intuitive since Budget Balance establishes that internal transfers net-out and Atomicity ensures that a single prosumer's deviation has a negligible effect on the prices faced by others, effectively eliminating price externalities. This potential game structure guarantees no efficiency loss due to decentralization (i.e., a Price of Anarchy of 1).

The necessity direction of \cref{prop-welfare-equilvanence} ($\text{Max Welfare} \implies \text{BNE}$) is slightly trickier to prove and relies on two additional axioms on top of those needed for proving sufficcy, namely Individual Rationality (Axiom \ref{ax:IR}) and Responsiveness (Axiom \ref{ax:resp}). These two axioms ensure that whenever internal clearing is not optimized, there exists a profitable arbitrage opportunity on AMM prices that prosumers can be exploit to increase profits.

It is important to notice that \cref{prop-welfare-equilvanence} holds \textit{regardless} of which internal price functions or trading function $\psi(\cdot)$ is used---as long as it is axiom-compliant and prosumers behave atomistically. In this sense, \cref{prop-welfare-equilvanence} provides us with a benchmark result. One way to pin down a specific curve or pricing mechanism would be by attributing priorities among prosumers, which would relax the linearity assumption on welfare. Doing so could be done for example by assigning different utility functions in accordance to assigned priorities (for example, the  hospital would have higher priority than residential households.) We will discuss more this aspect in the conclusion (\cref{sec:conclusion}). 

Besides of theoretical importance, \cref{prop-welfare-equilvanence} has also practical implication for computing equilibria. Thanks to it we can reduce a complex equilibrium problem to solving a Planner's problem and then decentralizing the solution satisfying individual feasibility (see \cref{sec:eq-computation}).

\subsubsection*{Dynamic Extension}

The stage-game analysis above focuses on a single epoch. When we consider how strategies evolve over time, we extend these efficiency results to the dynamic setting.

\begin{proposition}[Markov Perfect Equilibria]\label{prop-markov} 
The dynamic game exhibits a multiplicity of mixed-strategy MPEs. However, every equilibrium strategy profile $\boldsymbol{\sigma}^*$ implements the same unique aggregate outcome that maximizes the expected discounted profit of the local grid:
\begin{equation}
V^* = \max_{\boldsymbol{\sigma}} \sum_{e=1}^\infty \gamma^e \mathbb{E}_{\boldsymbol{\theta} \sim F(\cdot|z_e)} \left[ \sum_{t=1}^T \left( s_{At} \cdot \underline{\lambda}_t - d_{At} \cdot \overline{\lambda}_t \right) \right].
\end{equation}
\end{proposition}

This proposition confirms that the static welfare equivalence extends to the dynamic case: any MPE maximizes the long-term discounted net trade profits. 

While the aggregate value $V^*$ and the optimal path of grid trades are unique, the specific strategy profile $\boldsymbol{\sigma}^*$ is not. Because (in every interesting application) the number of degrees of freedom in the choice of the mixing weights  vastly exceeds the number of aggregate constraints, there exists a continuum of mixed strategies that all map to the exact same optimal aggregate outcome.

\subsection{Efficiency and Mechanism Comparison}
\label{sec:mechanism_comparison}

We benchmark the proposed AMM by comparing its welfare properties against the theoretical first-best outcome achievable by an omniscient social planner. While \cref{prop-welfare-equilvanence} establishes that the AMM maximizes the \textit{expected} welfare of the community, this is an \textbf{ex-ante} notion of efficiency, as prosumers optimize their strategies based on probabilistic beliefs about others. An ideal planner who observes the realization of all types $\boldsymbol{\theta}$ \textit{before} allocating resources could achieve a higher \textbf{ex-post} welfare by making the power flows contingent on the realization of the private state $(\boldsymbol{\theta}, \boldsymbol{b_0})$ (as well as the the public state). The relationship is captured by the following inequality:
\begin{equation}
\label{eq:welfare_hierarchy}
\mathcal{W}_{\text{VCG}} \equiv \mathbb{E}_{\boldsymbol{\theta}} \left[ \max_{\mathbf{x}} w(\mathbf{x}, \boldsymbol{\theta}) \right] \;\ge\; \max_{\boldsymbol{\sigma}} \mathbb{E}_{\boldsymbol{\theta}} \left[ w(\mathbf{x}(\boldsymbol{\sigma}), \boldsymbol{\theta}) \right] \equiv \mathcal{W}_{\text{AMM}}
\end{equation}
where $w(\cdot)$ denotes the realized global welfare for a specific type profile. The gap between these two values can be interpreted as the ``price'' of higher privacy and decentralization.

\paragraph{The Vickrey-Clarke-Groves (VCG) Benchmark.}
The canonical mechanism for implementing the ex-post social optimum ($\mathcal{W}_{\text{VCG}}$) is the Vickrey-Clarke-Groves (VCG) mechanism \citep{vickrey1961, clarke1971, groves1973}. In this direct-revelation setup, each agent $n$ reports their type $\hat{\theta}_n$. The mechanism then implements the socially optimal allocation $\mathbf{x}^*(\hat{\boldsymbol{\theta}})$ and charges each agent a payment equal to the externality they impose on others:
\[
P_n^{\text{VCG}}(\hat{\boldsymbol{\theta}}) = \left( \max_{\mathbf{x}_{-n}} \sum_{j \neq n} w_j(\mathbf{x}_{-n}, \hat{\theta}_j) \right) - \sum_{j \neq n} w_j(\mathbf{x}^*(\hat{\boldsymbol{\theta}}), \hat{\theta}_j).
\]

VCG is \textbf{strategy-proof} (truthful reporting is a dominant strategy) and achieves the unconstrained ex-post maximum welfare. However, its theoretical optimality is hardly practical for local energy communities. 

First, VCG is not budget-balanced (it generally runs a deficit), implying that the community would require an external subsidy to operate. Also, it requires prosumers to reveal their full private types (e.g., battery states and   consumption/generation profiles) to the Planner \textit{before} choosing their power flows. Besides raising possibly privacy concerns, it can be particularly inconvenient for  residential prosumers having to commit ahead of time to an energy schedule.

In contrast, the AMM described in this paper is an \textbf{indirect mechanism} that can better deal with these issues. It sacrifices strategy-proofness and ex-post optimality to guarantee the two properties that VCG lacks: By satisfying Budget Balance (Axiom \ref{BB}), the AMM ensures the market is self-sustaining without external subsidies. Furthermore, pricing is purely based on traded quantities rather than the full prosumer type, and  payments are determined \textit{after} recording the power flows.

For these reasons, the AMM is a practical alternative to a full VCG implementation: It is an optimal mechanism within the class of budget-balanced, anonymous, and responsive mechanisms, and  achieves the maximum possible welfare without subsidies and with lower informational requirements on the prosumers.

\subsection{Equilibrium Computation}\label{sec:eq-computation}

Computing the Markov Perfect Equilibrium (MPE) for a dynamic game with a continuum of heterogeneous agents can be challenging. However, the Welfare Equivalence result (\cref{prop-welfare-equilvanence}) provides a crucial simplification: instead of solving for the fixed point of $N$ coupled Bellman equations, we can simply solve a single concave Planner's Problem and then decentralize the allocation. We leverage this insight to construct the numerical procedure applied in \cref{quantitative} to data from metropolitan Paris.

The algorithm that we develop hereafter to approximate the infinite-horizon equilibrium proceeds epoch by epoch. In each epoch $e$, given the current empirical distribution of battery states $\nu_e=\frac{1}{N}\sum_{n=1}^N \delta_{b_{n0}(e)}$, we compute an approximate BNE over a lookahead horizon of $L$ epochs. This computation follows a four–stage procedure:

(i) the continuous population is quantized into \textbf{state-dependent representative types},

(ii) the dimensionality of the \textbf{action space is reduced} by restricting choices to a subset of salient strategies  for these types based on their current state,

(iii) a (potentially regularized) \textbf{planner problem} is solved to find the equilibrium strategies maximizing the mean-field welfare, and

(iv) the coordinated allocation is \textbf{decentralized} back to individuals via Euclidean projection.

While the equilibrium calculation considers the full $L$-epoch horizon to incorporate forward-looking behavior, only the actions corresponding to the \textit{first} epoch ($l=1$) are implemented. The resulting distribution of terminal battery states then determines the initial state distribution $\nu_{e+1}$ for the next epoch. Algorithm \ref{alg:equilibrium_computation} summarizes this rolling-horizon procedure that we will now outline in detail.

\paragraph{Population Approximation.} 
To begin with, the equilibrium solver algorithm needs a type population as input. We represent such population by approximating the type space $\Theta$ with a \emph{mixed histogram}. This structure enables the joint modeling of \textbf{discrete categories} $c \in \mathcal{C}$ (e.g., simple consumers and solar prosumers) and \textbf{continuous attributes} (e.g., capacities, generation profiles). For each category $c$, we partition the attribute space into cells $\{R_{c,j}\}_{j=1}^{J_c}$ with associated weights $w_{c,j}$, inducing a collection of uniform classes:
\[
\mathcal{H} = \big\{(c, R_{c,j}, w_{c,j}) \mid c \in \mathcal{C}, \, 1 \le j \le J_c \big\}.
\]
We then generate a  simulated population of $N$ agents by Monte-Carlo sampling from this structure:
\[
\theta_n \sim \mathrm{Unif}\big(\{c\} \times R_{c,j}\big) \quad \text{with probability } \omega_{c,j}.
\]

In each epoch $e$, an agent's full state consists of their static type $\theta_n$ and their initial state of charge $b_{n0}(e)$. To maintain tractability as battery levels evolve, we group the $N$ agents into $J$ \textbf{state-dependent representative types} corresponding to the static bins. The representative for bin $j$ in epoch $e$ is defined as the tuple $(\bar{\theta}_j, \bar{b}_j(e))$, where $\bar{\theta}_j$ is the midpoint (centroid) of cell $R_{c,j}$ and $\bar{b}_j(e)$ is the current average state of charge of all agents belonging to that bin.

\paragraph{Dimensionality Reduction of the Action Space.}
The feasible action set $\mathcal{A}_j^L(b_{j0})$ is depends on the agent's initial state of charge. Therefore, to ensure the BNE approximation remains accurate as the population state distribution $\nu_e$ evolves, the strategy banks that form the mixed-strategy support are \textbf{recomputed in every epoch $e$}. 

For each state-dependent representative type $(\theta_j, b_j(e))$, we generate a new strategy bank for a lookahead window of $L$ epochs: $\mathcal{B}_j^L(e) = \{\mathbf{a}_{jk}^{1:L}\}_{k=1}^{K_j} \subset \operatorname{ext}(\mathcal{A}_j^L(b_j(e)))$.
We take the support of the equilibrium mixed strategy $\sigma_j(e)$ to be a the strategy bank $\mathcal{B}_j^L(e)$ populated by solving the $L$-epoch best-response LP using \cref{alg:full_mpc} (with parameters $\theta_j$ and $b_j(e)$) under a large number of diverse price profiles. The resulting equilibrium strategy for type $j$ is a distribution over this bank, $\sigma_j^\star(e) \in \varDelta^{K_j}$, and the support for the overall mixed strategy profile $\boldsymbol{\sigma}(e)$ is the Cartesian product of these dynamic banks, $\mathcal{B}^L(e) = \times_{j=1}^J \mathcal{B}_j^L(e)$.

\paragraph{Planner Problem for Ex-Ante Welfare Maximization.}
With the strategy banks $\mathcal{B}_j^L$ populated, we can now determine how agents should mix among these candidate plans. We accomplish this by solving for the equilibrium strategy profile $\boldsymbol{\sigma}^\star$ that maximizes the \textbf{ex-ante expected community welfare}. 

To handle the non-differentiable absolute value in the welfare function (Equation \ref{eq:welfare-exp}), we introduce auxiliary variables for the aggregate export $\mathbf{X}^{+} \in \mathbb{R}^{TL}$ and import $\mathbf{X}^{-} \in \mathbb{R}^{TL}$. The problem of the Equilibrium Solver can thus be defined as maximizing the linear welfare term with an optional Tikhonov regularization term:\footnote{The regularization term increases the smoothness of the resulting equilibrium aggregate profiles.}

\begin{definition}[Equilibrium Solver]\label{eq_solver} The Equilibrium Solver is given by the following Quadratic Programming problem: 
    
\begin{equation}
\label{eq:planner-qp-types}
\max_{\boldsymbol{\sigma}, \mathbf{X}^+, \mathbf{X}^-} \quad \sum_{l=1}^L \gamma^{l-1} \left( \underline{\boldsymbol{\lambda}}^{(l)\top} \mathbf{X}^{+(l)} - \overline{\boldsymbol{\lambda}}^{(l)\top} \mathbf{X}^{-(l)} \right) - \eta \|\textsf{Diff} \, \bar{\mathbf{X}}(\boldsymbol{\sigma})\|_2^2
\end{equation}
subject to:
\begin{equation}
    \label{netput-aggregate}
\bar{\mathbf{X}}(\boldsymbol{\sigma}) = \sum_{j=1}^J \omega_j \left( S_j \sigma_j - D_j \sigma_j \right).
\end{equation}
\begin{align*}
\bar{\mathbf{X}}(\boldsymbol{\sigma}) - \mathbf{X}^+ + \mathbf{X}^- &= 0 && \text{(Net trade decomposition)} \\
\mathbf{X}^+ \ge 0, \quad \mathbf{X}^- &\ge 0 && \text{(Non-negativity)} \\
\mathbf{1}^\top \sigma_j &= 1 \quad \forall j && \text{(Probability simplex)} \\
\sigma_j &\ge 0 \quad \forall j && \text{(Non-negativity)}
\end{align*}

\end{definition}

\noindent{}In the objective function of the Solver, $\eta \geq 0$ is a scalar and $\textsf{Diff} \in \mathbb{R}^{(TL-1) \times TL}$ is the temporal difference matrix.\footnote{\textsf{Diff} has entries $\textsf{Diff}_{t,t} = -1$ and $\textsf{Diff}_{t,t+1} = 1$.}  The choice variable for type $j$ is the probability vector $\sigma_j \in \mathbb{R}^{K_j}$. The constraints enforce consistency between the net trade and the auxiliary variables and restrict the vector $\boldsymbol{\sigma} = (\sigma_1, \dots, \sigma_J) \in \mathbb{R}^{\sum K_j}$ so that  probabilities are correctly defined over the relevant simplices ($\sum_{k=1}^{K_j} \sigma_{j,k} = 1$). 

The connection between the decision variables $\boldsymbol{\sigma}$ and the action aggregates is given by the aggregate netput vector $\bar{\mathbf{X}}(\boldsymbol{\sigma}) \in \mathbb{R}^{TL}$. We will now see how to compute it: Let $K_j$ be the number of strategies in the bank $\mathcal{B}_j^L$ for representative type $j$.
Each plan $k \in \{1, \dots, K_j\}$ in the bank induces specific supply and demand trajectories, denoted $\mathbf{s}_{j,k} \in \mathbb{R}^{TL}$ and $\mathbf{d}_{j,k} \in \mathbb{R}^{TL}$. We construct the supply matrix $S_j \in \mathbb{R}^{TL \times K_j}$ and the demand matrix $D_j \in \mathbb{R}^{TL \times K_j}$ by stacking these trajectories as columns:
\[
S_j = \begin{bmatrix} | & & | \\ \mathbf{s}_{j,1} & \dots & \mathbf{s}_{j,K_j} \\ | & & | \end{bmatrix}, \quad
D_j = \begin{bmatrix} | & & | \\ \mathbf{d}_{j,1} & \dots & \mathbf{d}_{j,K_j} \\ | & & | \end{bmatrix}.
\] 
The resulting aggregate netput in \cref{netput-aggregate} is thus the sum of these expected power (net) injections weighted by $\omega_j = \mathbb{P}(\theta \in \text{bin } j | z_e)$ across all $J$ bins.

\paragraph{Decentralization and Feasibility Projection.}
Finally, to implement the equilibrium, we decentralize the aggregate solution $\boldsymbol{\sigma}^\star$ via Monte-Carlo assignment. 

Specifically, each agent $n$ (belonging to bin $j$) independently samples a candidate plan $\mathbf{a}_{n} \sim \mathrm{Cat}(\sigma_j^\star)$ from the bank. Then, since the agent's realized state $(\theta_n, b_{n0})$ generally deviates from the bin archetype, we enforce feasibility by mapping the candidate to the closest feasible action via Euclidean projection: $\widetilde{\mathbf{a}}_n = \operatorname{Prj}(\mathbf{a}_{n}; \theta_n, b_{n0})$. This step restores feasibility while preserving approximate incentive compatibility. 

The procedure concludes by executing the first epoch of $\widetilde{\mathbf{a}}_n$, determining the initial state distribution $\nu_{e+1}$ for the subsequent loop.

\begin{algorithm}[h!]
\caption{Rolling-Horizon Equilibrium Solver}
\label{alg:equilibrium_computation}
\small 
\begin{spacing}{0.9} 
\begin{algorithmic}[1]
\Require Mixed Histogram $\mathcal{H}$, Grid prices $\underline{\boldsymbol{\lambda}}, \overline{\boldsymbol{\lambda}}$, Lookahead window $L$, Epoch-length $T$, Total epochs $H$.
\Ensure Equilibrium strategies $\boldsymbol{\sigma}^\star(e)$, realized plans $\{\widetilde{\mathbf{a}}_n(e)\}$, state paths $\{b_{n0}(e)\}$.

\State \textbf{Init:} Sample $\theta_n \sim \mathcal{H}$; Init states $b_{n0}(1)$; Define bin centroids $\{\bar{\theta}_j\}_{j=1}^J$.

\For{$e = 1$ \textbf{to} $H$} \Comment{Rolling Horizon}
    \State \textbf{(i) Population Clustering}
    \For{$j = 1$ \textbf{to} $J$}
        \State Compute weight $\omega_j \gets |\mathcal{N}_j| / N$ and rep. state $\bar{b}_j(e) \gets \text{mean}(b_{n0})$ for $n \in \text{Bin } j$.
    \EndFor

    \State \textbf{(ii) Strategy Bank Generation} \Comment{Parallelizable}
    \For{$j = 1$ \textbf{to} $J$}
        \State $\mathcal{B}_j^L(e) \gets \emptyset$; Parameters $\chi_j \gets (\bar{\theta}_j, \bar{b}_j(e))$.
        \For{$k = 1$ \textbf{to} $K_{\text{samples}}$}
            \State Generate prices $\hat{\mathbf{r}}_k, \hat{\mathbf{c}}_k$. Solve LP: $\mathbf{a}_{j,k} \gets \arg\max \pi(\cdot; \chi_j, \hat{\mathbf{r}}_k, \hat{\mathbf{c}}_k)$.
            \State $\mathcal{B}_j^L(e) \gets \mathcal{B}_j^L(e) \cup \{ \mathbf{a}_{j,k} \}$ (retain extreme points).
        \EndFor
    \EndFor

    \State \textbf{(iii) Planner Optimization}
    \State Construct supply/demand matrices $S_j, D_j$ from banks $\mathcal{B}_j^L(e)$.
    \State Solve QP \eqref{eq:planner-qp-types} for mixing weights:
    \State \quad $\boldsymbol{\sigma}^\star(e) \gets \arg\max_{\boldsymbol{\sigma}} [ \mathcal{W}_{\text{Ex-Ante}}(\boldsymbol{\sigma}) - \eta \|\textsf{Diff}\, \bar{\mathbf{X}}(\boldsymbol{\sigma})\|_2^2 ]$
    \State Extract distributions $\sigma_j^\star(e)$ from $\boldsymbol{\sigma}^\star(e)$.

    \State \textbf{(iv) Decentralization \& State Update} \Comment{Parallelizable}
    \For{$n = 1$ \textbf{to} $N$}
        \State Identify bin $j$. Sample $\mathbf{a}_{n} \sim \text{Cat}(\sigma_j^\star(e))$.
        \State \textbf{Project:} $\widetilde{\mathbf{a}}_n \gets \arg\min_{\mathbf{a} \in \mathcal{A}_n^L} \|\mathbf{a} - \mathbf{a}_{n}\|_2$.
        \State Implement first epoch $\widetilde{\mathbf{a}}_n^{(1)}$. Update state: $b_{n0}(e+1) \gets \widetilde{b}_{nT}$.
    \EndFor
\EndFor
\State \Return $\{\widetilde{\mathbf{a}}_n(e), b_{n0}(e)\}_{n,e}$ and $\{\boldsymbol{\sigma}^\star(e)\}_e$.
\end{algorithmic}
\end{spacing}
\end{algorithm}

\subsection{Theoretical Guarantees for the Equilibrium Solver}

The careful reader will have noticed that the Solver in \cref{eq:planner-qp-types} maximizes the \textit{welfare at the average} trade---evaluating welfare as if the entire community were composed of $N$ identical ``average'' agents, each executing the same action---whereas the true objective (Eq. \ref{eq:welfare-exp}) is to maximize the \textit{expected total welfare} of the community. This is a crucial simplification since it allows us to optimize a deterministic proxy rather than integrating over the intractable $N$-dimensional joint distribution of types $F(\boldsymbol{\theta}|z_e)$, which would be required to compute the exact expectation of the piecewise-linear welfare function. We now explain why this approach is valid.

First, thanks to the Homogeneity Axiom (\cref{ax:homo}), the welfare function \textbf{scales linearly in aggregate netput} (i.e., it is 1-homogeneous). This allows us to relate total welfare and welfare at the average trade: Let $\mathbf{X}_N$ be the total realized net trade and $\bar{\mathbf{x}}_N = \mathbf{X}_N/N$ be the \emph{realized average} net trade per agent. The total welfare is exactly $N$ times the welfare of the average:
\[
W_{\text{total}}(\mathbf{X}_N) = N \cdot W_{\text{avg}}(\bar{\mathbf{x}}_N),
\]
where $W_{\text{avg}}(\cdot)$ is simply the standard welfare function defined in \cref{eq:welfare-netput} applied to unit-normalized quantities. Because $N$ is a positive constant, maximizing the expected total welfare is equivalent to maximizing the expected average welfare:
\[
\max_{\boldsymbol{\sigma}} \quad \mathbb{E}_{\boldsymbol{\theta}|z_e} \left[ W_{\text{total}}(\mathbf{X}_N(\boldsymbol{\sigma})) \right] \quad \iff \quad \max_{\boldsymbol{\sigma}} \quad \mathbb{E}_{\boldsymbol{\theta}|z_e} \left[ W_{\text{avg}}(\bar{\mathbf{x}}_N(\boldsymbol{\sigma})) \right].
\]

Nevertheless we did not eliminate yet the integration problem. Since $W_{\text{avg}}$ is concave and the realized average $\bar{\mathbf{x}}_N$ is a random variable, and by Jensen's Inequality,  $\mathbb{E}[W_{\text{avg}}(\bar{\mathbf{x}}_N)] \le W_{\text{avg}}(\mathbb{E}[\bar{\mathbf{x}}_N])$. The solver optimizes the right-hand side (the certainty-equivalent welfare of the expectation), which overestimates the welfare function.

However, this ``Jensen gap'' is driven by the variance of the realized average, $\mathrm{Var}(\bar{\mathbf{x}}_N \mid z_e)$. As the population size $N \to \infty$, the Strong Law of Large Numbers (SLLN) ensures that the realized average $\bar{\mathbf{x}}_N$ converges almost surely to the sample mean  $\bar{\mathbf{X}}$. Consequently, the variance vanishes, and by continuity and boundedness of the welfare function, the welfare of the stochastic average converges to the welfare of the deterministic mean field approximation.

\begin{proposition}[Convergence to the Mean-Field Welfare Limit] \label{prop:mfg-welfare}

Let $W_{\text{avg}}$ be the continuous welfare function and let individual actions $\mathbf{y}_n$ be conditionally i.i.d. and bounded given $z_e$. Let $\bar{\mathbf{x}}_N = \frac{1}{N}\sum_{n=1}^N \mathbf{y}_n$ be the realized average and $\bar{\mathbf{X}} = \mathbb{E}[\mathbf{y}_n \mid z_e]$ be the theoretical mean field. Then, the true expected welfare converges to the welfare of the mean field:

$$\lim_{N\to\infty} \mathbb{E}\left[W_{\text{avg}}(\bar{\mathbf{x}}_N) \mid z_e\right] = W_{\text{avg}}(\bar{\mathbf{X}}).$$

\end{proposition}

\paragraph{Approximation Error and $\epsilon$-Incentive Compatibility.}
Finally, we note that quantizing the continuous type space into $J$ discrete bins and ensuring feasibility by projection $\widetilde{\mathbf{a}}_n = \operatorname{Prj}(\mathbf{a}; \theta_n)$ can result in a strictly positive \textit{regret} relative to the true optimal strategy:
$$ 
\text{Regret}_n = \max_{\mathbf{a} \in \mathcal{A}(\theta_n)} \pi_n(\mathbf{a}) - \mathbb{E}_{\widetilde{\mathbf{a}}_n \sim \sigma_j^\star, \operatorname{Prj}} \left[ \pi_n(\widetilde{\mathbf{a}}_n) \right]$$

However, as the prosumer's individual optimization problem (Eq. \ref{eq:prosumer_lp_objective}) is a linear program, their optimal value functions and the feasible set boundaries are Lipschitz continuous with respect to the elements of $\theta$ (e.g., battery capacity, grid limits). Consequently, the regret is bounded by the distance between the agent's true type and the bin centroid. 

As the histogram mesh becomes finer, this error vanishes linearly, ensuring that the mechanism is asymptotically Incentive Compatible.\footnote{Notice that the regret bound is stronger than payoff continuity as the latter only implies $|\max_{\sigma} \mathbb{E}[\pi(\sigma;\theta_n)] - \max_{\sigma} \mathbb{E}[\pi(\sigma;\bar{\theta}_j)]| \le \kappa \delta$ while the regret bound ensures the specific mixed strategy $\sigma_j^\star$ (optimized for $\bar{\theta}_j$) yields an expected payoff close to the true optimum when played by type $\theta_n$.}

\begin{proposition}[Vanishing Regret] \label{prop:vanishing-regret}
Let $\delta = \max_{j} \operatorname{diam}(R_j)$ be the maximum diameter of the type bins. Under the assumption that the feasible set correspondence $\theta \rightrightarrows \mathcal{A}(\theta)$ is Lipschitz continuous, the expected regret for any agent $n$ is bounded linearly by the quantization error:
\[
\text{Regret}_n  \le \kappa \cdot \delta
\]
where $\kappa$ is a problem-dependent Lipschitz constant. Thus, as the population approximation becomes exact ($\delta \to 0$), the regret vanishes and the equilibrium becomes exact.
\end{proposition}

\section{Quantitative Experiments}
\label{quantitative}

Our goal in this section is to build simulations to quantify gains with our approach. We explore two scales with this model, both advantages in the prosumer and in the community level.

\subsection{Data description and source}

\begin{figure}
\centering
\begin{subfigure}{.5\textwidth}
  \centering
  \includegraphics[width=.8\linewidth]{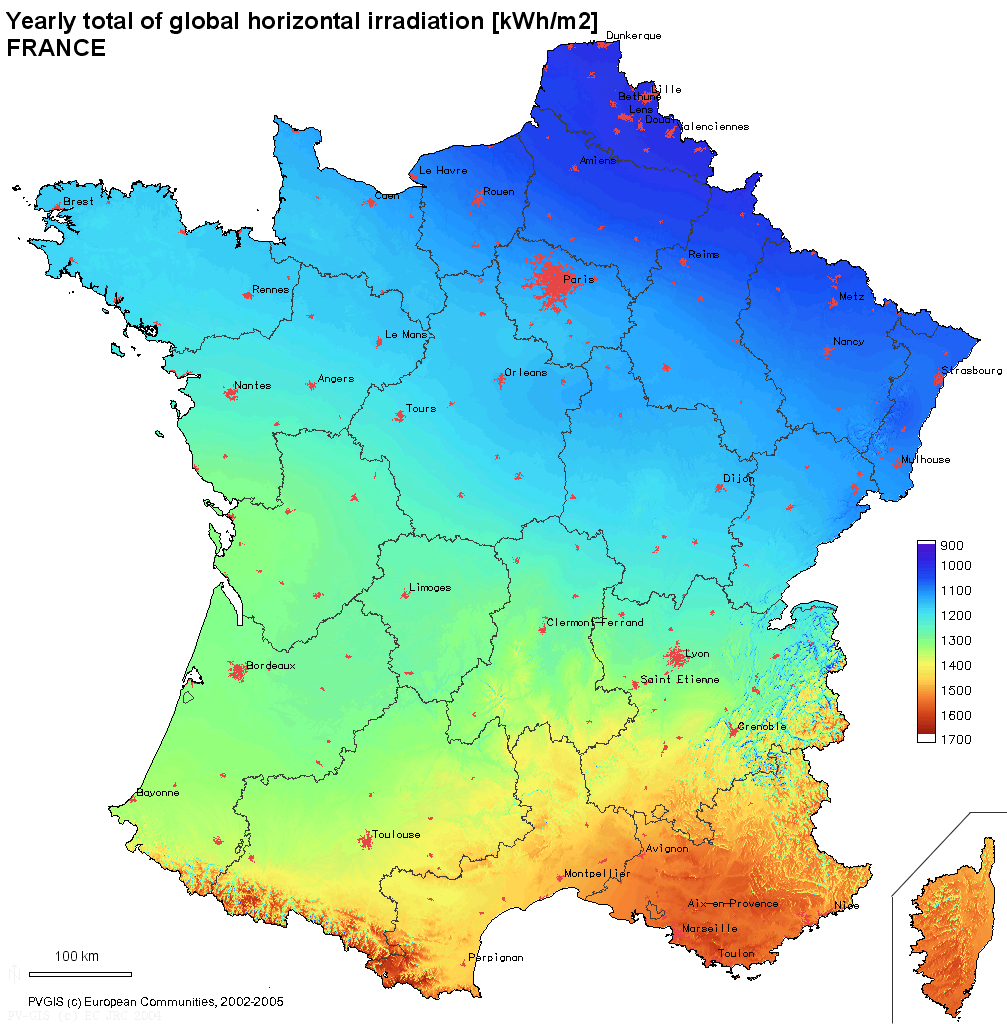}
  \caption{Average solar irradiation.}
  \label{fig:solarCarteFrance}
\end{subfigure}%
\begin{subfigure}{.5\textwidth}
  \centering
  \includegraphics[width=\linewidth]{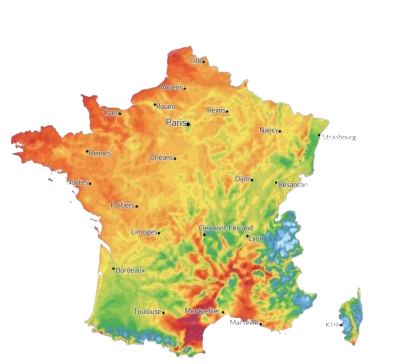}
  \caption{Average wind Speed at 100 meter.}
  \label{fig:windCarteFrance}
\end{subfigure}
\caption{Weather characteristics for solar and eolic energy production in metropolitan France.}
\label{fig:SW_CarteFrance}
\end{figure}

Two datasets are used and explored in this quantitative simulations. The first dataset contains historical weather information from Paris during the year of 2023. Paris was chosen as the base of our simulation due to its economical importance and the access to information such as energy consumption, demographic, solar irradiance and eolic power potential \cite{hakkarainen2015role}. The contrast between different regions in France for energy potential is shown in Figure \ref{fig:solarCarteFrance} \cite{solargris}, indicating differences in solar energy production potential in relevant regions, while Figure \ref{fig:windCarteFrance} shows the average wind speed at 100 meters which is the basis for eolic energy.

Historical data is taken for open-meteo weather \cite{weather2023}, containing information which impacts energy production, such as solar irradiance, cloud cover percentage, and wind speed. These values are measured and recorded hourly. Irradiance is measured as the total radiation received in the preceding hour per meter squared, $\frac{W}{m^2}$, taking into consideration Earth's tilt for the non-normal incidence. Cloud cover is defined as the fraction of the sky that is covered in clouds, affecting the performance and efficiency of solar panels. Wind speed is measured in $km\cdot h^{-1}$ at an altitude of 100 meters from the ground.

Energy consumption is taken from France's Transmission System Operator (RTE) \cite{RTE2023} from 2023, the same time period as the weather data, matching consumption and energy production, however consumption data is recorded every 15 minutes while weather data is hourly.  Household demand is computed as the total consumption in the city scaled by the number of buildings in a community, the average amount of people per building and the city population. Consumption information is scaled down based on a population of around 12 million inhabitants in Paris, obtained from the french population census.

Both datasets where broken down in 15 minutes steps, totaling 96 steps in each epoch or each day. As weather data is hourly recorded, cleaning, processing and inferring values in each timestep steps depends on the specific variable being analyzed. For solar energy, four main features are taken into account: consumption, irradiance, cloud cover, and wind speed. For consumption, irradiance and wind speed, non measured data or errors in measurement (NaN or zeros) are approximated as the mean between the following and the previous values. If this assumption is not possible, considering errors in whole the daily measure, they are approximated using previous day measurements. Values are broken down equally for each step, maintaining the same total hourly consumption. Filling data for and for wind speed is the same as for the other features, prioritizing the mean between previous and following values. On the other hand, breaking down values for time steps is done by maintaining the same value as previously measured, so all values in one hour are the constant.

\medskip 

Solar energy production is computed using irradiance ($G$), cloud cover ($C$), and an average efficiency ($e$) for solar panel. Efficiency is set to 15\% and a solar panel is considered to have 1.6 $m^2$, the energy production per panel ($E_s$) is computed as $E_s=G\cdot C\cdot e$. For eolic energy, its production is estimated as $E_e=0.5\cdot C_p\cdot \rho\cdot \pi\cdot R^2\cdot V^3$ for a wind turbine, where $c_p$ is  performance coefficient set to $0.4$, $\rho$ is air density as $1.293 \frac{kg}{m^3}$, $R$ is the blade (rotor) length approximated to $1.9$ meters, and $V$ the wind speed in $\frac{m}{s}$. To compute the demand and the supply in Paris, its population was fixed in 12 million and 32 thousand solar panels and wind turbines, the later was estimated based on the amount of private owned buildings in Paris. These values where scaled down for a representative community of 1,000 buildings with an average of 5 people per building. 

\medskip

\begin{figure}
    \centering

    \includegraphics[width=1\linewidth]{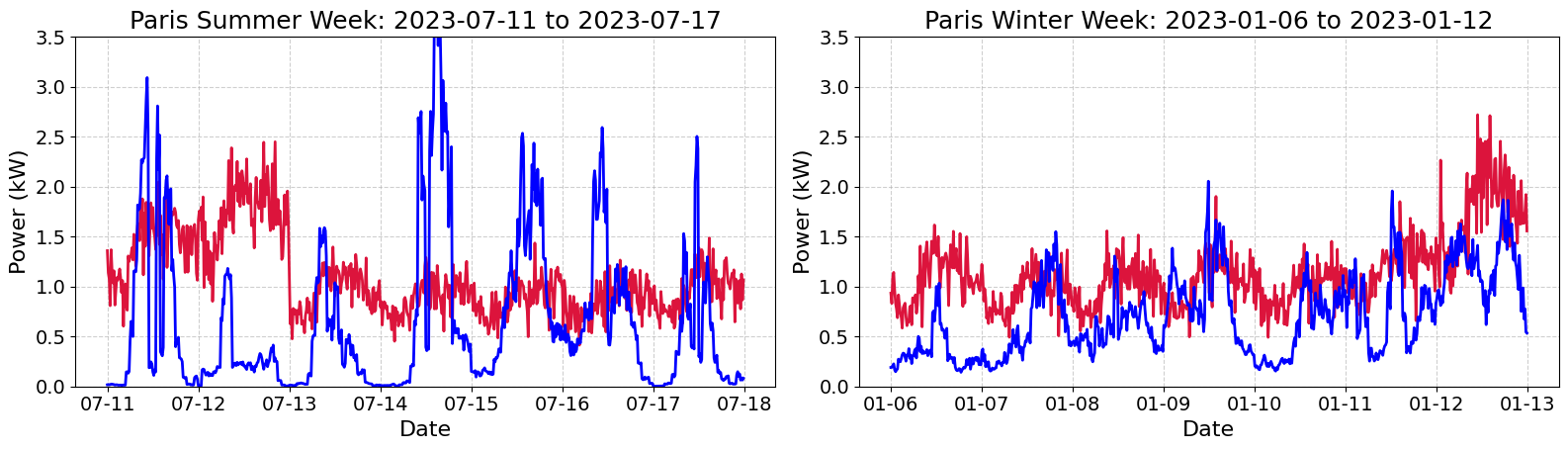}
    \caption{Community household energy profiles for demand and solar supply for prosumers in Paris during a summer week (07/10/2023 to 07/16/2023) on the left side, and a winter week (01/06/2023 to 01/12/2023) on the right side. Red lines indicate community demand, while blue and black lines represent solar and Eolic energy supply respectively.}
    \label{fig:energyProfComSummer}
\end{figure}

Communities demand and energy production are computed using the scaling factor previously described, accounting for a subset of the population with solar and eolic energy production and consumption. A sample of the results are displayed in Figure \ref{fig:energyProfComSummer}, showing a summer week on the left side and a winter week on the right side. In both cases the dependence and volatility in energy production based on local weather is visible, being a byproduct of factors such as hour of the day and time of the year.

\begin{figure}
    \centering
    \includegraphics[width=1 \linewidth]{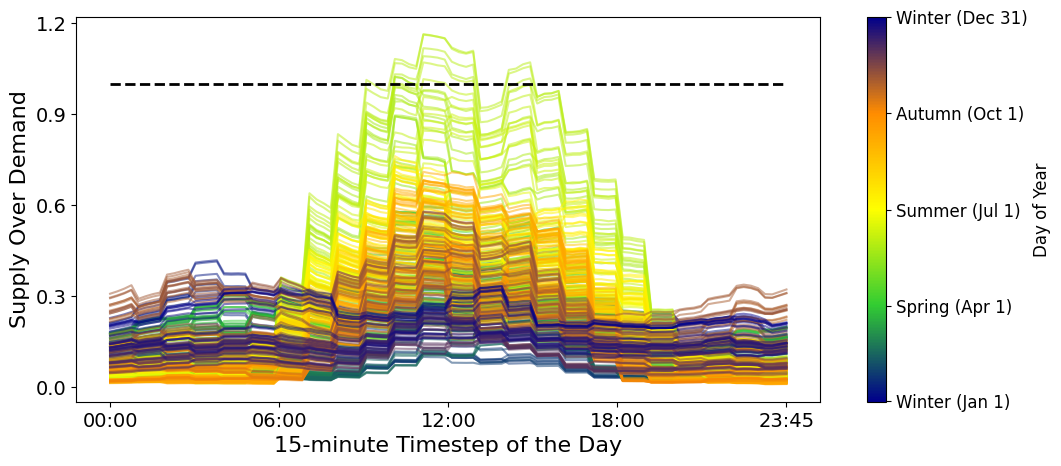}
    \caption{Yearly supply over demand for Paris. Each line represents the average over two weeks and their colors indicate the season. A dashed line represents the equality between supply and demand.}
    \label{fig:yearSupplyOverDemandProf}
\end{figure}

Supply and demand fluctuate depending on the season, daily supply over demand is displayed in Figure \ref{fig:yearSupplyOverDemandProf} showing Paris information and how it changes during the day and the seasons. Each line corresponds to a two weeks moving average accounting for one week prior and one week posterior to the specific day. A dashed line is displayed highlighting when supply equals demand. During summer and spring, supply overcomes demand, mostly on the solar energy production hours.

These pieces of information are used to simulate gains in the prosumer level using our AMM approach, clarifying advantages and benefits with a real data example.

\subsection{Prosumer-Level Gains from Decentralization}
\label{prosumer-level}

In this part, we explore monetary gains due to the implementation of AMM using solar panels and wind turbines for energy production with batteries for energy storage, aiming in analyzing profit differences a prosumer can achieve with and without AMM usage. Improving prosumer gains and energy allocation based on pricing functions help the promotion of decentralized clean energy production not only in an environmental point but also in a monetary side, decreasing the break even time for equipment investments and increasing their acquisition incentives.

\medskip

Simulations are made using France energy price as a base with the distinction between peak hour price $\bar{\lambda}$ (moments which overall consumption are higher), off-peak hour price $\underline{\lambda}$ (when consumption is lower) and resale price for prosumer production to the grid. Peak hours are set from 8 am to 12 am and from 1 pm to 8 pm. Energy storage is done with a 20 kW battery, with a maximum charge/discharge power of 5 kW and a maximum energy transaction with the grid of 5 kW each 15 minutes timestep. Demand is broken down into two categories and their percentages are based on Enedis information \cite{enedis} about household items utilization and consumption optimization literature in France \cite{agnetis2013load,de2017energy}, setting a base load equivalent to 70\% of daily consumption and 30\% as flexible load.

\medskip 

\begin{figure}
    \centering
    \includegraphics[width=1\linewidth]{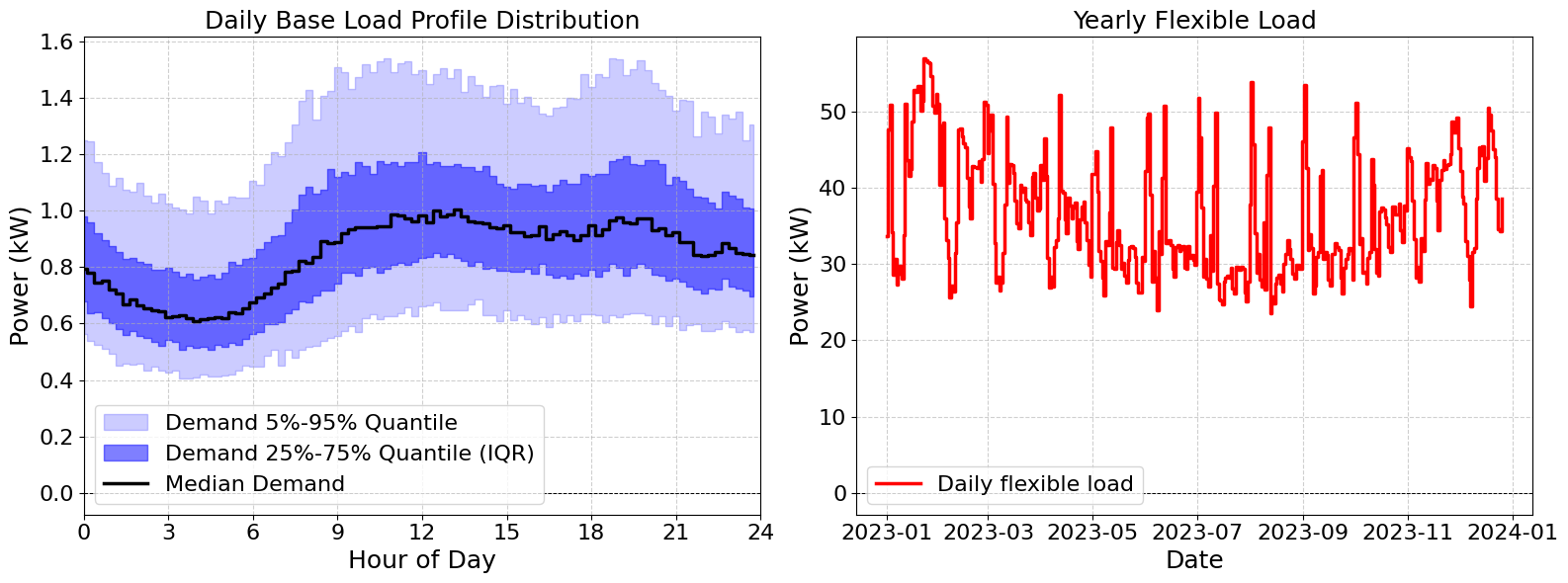}
        \caption{Base and Flexible Consumption profile for 2023. On the left side is the daily base load and its quantiles for the population distribution. On the right side the yearly yearly flexible load for the hole population. Flexible load represents a percentage of the base load.}
    \label{fig:daily_baseload_sim}
\end{figure}

Rolling-horizon simulation is performed over 360 days, with 15 minutes time steps and a fixed look ahead window of 4 days. For energy prices, $\bar{\lambda}$ is settled to 21.46 c€ per kWh, $\underline{\lambda}$ to 16.96 c€ per kWh, and resale price is set to 8.86 c€ per kWh. The main pieces of information are recorded during simulation, namely as battery information such as charge/discharge and trade profile, monetary transactions are also monitored during simulation. Two distinct profiles are used for simulations, the consumer and the prosumer, where the prosumer produces and offers energy supply, solar or wind, besides only demanding it. Solar and eolic energy production and quantities follow the same conditions as before. Consumption is broken down into a daily base load and a flexible amount, the later can be allocated during different moments of the day. Daily base load and yearly flexible loads are shown in Figure \ref{fig:daily_baseload_sim}, where the hourly base load distribution is displayed in the left side, and the right side indicates daily flexible.

\medskip

\begin{figure}
    \centering
    \includegraphics[width=.8 \linewidth]{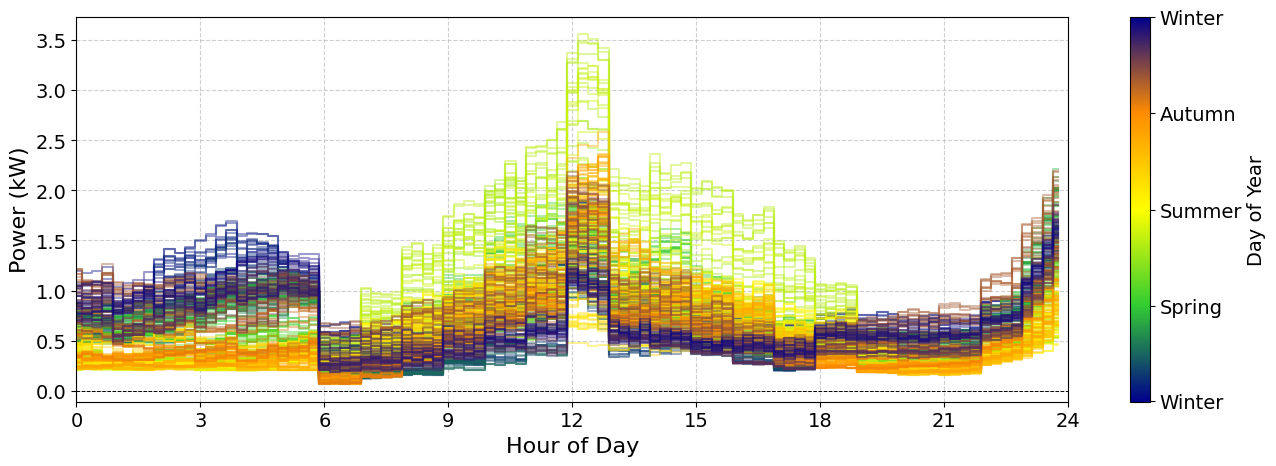}
    \caption{Optimized consumption taking into account flexible load allocation. Each line represents the two weeks average value, with one week prior and one week posterior to the analyzed curve. Colors represent the day of the year and its season.}
    \label{fig:optCons}
\end{figure}

Household appliances scheduling, known as demand-side management (DSM), is a strategy for allocating flexible energy consumption based on energy price and usage plausibility \cite{agnetis2013load, warren2014review, strbac2008demand, arteconi2012state}. In our framework, optimized consumption is dynamically allocated and Figure \ref{fig:optCons} shows a two weeks moving average for dynamic allocation results, so each line represents the average allocation accounting for one week prior and one week posterior the the analyzed day, their color represents the day of the year. Allocation depends on weather, energy production and price, indicating a seasonal dependency and an hourly dependency, where a more predominant part of the allocation is performed during sunny hours for summer while more prominent consumption is donne during off peak hours for winter months. 

\begin{figure}
    \centering
    \includegraphics[width=1\linewidth]{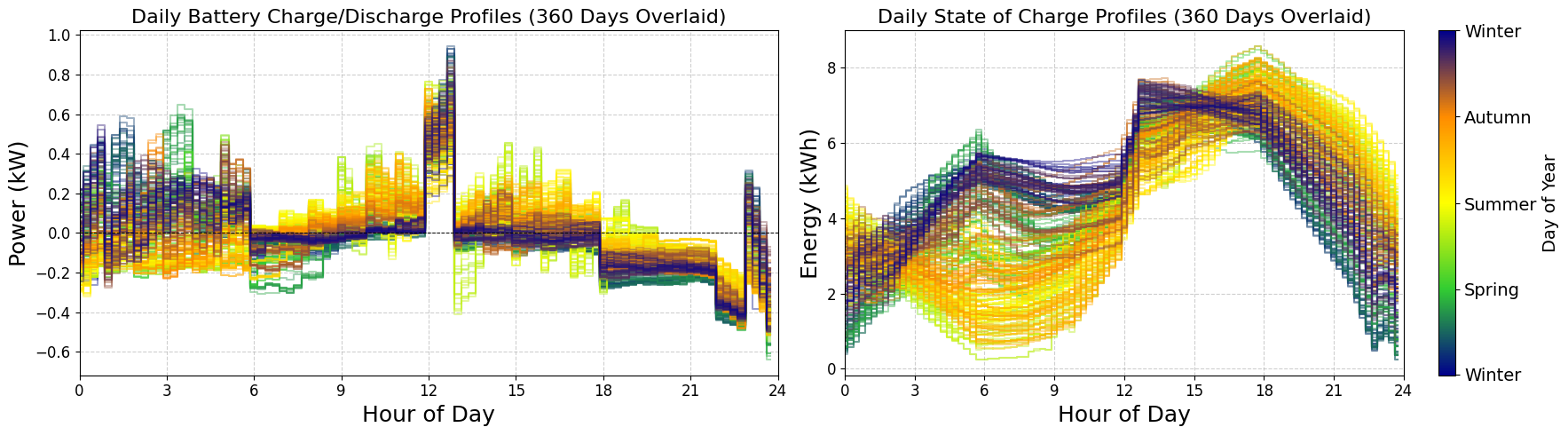}
    \caption{Daily battery profile throughout the year. Battery charge and discharge profile on the left graph, state of charge on the right graph. Curves represent a two weeks average value and their colors the season of the year.}
    \label{fig:BatChDis}
\end{figure}

Besides consumption allocation, battery usage is a key factor for P2P energy trading and financial gains. Figure \ref{fig:BatChDis} shows a two weeks average battery charge/discharge profile on the left side and SoC on the right side based on the hour of the day and the season of the year. In both cases seasonal characteristics are identifiable, but SoC reflects a greater sensibility with more distinguishable curves for summer and winter, where the shape of the curves differs mostly based on the season.

\begin{figure}
    \centering
    \includegraphics[width=1\linewidth]{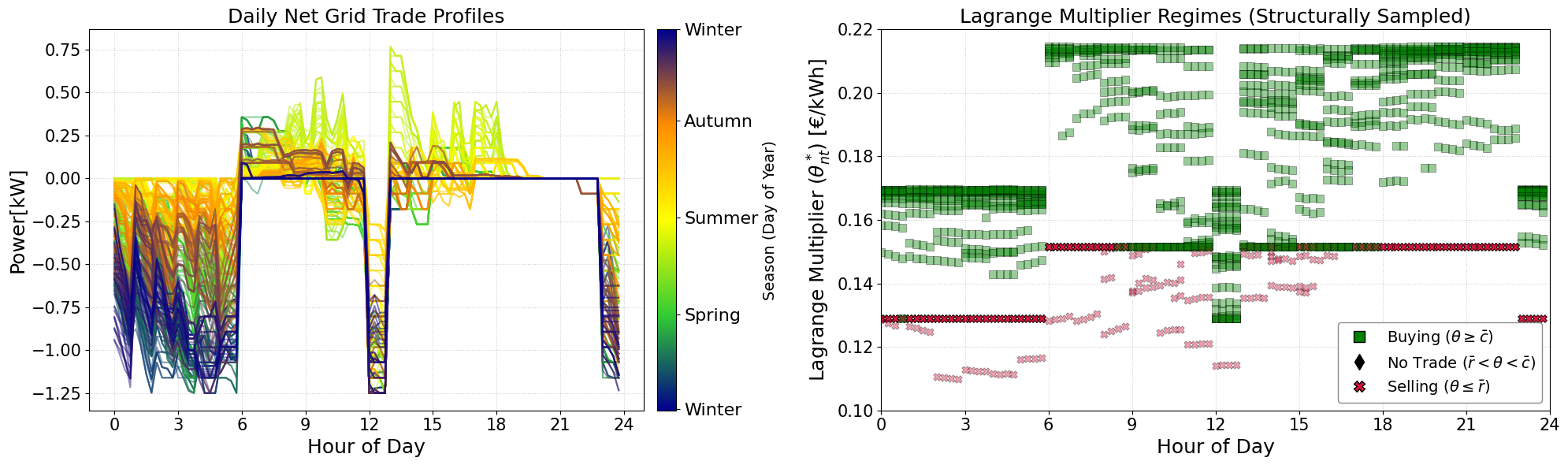}
    \caption{Daily grid trade profiles and Lagrange multiplier regimes during 2023. Net grid trade is shown on the left side and the Lagrange multiplier on the right side. Colors represent the time of the year for both images, shapes represent the trading regime (Buy, Sell, no-Trade).}
    \label{fig:XNet_LM}
\end{figure}

Trades in the community, described in equation \ref{eq:trading_rule_detailed}, are dictated by the prosumer's internal value perception $\theta_{nt}^*$ (the multiplier on the resource constraint for Eq. \ref{eq:prosumer_lp_objective}). The net grid trade profile and the internal price are shown in Figure \ref{fig:XNet_LM}, where the trading profiles are displayed on the left side and the Lagrange Multiplier on the right side for the whole year simulation.

\begin{figure}
    \centering
    \includegraphics[width=1\linewidth]{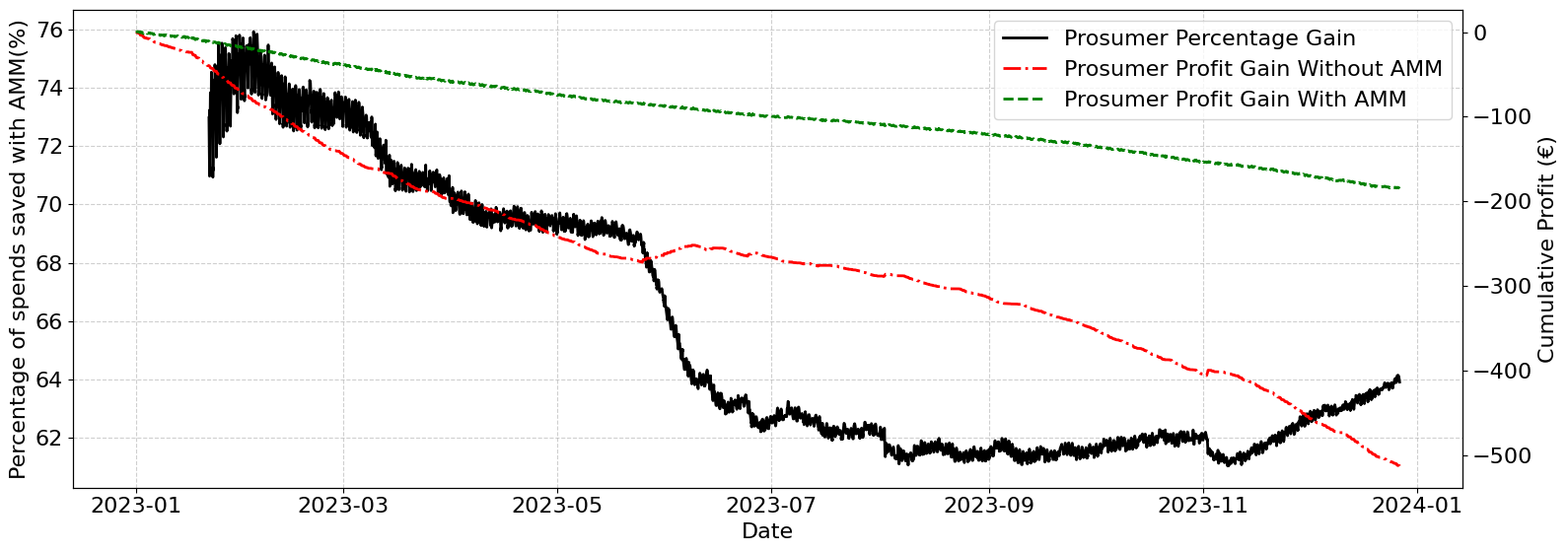}
    \caption{Cumulative gain using linear pricing function with and without AMM usage. Red line represents the cumulative profit without AMM, the green line represents the cumulative profit with AMM, and the  black line the percentage gain considering both scenarios.}
    \label{fig:CumProft}
\end{figure}

Using the linear pricing function, cumulative profits are shown in Figure \ref{fig:CumProft}, indicating the cumulative gain for the prosumer using both approaches, with and without AMM, on the right axis and on the left axis the percentage indicating the difference between both models over the minimal cumulative profit. These results highlight not only an absolute monetary gain, but it also reflects a decrease in the break even time for solar panels and wind turbines investments. Compared with the benchmark approach, AMM provides a monetary gain around 60\% with respect to the simulation without AMM.

\subsection{Grid-Level Gains from Decentralization}

\begin{figure}
    \centering
    \includegraphics[width=1 \linewidth]{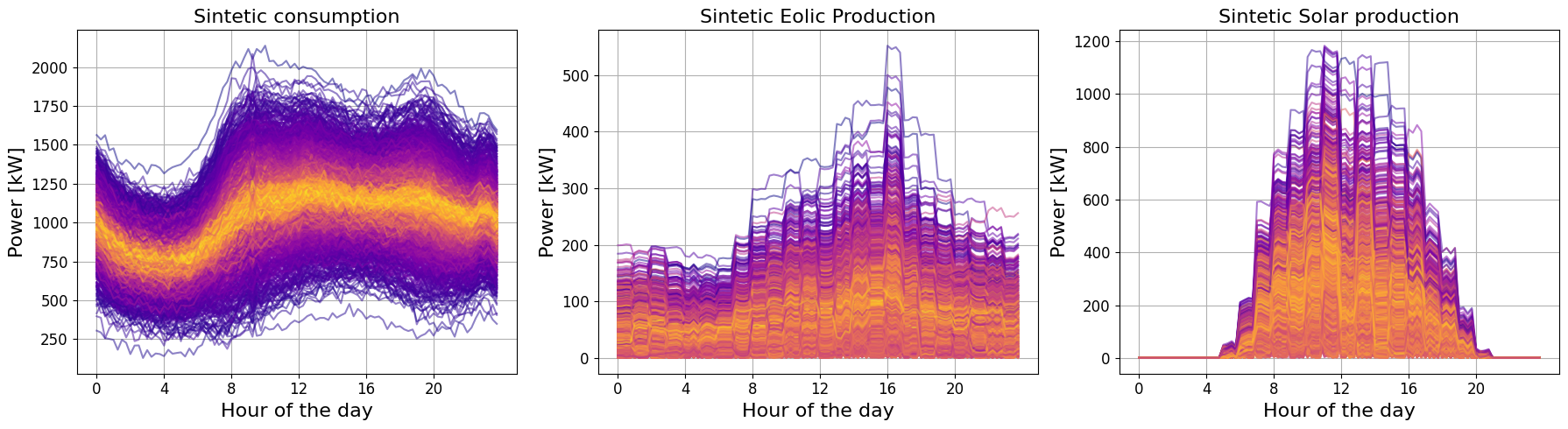}
    \caption{Simulated curves for consumption and energy production during summer. Upper side shows daily energy consumption, lower left side eolic energy supply and lower right side solar supply. Colors represent their respective depth values with respect to the ensemble.}
    \label{fig:SintDepth}
\end{figure}

Following prosumer level gains, we aim in this part to model grid-level gains from decentralization in a community. We implement a simulation framework to model and analyze the behavior of diverse energy agents interacting within a local energy market, facilitated by an Automated Market Maker. For these simulations, three different profiles of agents are used: a pure consumer, a solar prosumer, and a wind prosumer. All agents possess batteries, as well as a base load and a flexible load, which are all randomly chosen. Solar and wind prosumers supply solar and eolic energy respectively besides their energy consumption. For each profile, synthetic daily curves for energy supply and consumption are independently generated based on real data used on previous studies for Paris. Curves are generated for each of the four seasons of the year, yet we focus her on summer during our simulations, creating a set of distinct profiles describing a behavioral distribution. Used data is shown in Figure \ref{fig:SintDepth}, where the synthetic consumption is places in the upper part, eolic supply is displayed in the lower left part, and solar supply in the lower right panel. To select the data from a distribution we use the functional concept of data-depth, and the color of each curve represents their depth value, determining the centrality and likelihood of a behavior \cite{gijbels2017general, mozharovskyi2020nonparametric}. Depth is calculated using projection depth ($D_P$) following new developments for its computational aspects \cite{zuo2000general, zuo2003projection, zuo2011exact, mosler2022choosing, leone2025massive} for function data \cite{gijbels2017general}, where the (projection) depth of a curve $X_i$ with respect to all $n$ curves $X=\left(X_1, \dots, X_n\right)$ is written as $D_P\left(X_i\middle|X\right)$.


\begin{figure}[h!]
    \centering
    \includegraphics[width=1 \linewidth]{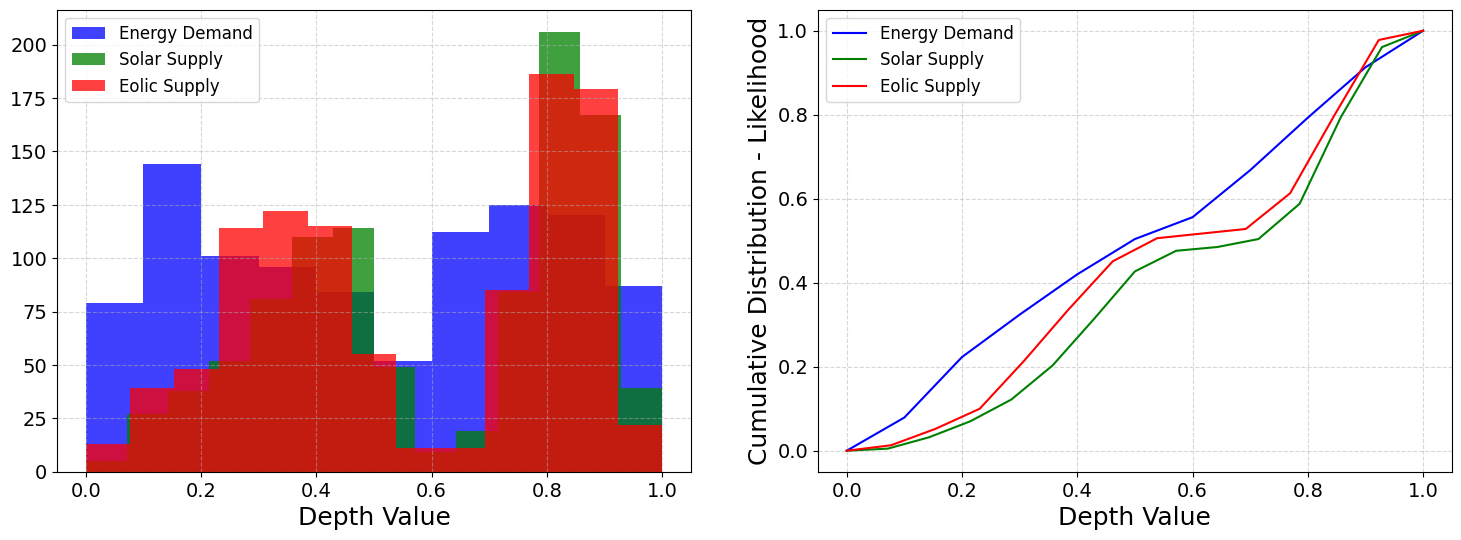}
    \caption{Data-depth distribution histogram and cumulative values for representative synthetic data for Paris. Each color represents a synthetic dataset depth value distribution. Depth values are normalized to stretch between 0 and 1.}
    \label{fig:depthDist}
\end{figure}

Depth values are used to create a way to select one out of all synthetic curves from a uniform distribution, following ideas described in \cite{mozharovskyi2020nonparametric}, such selection is donne using the inverse mapping between the cumulative distribution of the depth values $F_{D_P\left(X_i\middle|X\right)}$ and a uniform distributed random variable $U\sim \mathcal{U}_{[0,1]}$. Depth distribution is shown in Figure \ref{fig:depthDist}, on the left side the histogram for each synthetic dataset and on the right the cumulative distribution. Depth values are normalize between $[0,1]$ keeping the underlying distribution the same. As we aim to use a mixed strategy framework, the number of bins are computed using the Freedman-Diaconis rule \cite{diaconis1980finite}, where number of bins $N$ is selected for each distribution using $N=\frac{max(D(\cdot\mid X))-min(D(\cdot\mid X))}{2 \times IQR\times n^{1/3}}$, where $D(\cdot\mid X)$ represents all depth values, $IQR$ is the interquartile range, and $n$ is the number of points in the distribution.

\begin{figure}[h!]
    \centering
    \includegraphics[width=1\linewidth]{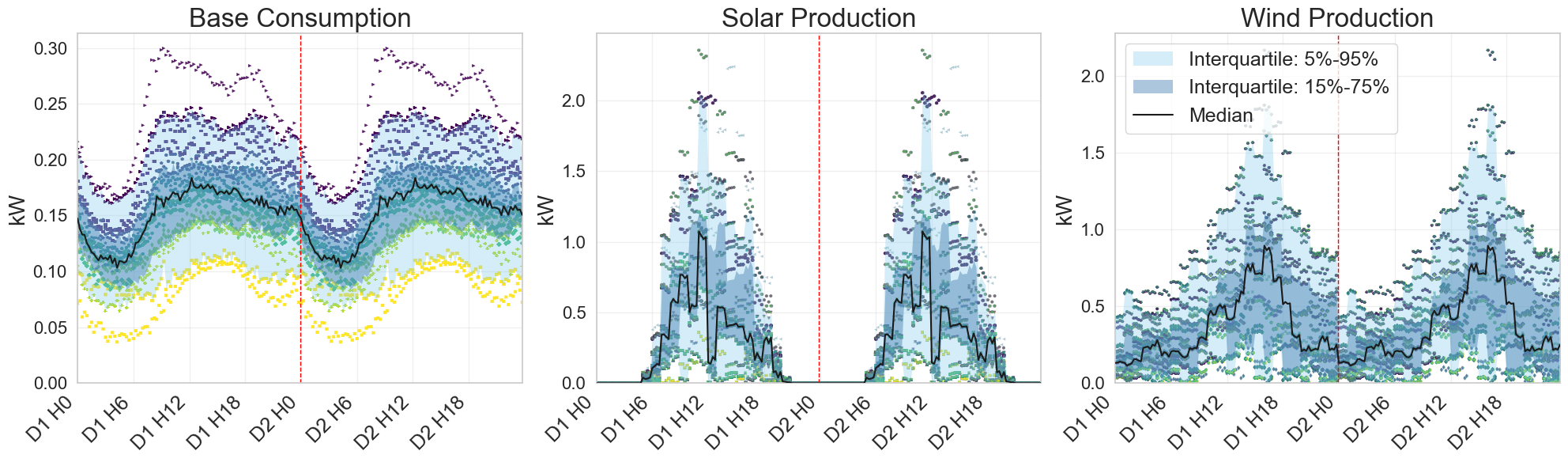}
    \caption{Profile distribution for base supply and demand during first two simulated summer days. On the left is the base load consumption accounting all agents, in the middle the solar production, and the right eolic energy production. Continuous black line is the median of the distribution, blue regions represent quantiles from 5\% to 95\% in light blue and 25\% to 75\% in dark blue, and scattered dots are the individual profiles.}
    \label{fig:gridProfConProd}
\end{figure}

Simulations are performed with a total of 1,000 agents, where 30\% are solar prosumers, 30\% wind prosumers, and 40\% are consumers, all of them with a two look-ahead days. Initial battery levels and total battery capacity are randomly chosen for each agent. Flexible load is formulated from base load and a random noise component, differentiating possible agents with same consumption curves. For the mixed strategy, the profile representing the center of each bin is chosen to be the profile whose depth value is the closest to the center depth, we also distinguish behaviors above and bellow the median curve for the representative behavior. Simulation is performed for two days. Consumption and production profiles for solar and wind energy are in Figure \ref{fig:gridProfConProd}, where full lines represent the mean profile, light blue regions represent the interquartile region from 5\% to 95\%, in dark blue the 25\% to 75\% region and the scattered dots the individual profiles. Scattered points are the individual profiles used in the simulation, indicating aspects about the distribution of behaviors and the individual profiles of agents.

\begin{figure}[h!]
    \centering
    \includegraphics[width=1\linewidth]{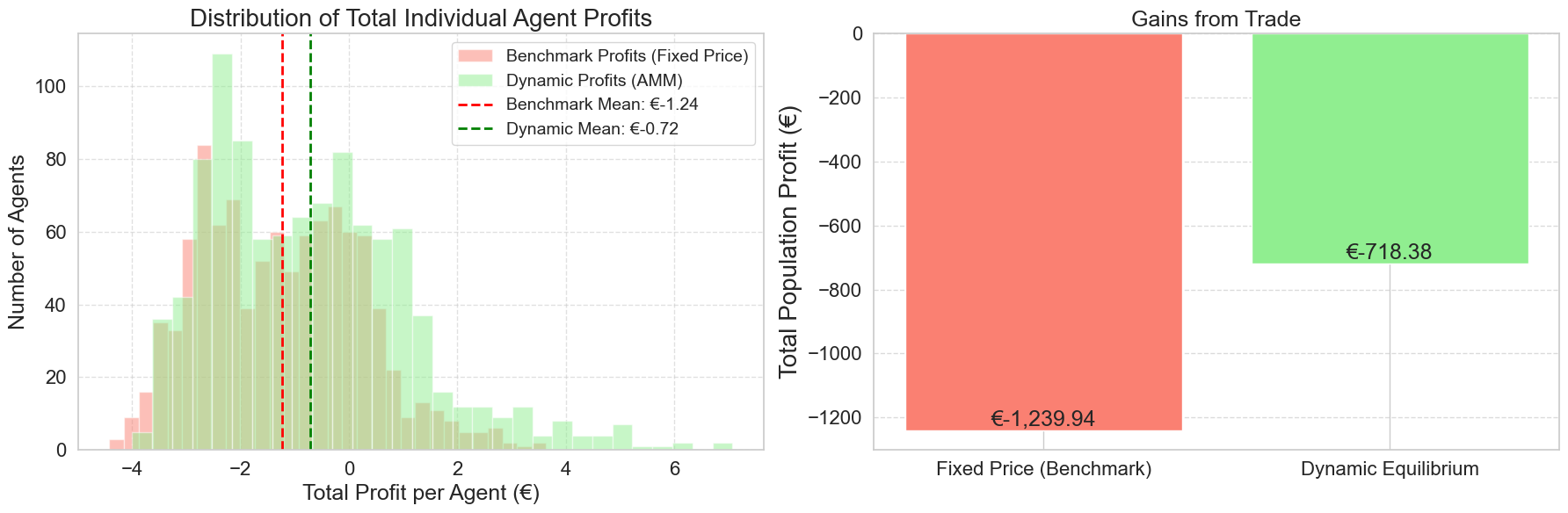}
    \caption{Distribution of profits and gains from trading comparing approaches with and without AMM. On the left side the distribution of profit for individual agents, on the right the total gain from trades comparing both approaches.}
    \label{fig:DistProfIndAgent}
\end{figure}

During these simulations the profit of each individual agent is recorded and compared with the benchmark profit using fixed prices with no dynamic approach. Both profit profiles and distributions are shown in the left part of Figure \ref{fig:DistProfIndAgent}, where green bars indicate the profits with AMM  and red ones the benchmark distribution. The total gains for the community is displayed on the right side of Figure \ref{fig:DistProfIndAgent} accounting for all agent profiles, indicating a 42\% gain with respect to the benchmark when using our dynamic equilibrium approach.

\begin{figure}
    \centering
    \includegraphics[width= \linewidth]{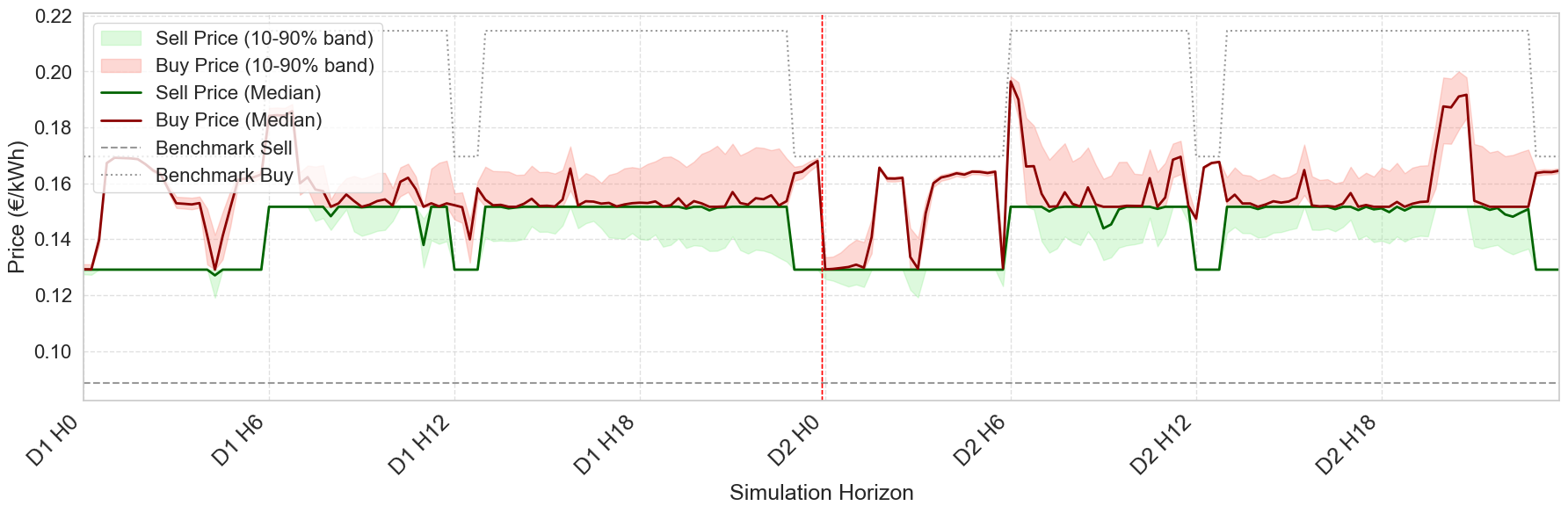}
    \caption{Simulated AMM price bands for all epochs. Full red lines represent the median buying price and red regions the 10\% to 90\% interquartile values, green lines represent the median selling price and green regions the 10\% to 90\% interquartile values. Dotted and dashed lines are the sell and buy benchmark prices for the grid, following the right axis.}
    \label{fig:AmmPriceBands}
\end{figure}

To further explore these results we can analyze the trading prices during the day in Figure \ref{fig:AmmPriceBands}, where buying and selling prices for AMM and for the benchmark are indicated in full lines and dashed lines, respectively. Buy and sell prices are subject to changes due to different aspects like energy supply, demand, and grid prices. Overall selling prices follow the benchmark trend for sell prices, where agent effort to match supply and demand is reflected on price changes.

\begin{figure}[h!]
    \centering
    \includegraphics[width=\linewidth]{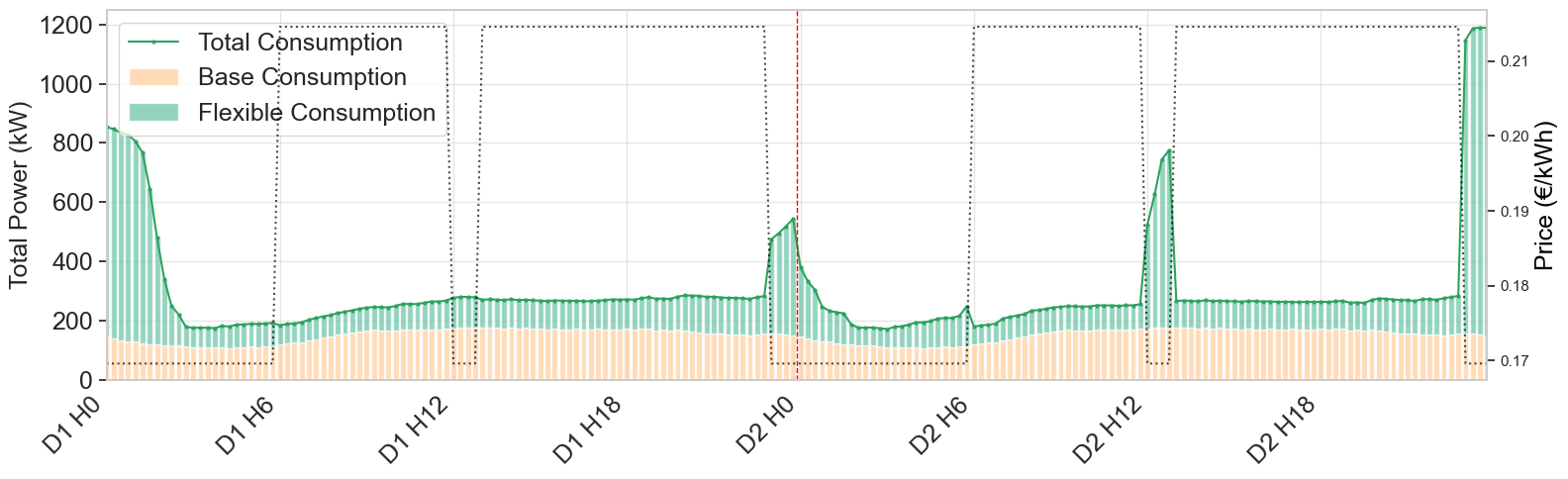}
    \caption{Aggregated community consumption. Peach color bars represent the base load consumption, green bars the flexible load and the full green line the total consumption considering both pieces. Benchmark grid prices are indicated by the dashed gray line following the right axis.}
    \label{fig:AggCoCuns}
\end{figure}

These price differences, mainly the benchmark one, interfere in the flexible load consumption allocation, as indicated in Figure \ref{fig:AggCoCuns}. Dynamic allocation favors assigning more consumptions in low price regions. Dynamic pricing seems to cause a lower impact in the placing of the flexible consumption, being mostly dictated by the (fixed) grid price, once it also impact and dictates community buy and sell prices.

\begin{figure}[h!]
    \centering
    \includegraphics[width= \linewidth]{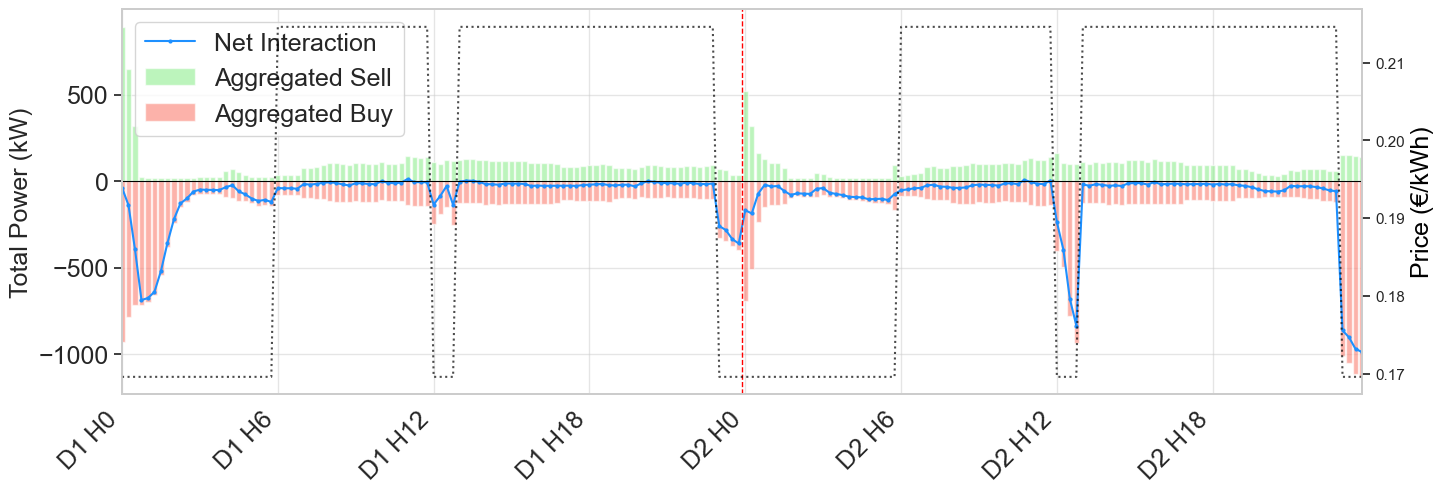}
    \caption{Aggregated grid interaction between grid and prosumers. Green bars represent the aggregated energy sold to the grid and red bars the aggregated energy bought for the community. Blue line indicate the net interaction between grid and community.}
    \label{fig:AggGridInter}
\end{figure}

We can see the impact of the grid interaction with the community in Figure \ref{fig:AggGridInter}, where green bars represent the community selling energy to the grid and red bars energy purchase by the community. The blue line represents the net interaction between both agents, grid and prosumers, which is overall the community buying from the grid to match supply and demand. This net interaction is also impacted mostly by the grid price change, where agents are more prone to buy from the grid once it is cheaper, influencing other aspects like demand allocation and battery charge.

\begin{figure}[h!]
    \centering
    \includegraphics[width= \linewidth]{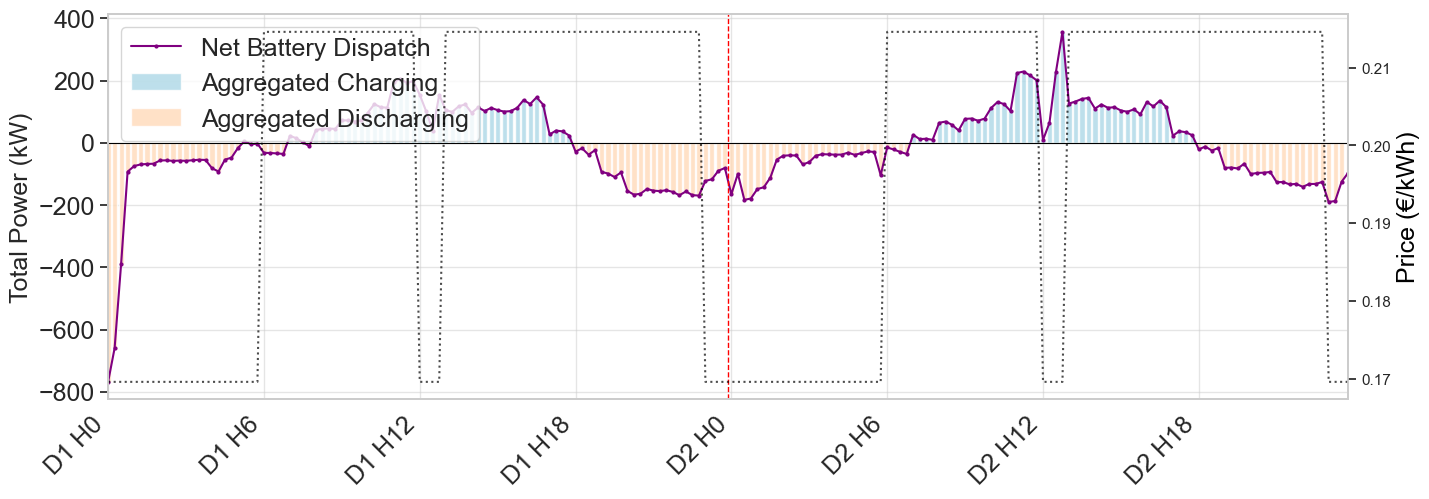}
    \caption{Aggregated battery dispatch profile. Blue bars indicate battery charging and peach bars battery discharge. Purple line represents the net dispatch profile of the community.}
    \label{fig:AggBatDis}
\end{figure}

Battery dispatch follows a different trend than demand allocation and net interaction. While both parameters are more sensible to the grid price, battery dispatch, in Figure \ref{fig:AggBatDis}, is more sensible to the internal community energy supply, where charging occurs mostly during solar hours and discharging when production is lower. Even thought another aspect impacts the charging and discharging profiles, grid price decreases affect the charging characteristic. In the image, blue bars represent moments where battery is being charged, while peach bars when battery is being discharged, and the purple line is the outline of the aggregated dispatch profile.

\begin{figure}
    \centering
    \includegraphics[width=1 \linewidth]{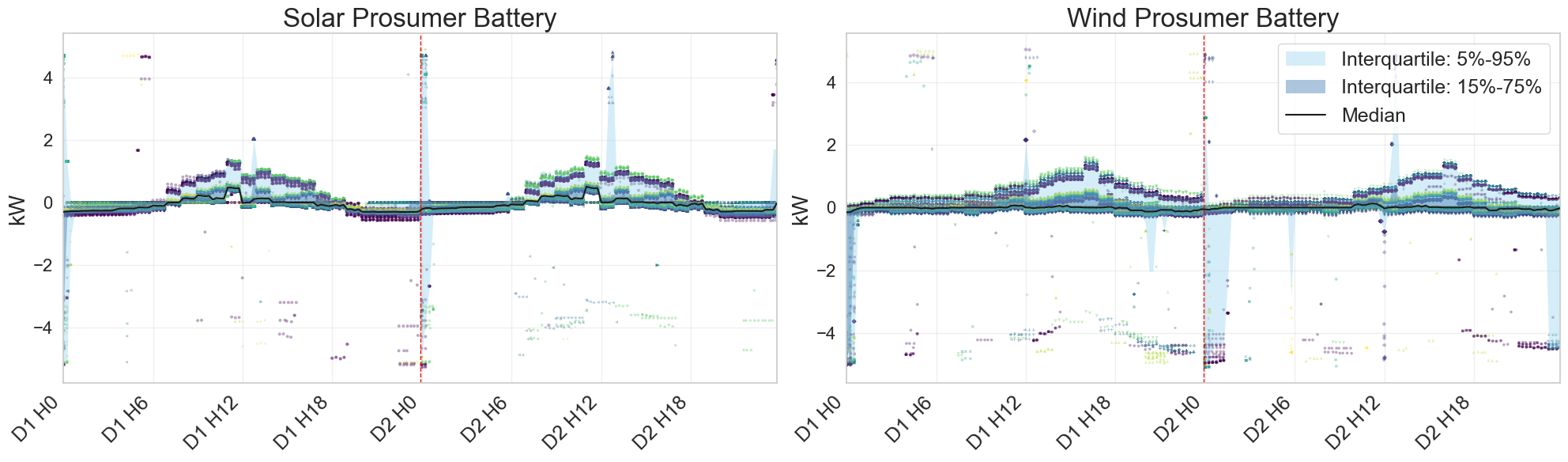}
    \caption{Aggregated battery state for distinct prosumer profiles. Left image is the battery state for solar prosumers, and the right image the battery state for wind prosumers. For all images, black line represents the median value, light blue regions represent the 5\% to 95\% interquartile range, dark blue regions represent the 25\% to 75\% interquartile range.}
    \label{fig:AggBatteryInd}
\end{figure}

To deepen the visualization of the results of the mixed strategy, aggregated results for each type of agent are analyzed. First, Figure \ref{fig:AggBatteryInd} show the aggregated battery charge/discharge for solar prosumers and wind prosumers on the left and right image respectively. These results reflect the sensibility of each prosumer type regarding weather conditions, where battery charging for solar prosumers is concentrated during sunny hours, and discharging otherwise, while wind prosumers do not have such constrains and can charge regardless. Scattered dots indicate individual behaviors during the simulation and the colored regions are the interquartile ranges, indicting these distinctions in the mixed strategy. 

\begin{figure}
    \centering
    \includegraphics[width=1 \linewidth]{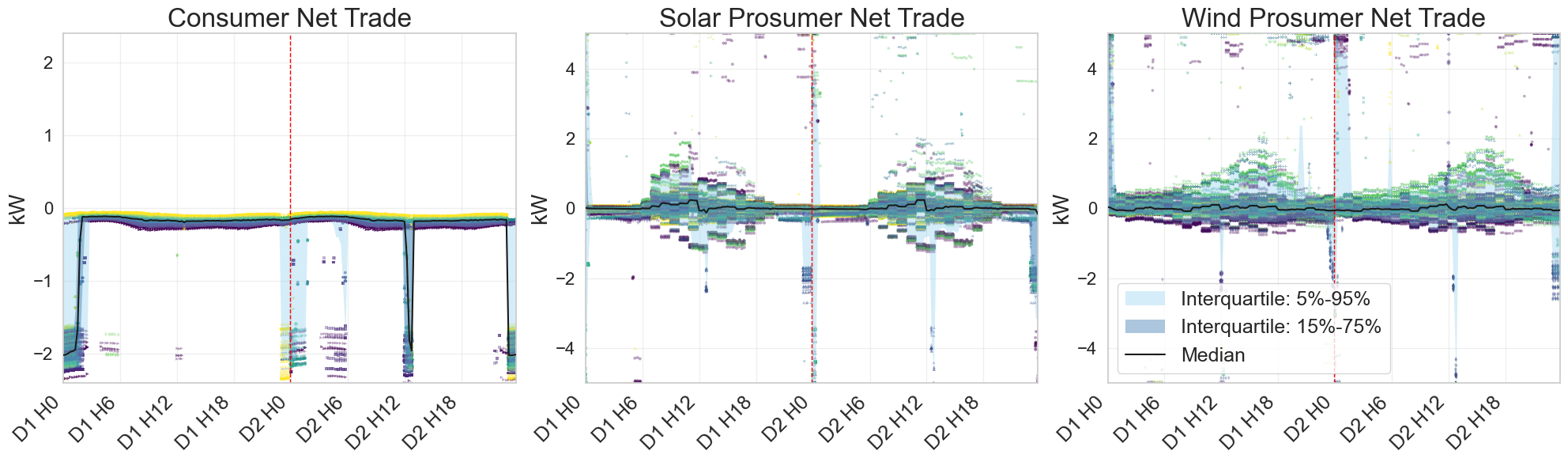}
    \caption{Aggregated net trading profiles for distinct agent types. Left image represents consumer net trade, middle image is the net trade for solar prosumers, and the right image net trade for wind prosumers. For all images, black line represents the median value, light blue regions represent the 5\% to 95\% interquartile range, dark blue regions represent the 25\% to 75\% interquartile range.}
    \label{fig:AggNetTradeProfile}
\end{figure}

Second, net trading profiles, in Figure \ref{fig:AggNetTradeProfile}, illustrate the differences between pure consumers on the left graph, solar prosumers on the center, and wind prosumers on the right. Some similar hourly trends seen in the battery state are present for the trading profile, where solar prosumers have their trades concentrated during sun hours and wind prosumers have more disperse profiles during the day. Pure consumer agents present a more grid price sensible profile, where most of the trades occur during off-peak hours, both during night time and at noon. Even thought off-peak prices play a role in the net trading of energy supplying prosumers, its importance is greater for pure consumers, not being as robust as suppliers.

\section{Conclusion}
\label{sec:conclusion}

In this paper, we have provided a decentralized market clearing mechanism for local power grids based on the design principles of Automated Market Makers in decentralized finance. We started posing axioms that guarantee an ideal functioning of the market. We then showed that our axioms are satisfied by an AMM based on batch execution, proportional payments, and concentrated liquidity. The market so designed implements a potential game, enabling an exact welfare equivalence between decentralized equilibrium and the solution of a social planner's problem. Numerical experiments based on  data from the Paris metropolitan region illustrate substantial gains from trade relative to grid-only benchmarks.

Several promising research directions extend naturally from our analysis.  
First, one may endow the AMM with its own inventory: a community battery. This introduces the AMM's battery level as a new state variable. The pricing functions can be defined as using the state of the battery rather than the energy supplied in a given time slot. Besides exporting on importing energy, the community will have to decide on how much energy to charge or discharge. Second, the aggregation and routing of \emph{multiple interacting microgrids} through a hierarchy of AMMs opens the door to scalable regional coordination. The freamework can later be extended to perform systemic risk analysis and compare shock propagation  between a centralized energy infrastructure and a decentralized one made of interconnected local grids operating through AMMs.

Finally, while our potential-game characterization holds for any axiom-compliant curve assuming linear prosumer payoffs, one can define a Planner with distributional objectives and go beyond the linear model. This opens the question of how to design trading functions or dynamic prices that encode such distributional preferences. Understanding how to extend the our results in this direction---while retaining tractability---is an important direction for future work.

\bibliography{bibliography}
\bibliographystyle{apalike}
\addcontentsline{toc}{section}{\refname}
\appendix

\newpage
\section{Notation Table}

\begin{table}[h!]
\centering
\caption{Notation for Prosumer Optimization Model}
\label{tab:prosumer_notation}
\begin{adjustbox}{width=\textwidth} 
\begin{threeparttable}
\begin{tabular}{@{}llllcl@{}}
\toprule
Symbol & Description & Domain & Dim. & Unit & Type \\
\midrule
\multicolumn{6}{l}{\textit{Indices}} \\
$e$ & Epoch index & $\mathbb{N}^+$ & Scalar & -- & Index \\
$t$ & Timestep index within epoch & $\{1, \dots, T\}$ & Scalar & -- & Index \\
$\Delta$ & Physical timestep duration & $\mathbb{R}^+$ & Scalar & h & Parameter \\
$T$ & Number of timesteps per epoch & $\mathbb{N}^+$ & Scalar & -- & Parameter \\
\addlinespace
\multicolumn{6}{l}{\textit{Prosumer Parameters}} \\
$B_n$ & Battery energy capacity & $\mathbb{R}^+_0$ & Scalar & kWh & Parameter \\
$K_n$ & Battery power limit & $\mathbb{R}^+_0$ & Scalar & kW & Parameter \\
$X_n$ & Grid connection limit & $\mathbb{R}^+_0$ & Scalar & kW & Parameter \\
$\omega_{nt}$ & Local generation & $\mathbb{R}^+_0$ & $T \times 1$ & kW & Parameter \\
$\alpha^{\text{base}}_{nt}$ & Baseline consumption & $\mathbb{R}^+_0$ & $T \times 1$ & kW & Parameter \\
$\alpha^{\text{flex}}_{n}$ & Flexible energy requirement & $\mathbb{R}^+_0$ & Scalar & kWh & Parameter \\
\addlinespace
\multicolumn{6}{l}{\textit{Control Variables}} \\
$s_{nt}$ & Power sold to market & $[0, X_n]$ & $T \times 1$ & kW & Variable \\
$d_{nt}$ & Power bought from market & $[0, X_n]$ & $T \times 1$ & kW & Variable \\
$k_{nt}$ & Battery power flow (charge $>0$) & $[-K_n, K_n]$ & $T \times 1$ & kW & Variable \\
$p_{nt}$ & Total consumption power & $\mathbb{R}^+_0$ & $T \times 1$ & kW & Variable \\
$x_{nt}$ & Net power traded ($s_{nt}-d_{nt}$) & $[-X_n, X_n]$ & $T \times 1$ & kW & Derived \\
$b_{nt}$ & Battery state of charge (end of $t$) & $[0, B_n]$ & $T \times 1$ & kWh & Derived \\
\addlinespace
\multicolumn{6}{l}{\textit{Prices}\tnote{*}} \\
$\underline{\lambda}_t$ & Effective Grid Feed-in Price & $\mathbb{R}^+_0$ & $T \times 1$ & €/kW & Parameter \\
$\overline{\lambda}_t$ & Effective Grid Retail Price & $\mathbb{R}^+_0$ & $T \times 1$ & €/kW & Parameter \\
$\bar{r}_t$ & Effective Market Sell Price & $\mathbb{R}^+_0$ & $T \times 1$ & €/kW & Parameter \\
$\bar{c}_t$ & Effective Market Buy Price & $\mathbb{R}^+_0$ & $T \times 1$ & €/kW & Parameter \\
\addlinespace
\multicolumn{6}{l}{\textit{Objective Function}} \\
$P_n(\bm{x}_t)$ & Net payment for timestep $t$ & $\mathbb{R}$ & $T \times 1$ & € & Derived \\
$\Pi_n(\cdot)$ & Prosumer value function & $\mathbb{R}$ & Scalar & € & Derived \\
$\gamma$ & Inter-epoch discount factor & $[0, 1)$ & Scalar & -- & Parameter \\
\bottomrule
\end{tabular}
\begin{tablenotes}
  \item[*]\footnotesize{All price parameters in the optimization problem are scaled by the timestep duration $\Delta$(hours). This allows costs to be computed directly as the dot product of price and power: $[\text{€/kW}] = [\text{€/kWh (Raw)}] \times [\text{h}]$.}
\end{tablenotes}
\end{threeparttable}
\end{adjustbox}
\end{table}

\newpage\section{Prosumer's Best Response in Matrix Form and Lagrangian}
\label{sec:KKT-prosumer-matrix}

Condier the prosumer's single-epoch optimization problem of maximizing profit given forecasted prices $(\bar{\mathbf{r}}, \bar{\mathbf{c}})$.  Let the stacked decision vector be $\mathbf{z} \in \mathbb{R}^{5T}$, where $\mathbf{z}^\top = \begin{pmatrix} \mathbf{s}^\top & \mathbf{d}^\top & \mathbf{k}^\top & \mathbf{p}^\top & \mathbf{b}^\top \end{pmatrix}$. The marginal effects of the controls on the objective are given by the vector $\mathbf{c}_{\pi} \in \mathbb{R}^{5T}$, $\mathbf{c}_{\pi}^\top = \begin{pmatrix} \bar{\mathbf{r}}^\top & -\bar{\mathbf{c}}^\top & \mathbf{0}^\top & \mathbf{0}^\top & \mathbf{0}^\top \end{pmatrix}$. The optimization problem in matrix form is:
\begin{align*}
\max_{\mathbf{z}} \quad & \mathbf{c}_{\pi}^\top \mathbf{z}  + \gamma \bar{\Pi}(b_T) \\
\text{s.t.} \quad & \mathbf{A}_{\text{eq}} \mathbf{z} = \mathbf{b}_{\text{eq}} \\
& \mathbf{A}_{\text{ineq}} \mathbf{z} \le \mathbf{b}_{\text{ineq}}
\end{align*}
where $b_T$ is the last element of the $\mathbf{b}$ sub-vector within $\mathbf{z}$. The constraint matrices $\mathbf{A}_{\text{eq}} \in \mathbb{R}^{(2T+1) \times 5T}$, $\mathbf{A}_{\text{ineq}} \in \mathbb{R}^{10T \times 5T}$ and vectors $\mathbf{b}_{\text{eq}} \in \mathbb{R}^{2T+1}$, $\mathbf{b}_{\text{ineq}} \in \mathbb{R}^{10T}$ are defined as:
\[
\mathbf{A}_{\text{eq}} =
\begin{pmatrix}
-\mathbf{I} & \mathbf{I} & -\mathbf{I} & -\mathbf{I} & \mathbf{0} \\  
\mathbf{0}^\top & \mathbf{0}^\top & \mathbf{0}^\top & \mathbf{1}^\top & \mathbf{0}^\top \\ 
\mathbf{0} & \mathbf{0} & -\mathbf{L} & \mathbf{0} & \mathbf{I} 
\end{pmatrix}
\quad
\mathbf{b}_{\text{eq}} =
\begin{pmatrix}
-\boldsymbol{\omega} \\
\alpha^{\text{flex}} + \mathbf{1}^\top \boldsymbol{\alpha}^{\text{base}} \\
b_0 \mathbf{1}
\end{pmatrix}
\]
\[
\mathbf{A}_{\text{ineq}} =
\begin{pmatrix}
-\mathbf{I} & \mathbf{0} & \mathbf{0} & \mathbf{0} & \mathbf{0} \\ 
\mathbf{I} & \mathbf{0} & \mathbf{0} & \mathbf{0} & \mathbf{0} \\  
\mathbf{0} & -\mathbf{I} & \mathbf{0} & \mathbf{0} & \mathbf{0} \\ 
\mathbf{0} & \mathbf{I} & \mathbf{0} & \mathbf{0} & \mathbf{0} \\  
\mathbf{0} & \mathbf{0} & -\mathbf{I} & \mathbf{0} & \mathbf{0} \\ 
\mathbf{0} & \mathbf{0} & \mathbf{I} & \mathbf{0} & \mathbf{0} \\  
\mathbf{0} & \mathbf{0} & \mathbf{0} & -\mathbf{I} & \mathbf{0} \\ 
\mathbf{0} & \mathbf{0} & \mathbf{0} & \mathbf{0} & -\mathbf{I} \\ 
\mathbf{0} & \mathbf{0} & \mathbf{0} & \mathbf{0} & \mathbf{I}     
\end{pmatrix}
\quad
\mathbf{b}_{\text{ineq}} =
\begin{pmatrix}
\mathbf{0} \\
X \mathbf{1} \\
\mathbf{0} \\
X \mathbf{1} \\
K \mathbf{1} \\
K \mathbf{1} \\
-\boldsymbol{\alpha}^{\text{base}} \\
\mathbf{0} \\
B \mathbf{1}
\end{pmatrix}
\]

\paragraph{Lagrangian.}
Let $\boldsymbol{\chi} \in \mathbb{R}^{2T+1}$ be the Lagrange multipliers for the equality constraints, partitioned as $\boldsymbol{\chi} = (\boldsymbol{\theta}^\top, \nu, \boldsymbol{\lambda}^\top)^\top$. Let $\boldsymbol{\phi} \in \mathbb{R}^{9T}$, $\boldsymbol{\phi} \ge \mathbf{0}$ be the multipliers for the inequality constraints, partitioned as $\boldsymbol{\phi} = ((\boldsymbol{\zeta}^s)^\top, (\boldsymbol{\xi}^s)^\top, (\boldsymbol{\zeta}^d)^\top, (\boldsymbol{\xi}^d)^\top, (\boldsymbol{\kappa}^L)^\top, (\boldsymbol{\kappa}^U)^\top, \boldsymbol{\mu}^\top, (\boldsymbol{\iota}^L)^\top, (\boldsymbol{\iota}^U)^\top)^\top$. The Lagrangian is given by:
\[
\mathcal{L}(\mathbf{z}, \boldsymbol{\psi}, \boldsymbol{\phi}) = \left[ \mathbf{c}_{\pi}^\top \mathbf{z} + \gamma \bar{\Pi}(b_T) \right] + \boldsymbol{\chi}^\top (\mathbf{b}_{\text{eq}} - \mathbf{A}_{\text{eq}} \mathbf{z}) + \boldsymbol{\phi}^\top (\mathbf{b}_{\text{ineq}} - \mathbf{A}_{\text{ineq}} \mathbf{z})
\]

\paragraph{KKT Stationarity Condition.}
The gradient of the Lagrangian with respect to the primal variables $\mathbf{z}$ must be zero at the optimum $(\mathbf{z}^*, \boldsymbol{\chi}^*, \boldsymbol{\phi}^*)$. Let $\mathbf{e}_T = (0, \dots, 0, 1)^\top \in \mathbb{R}^T$ select the last element, and define the corresponding selector vector in $\mathbb{R}^{5T}$ as $\mathbf{e}_{5T}^\top = (\mathbf{0}^\top, \mathbf{0}^\top, \mathbf{0}^\top, \mathbf{0}^\top, \mathbf{e}_T^\top)$. Let $\pi^b \in \partial \bar{\Pi}(b_T^*)$ be an element of the supergradient of the continuation value function evaluated at the optimal terminal state $b_T^*$. The stationarity condition $\nabla_{\mathbf{z}} \mathcal{L} = \mathbf{0}$ yields:
\[
\mathbf{c}_{\pi} + \gamma \pi^b \mathbf{e}_{5T} - \mathbf{A}_{\text{eq}}^\top \boldsymbol{\chi}^* - \mathbf{A}_{\text{ineq}}^\top \boldsymbol{\phi}^* = \mathbf{0}
\]
Expanding this matrix equation by substituting the definitions of $\mathbf{A}_{\text{eq}}$, $\mathbf{A}_{\text{ineq}}$, $\boldsymbol{\chi}^* = (\boldsymbol{\theta}^{*\top}, \nu^*, \boldsymbol{\lambda}^{*\top})^\top$, and the partitioned $\boldsymbol{\phi}^*$ gives the first-order conditions for each block of variables:
\begin{align*}
\nabla_{\mathbf{s}} \mathcal{L} &= \bar{\mathbf{r}} + \boldsymbol{\theta}^* - \boldsymbol{\zeta}^{s*} + \boldsymbol{\xi}^{s*} = \mathbf{0} \\
\nabla_{\mathbf{d}} \mathcal{L} &= -\bar{\mathbf{c}} + \boldsymbol{\theta}^* - \boldsymbol{\zeta}^{d*} + \boldsymbol{\xi}^{d*} = \mathbf{0} \\
\nabla_{\mathbf{k}} \mathcal{L} &= \boldsymbol{\theta}^* +  \mathbf{L}^\top \boldsymbol{\lambda}^* - \boldsymbol{\kappa}^{L*} + \boldsymbol{\kappa}^{U*} = \mathbf{0} \\
\nabla_{\mathbf{p}} \mathcal{L} &= \boldsymbol{\theta}^* - \nu^* \mathbf{1} - \boldsymbol{\mu}^* = \mathbf{0} \quad (\text{assuming } \boldsymbol{\zeta}^{p*} = \mathbf{0}) \\
\nabla_{\mathbf{b}} \mathcal{L} &= \gamma \pi^b \mathbf{e}_T - \boldsymbol{\lambda}^* + \boldsymbol{\iota}^{L*} - \boldsymbol{\iota}^{U*} = \mathbf{0}
\end{align*}
These equations define the relationships between the optimal primal variables (implicitly through complementary slackness) and the optimal dual variables (shadow prices). For instance, the condition for $\mathbf{p}$ implies $\boldsymbol{\theta}^* = \nu^*  \mathbf{1} + \boldsymbol{\mu}^*$, equating the marginal value of energy to its marginal value in consumption. Similarly, combining the conditions for $\mathbf{k}$ and $\mathbf{b}$ relates $\boldsymbol{\theta}^*$ to the discounted continuation value and the shadow prices of battery constraints over time. The conditions for $\mathbf{s}$ and $\mathbf{d}$ directly link $\boldsymbol{\theta}^*$ to the market prices, adjusted by the multipliers for binding non-negativity ($\boldsymbol{\zeta}^*$) or grid capacity ($\boldsymbol{\xi}^*$) constraints.

\paragraph{Other KKT Conditions.}
In addition to stationarity, the optimal solution must satisfy:
\begin{itemize}[noitemsep]
    \item \textbf{Primal Feasibility:} $\mathbf{A}_{\text{eq}} \mathbf{z}^* = \mathbf{b}_{\text{eq}}$ and $\mathbf{A}_{\text{ineq}} \mathbf{z}^* \le \mathbf{b}_{\text{ineq}}$.
    \item \textbf{Dual Feasibility:} $\boldsymbol{\phi}^* \ge \mathbf{0}$.
    \item \textbf{Complementary Slackness:} $\boldsymbol{\phi}^{*\top} (\mathbf{b}_{\text{ineq}} - \mathbf{A}_{\text{ineq}} \mathbf{z}^*) = 0$. This implies $\phi_i^* = 0$ for any non-binding inequality constraint $i$.
\end{itemize}
These conditions collectively characterize the optimal primal-dual solution $(\mathbf{z}^*, \boldsymbol{\chi}^*, \boldsymbol{\phi}^*)$ to the prosumer's best-response LP.

\section{Proofs}

\subsubsection*{Proof of \cref{prop:anonimity}}

\begin{proof}
\textbf{($\Rightarrow$) Anonymity implies the existence of $\Phi$.}
Assume the market mechanism is anonymous. Let $\bm{x}_{-n}$ and $\bm{x}'_{-n}$ be any two allocation vectors for the other prosumers that produce the same multiset, i.e., $\Sigma(\bm{x}_{-n}) = \Sigma(\bm{x}'_{-n})$. By definition, this implies that $\bm{x}'_{-n}$ is a permutation of $\bm{x}_{-n}$, which we denote by $\pi$. The definition of anonymity states that $P_n(x_n, \bm{x}_{-n}) = P_n(x_n, \pi(\bm{x}_{-n}))$, which is equivalent to $P_n(x_n, \bm{x}_{-n}) = P_n(x_n, \bm{x}'_{-n})$. Since the payment $P_n$ is invariant for any vector that generates the same multiset, it must be a function of the multiset itself. Thus, a function $\Phi$ as described exists and is well-defined.

\vspace{\baselineskip}

\noindent\textbf{($\Leftarrow$) The existence of $\Phi$ implies Anonymity.}
Conversely, assume there exists a function $\Phi$ such that $P_n(x_n, \bm{x}_{-n}) = \Phi(x_n, \Sigma(\bm{x}_{-n}))$. Let $\pi$ be any permutation of the other prosumers. A key property of multisets is that they are invariant to permutation, meaning $\Sigma(\bm{x}_{-n}) = \Sigma(\pi(\bm{x}_{-n}))$. This allows the following chain of equalities:
\begin{align*}
P_n(x_n, \pi(\bm{x}_{-n})) &= \Phi(x_n, \Sigma(\pi(\bm{x}_{-n}))) && \text{(by assumption)} \\
&= \Phi(x_n, \Sigma(\bm{x}_{-n})) && \text{(by multiset invariance)} \\
&= P_n(x_n, \bm{x}_{-n}) && \text{(by assumption)}
\end{align*}
The result satisfies the definition of anonymity.
\end{proof}

\subsubsection*{Proof of \cref{prop:affine-payment}}

\begin{proof}
\textbf{($\Rightarrow$) Coalition-Proofness implies the Linear Form.}
Assume the mechanism is coalition-proof. For a coalition of sellers $\mathcal{I}$, the definition requires $\sum_{n \in \mathcal{I}} \Psi(s_n, s, d) = \Psi(\sum_{n \in \mathcal{I}} s_n, s, d)$. For a fixed aggregate state $(s, d)$, let us define $f(\sigma) := \Psi(\sigma, s, d)$ for $\sigma>0$. The condition thus becomes $\sum_{n \in \mathcal{I}} f(s_n) = f(\sum_{n \in \mathcal{I}} s_n)$, which is Cauchy's functional equation. Its unique continuous solution is linear: $f(\sigma) = k \cdot \sigma$. Since the proportionality constant $k$ may depend on the state $(s,d)$, we define it as $k := r(s,d)$, yielding $\Psi(s_n, s, d) = s_n \cdot r(s,d)$ for sellers. A symmetric argument for a coalition of buyers yields $\Psi(-d_n, s, d) = -d_n \cdot c(s,d)$. Combining these cases for any prosumer establishes the linear form $P_n(\bm{x}) = s_n \cdot r(s,d) - d_n \cdot c(s,d)$.

\vspace{\baselineskip} 

\noindent\textbf{($\Leftarrow$) The Linear Form implies Coalition-Proofness.}
Conversely, assume the payment function has the linear form $\Psi(x_n, s, d) = s_n \cdot r(s,d) - d_n \cdot c(s,d)$. To verify the condition for a coalition of sellers $\mathcal{I}$ (where $d_n=0$ for all $n \in \mathcal{I}$), we show that both sides of the additivity requirement are equal. The left-hand side is $\sum_{n \in \mathcal{I}} \Psi(s_n, s, d) = \sum_{n \in \mathcal{I}} s_n \cdot r(s,d) = (\sum_{n \in \mathcal{I}} s_n) \cdot r(s,d)$. The right-hand side is $\Psi(\sum_{n \in \mathcal{I}} s_n, s, d) = (\sum_{n \in \mathcal{I}} s_n) \cdot r(s,d)$. Since the two sides are equal, the condition holds. A symmetric argument applies to buyers, completing the proof.
\end{proof}

\subsubsection*{Proof of \cref{prop:general-compliance}}
\begin{proof}

\textit{Preliminaries.}
Fix a session \(t\) and suppress \(t\) in notation when unambiguous. The pool is (re-)anchored at the start with reserves \((E_0,M_0)=(s,0)\) and level \(K=K(s,\underline{\lambda},\overline{\lambda})\).
The internal marginal price is the slope of the level set
\[
\rho(E,M)\;=\;\biggl|\frac{\partial_E \psi(E,M)}{\partial_M \psi(E,M)}\biggr|
\in(\underline{\lambda},\overline{\lambda})
\quad\text{by concentrated liquidity.}
\]
Let \(\Delta E=\min\{s,d\}\). The pool value gathered from internal trades is
\[
M(s,\Delta E)\;=\;\psi^{-1}_{\,E=E_0-\Delta E}\!\bigl(K(s,\underline{\lambda},\overline{\lambda})\bigr).
\]
Total costs/revenues and per-unit prices follow the construction in \S\ref{AMM-construction}:
\[
\begin{aligned}
&\text{If }d=s: && C_{\mathrm{tot}}=R_{\mathrm{tot}}=M(s,s),\\
&\text{If }d<s: && C_{\mathrm{tot}}=M(s,d),\quad R_{\mathrm{tot}}=M(s,d)+\underline{\lambda}(s-d),\\
&\text{If }d>s: && R_{\mathrm{tot}}=M(s,s),\quad C_{\mathrm{tot}}=M(s,s)+\overline{\lambda}(d-s),
\end{aligned}
\qquad
c(s,d)=\frac{C_{\mathrm{tot}}}{d},\quad r(s,d)=\frac{R_{\mathrm{tot}}}{s}.
\]
For prosumer \(n\) with \(x_n=s_n-d_n\), payment is \(P_n(\bm{x})=s_n r(s,d)-d_n c(s,d)\).

We now prove the \emph{if and only if} statement by showing both directions.

\bigskip
\noindent\textbf{(\(\Rightarrow\)) If the AMM is constructed from a trading function \(\psi\) with the stated properties, then the mechanism satisfies all axioms.}

\paragraph{Anonymity.}
By batch execution and aggregation, \(P_n\) depends only on \((x_n,s,d)\). Hence for any permutation \(\pi\) of \(\bm{x}_{-n}\),
\[
P_n(x_n,\bm{x}_{-n}) \;=\; \Psi(x_n,s,d) \;=\; P_n(x_n,\pi(\bm{x}_{-n})),
\]
which is Definition~\ref{def:anonymity}.

\paragraph{Coalition-Proofness.}
Since \(P_n(\bm{x})=s_n r(s,d)-d_n c(s,d)\) is linear in \((s_n,d_n)\), for any coalitions \(\mathcal{I}\subseteq\{x_n>0\}\), \(\mathcal{J}\subseteq\{x_n<0\}\),
\[
\sum_{n\in\mathcal{I}}\!\Psi(s_n,s,d)
= r(s,d)\sum_{n\in\mathcal{I}}s_n
= \Psi\!\left(\sum_{n\in\mathcal{I}}s_n,s,d\right),
\]
and analogously for buyers, which is exactly \eqref{eq:cproof}.

\paragraph{Monotonicity (positive prices).}
Strict monotonicity of \(\psi\) in each argument yields \(\partial_E\psi>0\) and \(\partial_M\psi>0\) on \(\mathbb{R}^2_+\).
Hence \(\rho(E,M)>0\) for internal trades; imbalance terms use \(\underline{\lambda},\overline{\lambda}>0\). Therefore \(r(s,d)>0\) and \(c(s,d)>0\) for all \(s,d>0\).

\paragraph{Individual Rationality.}
Concentrated liquidity enforces \(\rho(E,M)\in(\underline{\lambda},\overline{\lambda})\) internally. If \(s=d\), then
\(\underline{\lambda} < r(s,s)=c(s,s) < \overline{\lambda}\).
If \(s>d\) (surplus), the last \(s-d\) units clear at \(\underline{\lambda}\); proportional sharing implies
\(r(s,d)\to \underline{\lambda}\) as surplus grows and \(c(s,d)=M(s,d)/d<\overline{\lambda}\).
Thus
\[
r(s,d)=\underline{\lambda}\ \text{if } s\ge d,\qquad
c(s,d)<\overline{\lambda}\ \text{if } s>d,
\]
with symmetric statements when \(d\ge s\). This is precisely the IR boundary behavior.

\paragraph{No-Arbitrage.}
Internally, every marginal trade clears at \(\rho\in(\underline{\lambda},\overline{\lambda})\). Under proportional sharing, the average internal buy price does not fall below the average internal sell price; external balancing only widens the spread because \(\overline{\lambda}>\underline{\lambda}\). Hence \(r(s,d)\le c(s,d)\) for all \((s,d)\).

\paragraph{Homogeneity.}
Homotheticity of \(\psi\) implies \(\psi=\varphi\circ g\) for monotone \(\varphi\) and homogeneous \(g\) (degree \(k>0\)). Level sets of \(\psi\) are radial scalings; the slope \(\rho(E,M)\) depends only on \(M/E\). Under aggregation, \((s,d)\) enter via \(y=s/d\). Therefore
\[
r(\alpha s,\alpha d)=r(s,d),\qquad c(\alpha s,\alpha d)=c(s,d)\quad(\alpha>0),
\]
so both prices are homogeneous of degree zero.

\paragraph{Responsiveness.}
Quasi-concavity of \(\psi\) makes upper level sets \(\{\psi\geq K\}\) convex. This implies that their slope \(\rho\) is strictly decreasing along the energy axis (for fixed \(K\)) and strictly increasing along the money axis. Increasing \(s\) (holding \(d\) fixed) shifts execution toward portions with lower \(\rho\); increasing \(d\) shifts toward higher \(\rho\). Averaging via proportional sharing preserves these comparative statics:
\[
\frac{\partial r}{\partial s}<0,\quad \frac{\partial r}{\partial d}>0,\qquad
\frac{\partial c}{\partial s}<0,\quad \frac{\partial c}{\partial d}>0.
\]

\paragraph{Budget-Balance (accounting).}
By the above definitions and external legs, the operator’s net cash position is
\[
C_{\mathrm{tot}}
+\mathbf{1}_{\{s>d\}}\underline{\lambda}(s-d)
-R_{\mathrm{tot}}
-\mathbf{1}_{\{d>s\}}\overline{\lambda}(d-s)
=0.
\]
Equivalently,
\[
d\,c(s,d)-s\,r(s,d)
=
\mathbf{1}_{\{d>s\}}\overline{\lambda}(d-s)
-
\mathbf{1}_{\{s>d\}}\underline{\lambda}(s-d),
\]
so the mechanism is exactly balanced once the external trades are included, and the “self-sustaining” inequality in the axiom holds in the appropriate regions.

\bigskip
\noindent\textbf{(\(\Leftarrow\)) Conversely, suppose a market mechanism satisfies Anonymity, Coalition-Proofness, No-Arbitrage, Budget-Balance, Individual Rationality, Monotonicity, Homogeneity, and Responsiveness. Then it must arise from such an AMM with a trading function \(\psi\) satisfying the stated properties and with concentrated liquidity.}

\paragraph{Step 1: From axioms to batch execution and proportional payments.}
Anonymity implies prices cannot depend on order identity or timing, only on the multiset of allocations.
Coalition-proofness implies that splitting or merging trades cannot change total payments for a group of pure buyers or pure sellers. As in Proposition~\ref{prop:affine-payment}, this forces the payment to be linear in own quantity:
\[
P_n(\bm{x}) = s_n r(s,d) - d_n c(s,d),
\]
with uniform marginal prices \(r,c\) for all sellers and buyers. Thus trades must be executed in a single batch with proportional payments.

\paragraph{Step 2: Defining the demand curve.}
Fix a session \(t\) and a total initial energy supply  \(s>0\). Consider posting a scalar price \(\rho\) for energy. Let
\[
g_t(\rho)\in[0,s]
\]
denote the quantity of energy the mechanism is willing to \emph{continue to hold} at price \(\rho\); equivalently, \(s-g_t(\rho)\) is the quantity it is willing to sell at price \(\rho\). Formally, in the Myersonian framework of \citet{milionis2023myersonian},  \(g_t\) is the (right-continuous) inventory demand curve associated with the mechanism.

\paragraph{Claim 1.}
For all \(\rho<\underline{\lambda}_t\), we have \(g_t(\rho)=s\).

\emph{Proof.}
Suppose there exists \(\rho<\underline{\lambda}_t\) with \(g_t(\rho)<s\). Then at price \(\rho\) the mechanism is willing to sell \(s-g_t(p)>0\) units internally. A prosumer could buy this quantity from the AMM at price \(\rho\) and immediately resell it to the main grid at price \(\underline{\lambda}_t\), realizing a risk-free profit \((\underline{\lambda}_t-\rho)(s-g_t(\rho))>0\). This violates Individual Rationality and No-Arbitrage. Hence \(g_t(\rho)=s\) for all \(p<\underline{\lambda}_t\). \(\square\)

\paragraph{Claim 2}
The function \(g_t(\rho)\) is non-increasing in \(\rho\).

\emph{Proof.}
If \(g_t\) were not non-increasing, there would exist \(\rho_1<\rho_2\) with \(g_t(\rho_1)<g_t(\rho_2)\). Then an agent could buy at \(\rho_1\) when the mechanism is willing to sell more (lower energy reserves) and sell back at \(\rho_2\) when the mechanism is willing to hold more (lower energy reserves), generating internal cyclic arbitrage. This is precisely ruled out by the no-arbitrage characterization of inventory demand curves in \citet[Prop.~2.1]{milionis2023myersonian}. Hence \(g_t\) must be nonincreasing. \(\square\)

\paragraph{Claim 3}
For all \(\rho >\overline{\lambda}_t\), we have \(g_t(\rho)=0\).

\emph{Proof.}
Suppose \(g_t(\rho)>0\) for some \(\rho>\overline{\lambda}_t\). Consider a balanced market with \(s=d\). At price \(\rho\), the mechanism sells only \(s-g_t(\rho)\), leaving a residual inventory \(g_t(\rho)>0\) and the same aggregate demand. To clear the market while respecting IR, the AMM would have to sell \(g_t(\rho)\) units to the grid at \(\underline{\lambda}_t\) and simultaneously buy back the same amount from the grid at \(\overline{\lambda}_t\) to satisfy demand, which strictly violates Budget-Balance. Therefore \(g_t(\overline{\lambda}_t)=0\); by nonincreasingness, \(g_t(\rho)=0\) for all \(\rho>\overline{\lambda}_t\). \(\square\)

\paragraph{Step 3: Shape of \(g_t\) and existence of a quasi-concave potential.}
Claims 1–3 imply that \(g_t\) has the following form:
\[
g_t(\rho)=
\begin{cases}
s, & \rho<\underline{\lambda}_t,\\[4pt]
\text{strictly decreasing from }s\text{ to }0, & \rho\in[\underline{\lambda}_t,\overline{\lambda}_t],\\[4pt]
0, & \rho>\overline{\lambda}_t,
\end{cases}
\]
i.e., \(g_t\) is flat outside the interval \([\underline{\lambda}_t,\overline{\lambda}_t]\) and strictly decreasing on that interval. By construction, the support of trading prices lies in \([\underline{\lambda}_t,\overline{\lambda}_t]\); this is exactly ``concentrated liquidity''.

The Myersonian characterization result Thm.~3.4 of  \cite{milionis2023myersonian} states that there is a one-to-one correspondence between such arbitrage-free, budget-balanced, individually rational inventory demand curves \(g_t\) and AMMs implemented by strictly increasing, quasi-concave, homothetic trading potentials \(\psi\) whose level sets generate the same price schedule.\footnote{In our notation, the inventory axis corresponds to \(E\) and the monetary axis to \(M\). The slope of level sets of \(\psi\) recovers the price \(p\), and the induced inventory choices along those level sets recover \(g_t\).}

Thus the axioms imply the existence of a trading function \(\psi(E,M)\) that is (i) strictly increasing in \(E\) and \(M\), (ii) homothetic, and (iii) quasi-concave, and whose level sets implement precisely the same batch pricing rule with concentrated liquidity in \([\underline{\lambda}_t,\overline{\lambda}_t]\).

\medskip
Combining the forward and reverse directions, we conclude that a mechanism satisfies Anonymity, Coalition-Proofness, No-Arbitrage, Budget-Balance, Individual Rationality, Monotonicity, Homogeneity, and Responsiveness \emph{if and only if} it can be implemented by an AMM constructed from a trading function \(\psi\) that is strictly increasing, homothetic, quasi-concave, and has liquidity concentrated within \([\underline{\lambda}_t,\overline{\lambda}_t]\).
\qedhere

\end{proof}

\subsubsection*{Proof of \cref{lemma-potential-game}}

We show that the defintion of Potential Game applies. 

Budget balance implies $\sum_{i \in \mathcal{N}} \pi_i(\boldsymbol{\sigma}) = W(\boldsymbol{\sigma})$. 

For a unilateral deviation by agent $n$ from $\sigma_n$ to $\sigma'_n$, the change in global welfare is:
    $$
        \Delta W = \sum_{i \in \mathcal{N}} (\pi_i(\sigma'_n, \sigma_{-n}) - \pi_i(\sigma_n, \sigma_{-n}))
$$    
Under the Atomicity assumption, prices are invariant to deviations by a single prosumer, so $\pi_i(\sigma'_n, \sigma_{-n}) = \pi_i(\sigma_n, \sigma_{-n})$ for all $i \neq n$. The sum collapses to:
$$    
        W(\sigma'_n, \sigma_{-n}) - W(\sigma_n, \sigma_{-n}) = \pi_n(\sigma'_n, \sigma_{-n}) - \pi_n(\sigma_n, \sigma_{-n})
$$   
    Thus, $W$ is a potential function for the game.  \qed 

\subsubsection*{Proof of \cref{prop-welfare-equilvanence}}

\paragraph{Sufficiency.} We show that if \(\boldsymbol{\sigma}^*\) maximizes \(\bar{W}(\cdot)\), then  \(\boldsymbol{\sigma}^*\) must be a BNE.  \medskip{}

\noindent{}Suppose by contradiction that \(\boldsymbol{\sigma}^*\) is not a BNE. Then, there exists a prosumer \(i \in \mathcal{N}\) and an alternative strategy \(\sigma_i'\) such that  $\bar{\pi}_i(\sigma_i', \boldsymbol{\sigma}_{-i}^*) > \bar{\pi}_i(\boldsymbol{\sigma}^*).  $  Consider the strategy profile \(\tilde{\boldsymbol{\sigma}} = (\sigma_i', \boldsymbol{\sigma}_{-i}^*)\). 

By Atomicity  (\cref{as:mean_field}), $\bar{\pi}_n(\tilde{\boldsymbol{\sigma}}) = \bar{\pi}_n(\boldsymbol{\sigma}^*), \quad \forall n \neq i.$  

By Budget Balance (Axiom \ref{BB}), $\bar{W}(\boldsymbol{\sigma}) =  \sum_{n=1}^N \bar{\pi}_n(\boldsymbol{\sigma})$ for all $\boldsymbol{\sigma}$.

Combining the two properties: $\bar{W}(\tilde{\boldsymbol{\sigma}}) =  \sum_{n=1}^N \bar{\pi}_n(\tilde{\boldsymbol{\sigma}})  =  \bar{W}(\boldsymbol{\sigma}^*) + \left[ \bar{\pi}_i(\tilde{\boldsymbol{\sigma}}) - \bar{\pi}_i(\boldsymbol{\sigma}^*) \right].$ 

Since \(\bar{\pi}_i(\tilde{\boldsymbol{\sigma}}) - \bar{\pi}_i(\boldsymbol{\sigma}^*) > 0\), we must have   $\bar{W}(\tilde{\boldsymbol{\sigma}}) > \bar{W}(\boldsymbol{\sigma}^*).  $ 

This contradicts the initial assumption that \(\boldsymbol{\sigma}^*\) maximizes \(\bar{W}(\cdot)\). Therefore, no such profitable deviation \(\sigma_i'\) can exist, and \(\boldsymbol{\sigma}^*\) must be a Bayes-Nash Equilibrium.  \qed

\paragraph{Necessity.} We show that If \(\boldsymbol{\sigma}^*\) is a BNE, then \(\boldsymbol{\sigma}^*\)  maximizes \(\bar{W}(\cdot)\).  \medskip{}

\noindent{}Suppose not. Then there exists a mixed profile $ \boldsymbol{\sigma}'$ such that $\bar{W}(\boldsymbol{\sigma}') > \bar{W}(\boldsymbol{\sigma}^*)$.

Define the expected net trade vector $x(\boldsymbol{\sigma}) = \mathbb{E}_{\boldsymbol{\theta}, \boldsymbol{a} \sim \boldsymbol{\sigma}}[\mathbf{s}(\boldsymbol{a}) - \mathbf{d}(\boldsymbol{a})]$.

Let the welfare function be $\bar{W}(\mathbf{x}) = \sum_{t=1}^T [\underline{\lambda}_t x_t^+ - \overline{\lambda}_t x_t^-]$.\footnote{Recall that $x^+=\max(x,0)$ and $x^-=\max(-x,0)$.} 

Define also the piecewise‑linear function \(w_t(x) = \underline{\lambda}_t x^+ - \overline{\lambda}_t x^-\). 

Since \(\underline{\lambda}_t < \overline{\lambda}_t\), the piecewise‑linear function \(w_t(x) = \underline{\lambda}_t x^+ - \overline{\lambda}_t x^-\) is concave. Therefore, the welfare function   $\bar{W}(\mathbf{x}) = \sum_t w_t(x_t)$  is a concave function of the vector \((x_1,\dots,x_T)\).

Since $\bar{W}(\boldsymbol{\sigma}') > \bar{W}(\boldsymbol{\sigma}^*)$, concavity implies the existence of a subgradient $g \in \partial\bar{W}(x(\boldsymbol{\sigma}^*))$ such that $\langle g, x(\boldsymbol{\sigma}') - x(\boldsymbol{\sigma}^*) \rangle > 0$. Moreover, the components of the subgradient are given by  
\[
g_t =
\begin{cases}
\underline{\lambda}_t, & \text{if } x_t(\boldsymbol{\sigma}^*) > 0, \\[2pt]
\overline{\lambda}_t, & \text{if } x_t(\boldsymbol{\sigma}^*) < 0, \\[2pt]
\in [\underline{\lambda}_t, \overline{\lambda}_t], & \text{if } x_t(\boldsymbol{\sigma}^*) = 0.
\end{cases} \]

This implies the existence of periods $t, \tau$ where the direction of improvement is to move energy from $\tau$ to $t$. Formally:
$$g_t [x_t(\boldsymbol{\sigma}') - x_t(\boldsymbol{\sigma}^*)] + g_\tau [x_\tau(\boldsymbol{\sigma}') - x_\tau(\boldsymbol{\sigma}^*)] > 0.$$
This inequality requires $x_t(\boldsymbol{\sigma}') > x_t(\boldsymbol{\sigma}^*)$ (increasing net flow at $t$) and $x_\tau(\boldsymbol{\sigma}') < x_\tau(\boldsymbol{\sigma}^*)$ (decreasing net flow at $\tau$).

We analyze the three regimes in which the inequality can occur. In each, we show a profitable deviation exists for a set of agents with positive $F(\cdot \mid z_e)$-measure in the support of $\boldsymbol{\sigma}^*$.

\paragraph{Case 1: Surplus-Reallocation} ($x_t^* > 0, x_\tau^* > 0$). \\
    In this case, the subgradients are $g_t = \underline{\lambda}_t$ and $g_\tau = \underline{\lambda}_\tau$, with the condition $\underline{\lambda}_t > \underline{\lambda}_\tau$.
    By Individual Rationality (Axiom \ref{ax:IR}), the AMM sell prices are anchored at the grid floor: $\bar{r}_t = \underline{\lambda}_t$ and $\bar{r}_\tau = \underline{\lambda}_\tau$.
    
    Define the set of active sellers at $\tau$:     $\mathcal{S}_\tau = \{(n,\theta_n): s_{n\tau} > 0 \text{ under } \sigma_n^*\}.$
    
    Since $x_\tau^* > 0$, this set has positive measure. Any agent in $\mathcal{S}_\tau$ can profitably deviate by shifting supply to $t$, gaining $\Delta \pi = \bar{r}_t - \bar{r}_\tau = \underline{\lambda}_t - \underline{\lambda}_\tau > 0$.

\paragraph{Case 2: Deficit-Reallocation} ($x_t^* < 0, x_\tau^* < 0$). \\
    Here, $g_t = \overline{\lambda}_t$ and $g_\tau = \overline{\lambda}_\tau$, with $\overline{\lambda}_t < \overline{\lambda}_\tau$.
    By Individual Rationality, the AMM buy prices are anchored at the grid ceiling: $\bar{c}_t = \overline{\lambda}_t$ and $\bar{c}_\tau = \overline{\lambda}_\tau$.
    
    Define the set of active buyers at $\tau$: $    \mathcal{B}_\tau = \{(n,\theta_n): d_{n\tau} > 0 \text{ under } \sigma_n^*\}.$
    
    Since $x_\tau^* < 0$, this set has positive measure. Any agent in $\mathcal{B}_\tau$ can profitably deviate by shifting demand to $t$, saving $\Delta \pi = \bar{c}_\tau - \bar{c}_t = \overline{\lambda}_\tau - \overline{\lambda}_t > 0$.

\paragraph{Case 3: Cross-Reallocation} ($x_t^* < 0, x_\tau^* > 0$). \\
    Here, $g_t = \overline{\lambda}_t$ and $g_\tau = \underline{\lambda}_\tau$. The condition for welfare improvement is $\overline{\lambda}_t > \underline{\lambda}_\tau$.
    By Individual Rationality, prices are set at the unfavorable bounds: $\bar{c}_t = \overline{\lambda}_t$ and $\bar{r}_\tau = \underline{\lambda}_\tau$.
    
    Define the sets of active traders in the support of $\boldsymbol{\sigma}^*$: 
    
    $ \mathcal{B}_t = \{(n,\theta_n): d_{nt} > 0\}, \quad \mathcal{S}_\tau = \{(n,\theta_n): s_{n\tau} > 0\}.$
    
    Since active trade implies non-zero measure, both sets have positive measure given $x_t^* < 0$ and $x_\tau^* > 0$. Consider two potential deviations:
    \begin{itemize}
        \item \textbf{Demand Shift} (for $n \in \mathcal{B}_t$): Shift demand from $t$ to $\tau$. Gain: $\pi_D = \overline{\lambda}_t - \bar{c}_\tau$.
        \item \textbf{Supply Shift} (for $n \in \mathcal{S}_\tau$): Shift supply from $\tau$ to $t$. Gain: $\pi_S = \bar{r}_t - \underline{\lambda}_\tau$.
    \end{itemize}
    We need not verify which specific deviation is profitable, only that at least one of them must be. Summing the payoff differences:
    \[
    \pi_D + \pi_S = (\overline{\lambda}_t - \bar{c}_\tau) + (\bar{r}_t - \underline{\lambda}_\tau) = (\overline{\lambda}_t - \underline{\lambda}_\tau) + (\bar{r}_t - \bar{c}_\tau).
    \]
    The first term $(\overline{\lambda}_t - \underline{\lambda}_\tau)$ is strictly positive by the welfare improvement assumption.
    The second term $(\bar{r}_t - \bar{c}_\tau)$ represents the internal price spread. 
    
    Notice that Responsiveness (Axiom \ref{ax:resp}) combined with Individual Rationality imply that $\bar{r}_t > \underline{\lambda}_t$, and $\bar{c}_\tau < \overline{\lambda}_\tau$ (see  Eq.\ref{eq:prices_IR+RESP}).

    Consequently,  $\pi_D + \pi_S > 0$. This implies $\max(\pi_D, \pi_S) > 0$. So  almost surely there exists an agent (either in $\mathcal{B}_t$ or $\mathcal{S}_\tau$) with a strictly profitable deviation.

In all cases, the hypothesis that $\boldsymbol{\sigma}^*$ does not maximize welfare leads to the existence of a profitable unilateral deviation, contradicting the definition of a BNE. \qed

\subsubsection*{Proof of \cref{prop-markov}}

To prove existence and multiplicity of the MPE, we first aggregate the individual equilibrium conditions into a single Planner's Problem.

\paragraph{Equivalence to Planner's Problem.}
Let $(\boldsymbol{\sigma}^*, \bar{\boldsymbol{\rho}}^*, \Pi^*)$ be an MPE. Summing the individual Bellman equations for all $n \in \mathcal{N}$:
\[
    \sum_{n} \Pi_n(b, z) = \sum_{n} \max_{\sigma_n} \left\{ \pi_n(\sigma_n, \boldsymbol{\sigma}^*_{-n}, \bar{\boldsymbol{\rho}}^*) + \gamma \mathbb{E}[\Pi_n(b', z')] \right\}.
\]

By \cref{lemma-potential-game} (Potential Game), $\sum_n \pi_n = \mathcal{W}$.

By Atomicity, cross-terms in the prosumer optimization problems vanish. Thus,$$ \sum_{n} \max_{\sigma_n} \Big\{ \pi_n(\sigma_n, \boldsymbol{\sigma}^*_{-n}, \bar{\boldsymbol{\rho}}^*) + \gamma \mathbb{E}[\Pi_n(b', z')] \Big\} 
= 
\max_{\boldsymbol{\sigma}} \Big\{ \sum_{n} \pi_n(\boldsymbol{\sigma}, \bar{\boldsymbol{\rho}}^*) + \gamma \mathbb{E}[\Pi_n(b', z')] \Big\}.$$

Defining the aggregate value function $V(b, z) \equiv \sum_{n} \Pi_n(b, z)$, the sum of the prosumer Bellman equations yields the single Bellman equation of the Social Planner, where the strategy profile $\boldsymbol{\sigma}^*$ maximizes the aggregate objective:
\begin{equation} \label{eq:planner_bellman}
    V(b, z) = \max_{\boldsymbol{\sigma} \in \Delta\operatorname{ext}(\mathcal{A}(b))} \left\{ \mathcal{W}(\boldsymbol{\sigma}, b, z) + \gamma \mathbb{E}[V(b', z') \mid \boldsymbol{\sigma}] \right\}, \qquad  \mathcal{A}(b) \equiv \times_n \mathcal{A}_n(b_n).
\end{equation}
Thus, any MPE strategy profile $\boldsymbol{\sigma}^*$ must be an optimal policy for the Planner.

\paragraph{Existence.}
Let $\mathcal{B}(\mathcal{S})$ be the Banach space of bounded continuous functions on the state space of the game $\mathcal{S} \equiv \mathcal{B}_{\text{agg}} \times \mathcal{Z} \subset \mathbb{R}^{N + Z}$, where 
$\mathcal{B}_{\text{agg}} = \prod_{n=1}^N [0, B_n]$ is the hyperrectangle of joint battery states and $\mathcal{Z} \subseteq \mathbb{R}^Z$ is the state space of public variables (with $Z$ denoting the dimension of the public signal vector $z$).
Define the Bellman operator $\mathcal{T}: \mathcal{B}(\mathcal{S}) \to \mathcal{B}(\mathcal{S})$ associated with \eqref{eq:planner_bellman}:
\[
    (\mathcal{T}V)(b, z) = \max_{\boldsymbol{\sigma}} \left\{ \mathcal{W}(\boldsymbol{\sigma}, b, z) + \gamma \mathbb{E}[V(b', z')] \right\}.
\]
Since the stage welfare $\mathcal{W}$ is bounded (by constraints $B_n, X_n$) and $\gamma \in (0,1)$, $\mathcal{T}$ satisfies Blackwell’s sufficiency conditions \citep{blackwell1965}: \textit{monotonicity} holds ($V \le U \implies \mathcal{T}V \le \mathcal{T}U$) by the linearity of the expectation operator, and \textit{discounting} holds ($\mathcal{T}(V+c) = \mathcal{T}V + \gamma c$) due to the discount factor $\gamma \in [0,1)$. 

By the Banach Fixed Point Theorem \citep{banach1922}, $\mathcal{T}$ is a contraction mapping with a unique fixed point $V^*$. The set of optimal policies is non-empty, implying existence of an MPE.

\paragraph{Multiplicity.}
While the aggregate value function $V^*$ is unique, the strategy profile $\boldsymbol{\sigma}^*$ is not. Multiplicity arises from the many-to-one mapping from individual strategies to aggregate outcomes. 

Let $M_n \equiv |\operatorname{ext}(\mathcal{A}_n)|$ be the number of pure strategies for agent $n$ so that $\boldsymbol{\sigma}_n \in \varDelta^{M_n-1}$. The set of mixed strategy profiles implementing a given optimal aggregate path $(\bar{\mathbf{s}}, \bar{\mathbf{d}})$ is given by the inverse image of the linear aggregation map $\mathsf{A}: \prod_n \varDelta^{M_n-1} \to \mathbb{R}^{2T}$. That is, $$    \Sigma^*_{(\bar{\mathbf{s}}, \bar{\mathbf{d}})} = \left\{ \boldsymbol{\sigma} \in \prod_{n=1}^N \varDelta^{M_n-1} \;\middle|\; \mathsf{A}\boldsymbol{\sigma} = \begin{pmatrix} \bar{\mathbf{s}} \\ \bar{\mathbf{d}} \end{pmatrix} \right\} = \mathsf{A}^{-1} \begin{pmatrix} \bar{\mathbf{s}} \\ \bar{\mathbf{d}} \end{pmatrix}  \cap \mathcal{P},$$
where $\mathcal{P} = \prod_n \varDelta^{M_n-1}$ is the product of strategy simplices, and $\mathsf{A}$ is the linear aggregation operator:
$$    \mathsf{A}\boldsymbol{\sigma} \equiv \sum_{n=1}^N \sum_{k=1}^{M_n} \sigma_{n,k} \begin{pmatrix} \mathbf{s}_{n,k} \\ \mathbf{d}_{n,k} \end{pmatrix},$$
with $(\mathbf{s}_{n,k}, \mathbf{d}_{n,k})$ representing the $k$-th extreme point (pure strategy) for agent $n$.

Since $(\bar{\mathbf{s}}, \bar{\mathbf{d}})$ lies in the convex hull of the extreme points---i.e., $(\bar{\mathbf{s}}, \bar{\mathbf{d}}) \in  \sum_{n=1}^N \operatorname{conv}(\operatorname{ext}(\mathcal{A}_n))$---$ \Sigma^*_{(\bar{\mathbf{s}}, \bar{\mathbf{d}})}$  is a non-empty convex polytope. By the Fundamental Theorem of Linear Maps \citep[e.g., Theorem 3.21 in][]{axler2015linear}, we can lower-bound its dimension:
\[
    \dim(\Sigma^*) \ge \sum_{n=1}^N (M_n - 1) - \operatorname{rank}(\mathsf{A}).
\]
Unless the linear system $\mathsf{A}\boldsymbol{\sigma} = (\bar{\mathbf{s}}, \bar{\mathbf{d}})$ is over-determined ($2T \ge \sum_n (M_n - 1)$) and the extreme point images are linearly independent---a measure-zero event with heterogeneous prosumers---this dimension is \textit{strictly} positive. This confirms the existence of a continuum of payoff-equivalent equilibria. \qed

\subsubsection*{Proof of \cref{prop:mfg-welfare}}
The proof proceeds in two steps: first establishing almost sure convergence of the realized welfare via the Law of Large Numbers, and then proving the convergence of the expected welfare via the Dominated Convergence Theorem.

\paragraph{Almost Sure Convergence of the Realized Welfare.} Conditioned on the public state $z_e$, the sequence of realized net trade vectors $\{\mathbf{y}_n\}_{n \ge 1}$ is i.i.d. by construction. Since $\mathcal A(\theta_n)$ are compact sets, the support of $\mathbf{x}_n$ is uniformly bounded, ensuring that $\mathbb{E}\|\mathbf{x}_n\| < \infty$. Consequently, the SLLN applies component-wise, guaranteeing: $\bar{\mathbf{x}}_N \xrightarrow{a.s.} \bar{\mathbf{X}}$. 

Since the welfare function $W_{\text{avg}}(\cdot)$ is continuous, the Continuous Mapping Theorem implies that:
\begin{equation*}
    \bar{\mathbf{x}}_N \xrightarrow{a.s.} \bar{\mathbf{X}} \implies W_{\text{avg}}(\bar{\mathbf{x}}_N) \xrightarrow{a.s.} W_{\text{avg}}(\bar{\mathbf{X}}). \tag*{\qed}
\end{equation*}

\paragraph*{Convergence of the Expected Welfare}
From the previous part of the proof, we have almost sure convergence: $W_{\text{avg}}(\bar{\mathbf{x}}_N) \xrightarrow{a.s.} W_{\text{avg}}(\bar{\mathbf{X}})$. To interchange limit and expectation, we invoke the Dominated Convergence Theorem. 

The feasible domain lies in the convex hull of the action spaces $\mathcal{A}(\theta)$, a compact subset of $\mathbb{R}^{TL}$. Since $W_{\text{avg}}(\cdot)$ is continuous on this compact set, it is uniformly bounded by some finite constant $\kappa$. The sequence of random variables $\{W_{\text{avg}}(\bar{\mathbf{x}}_N)\}$ is thus dominated by $\kappa$. 

The DCT applies, yielding:
\begin{equation*}
  \lim_{N \to \infty} \mathbb{E}\left[ W_{\text{avg}}(\bar{\mathbf{x}}_N) \right] = \mathbb{E}\left[ \lim_{N \to \infty} W_{\text{avg}}(\bar{\mathbf{x}}_N) \right] = W_{\text{avg}}(\bar{\mathbf{X}}). \tag*{\qed}  
\end{equation*}

\subsubsection*{Proof of \cref{prop:vanishing-regret}}

Let $\theta_n$ be the agent's true type and $\bar{\theta}_j$ be the centroid of the bin $R_j$ containing $\theta_n$. The quantization error is $\|\theta_n - \bar{\theta}_j\| \le \delta$.

The regret is defined as the difference between the agent's true optimal value and the expected value obtained from the projected  strategy recommended by the solver:
$$\text{Regret}_n = V(\theta_n) - \mathbb{E}_{\mathbf{a} \sim \sigma_j^\star} \left[ \pi_n(\operatorname{Prj}(\mathbf{a}; \theta_n)) \right],$$
where $V(\theta) = \max_{\mathbf{a} \in \mathcal{A}(\theta)} \pi(\mathbf{a})$ is the optimal value function.

Since the prosumer's optimization problem is a Linear Program (LP), its value function $V(\theta)$ is Lipschitz continuous with respect to $\theta$. Thus, there exists a constant $L_1$ such that:
$$|V(\theta_n) - V(\bar{\theta}_j)| \le L_1 \|\theta_n - \bar{\theta}_j\| \le L_1 \delta.$$
Now let $\mathbf{a}_k$ be a pure strategy in the support of the equilibrium strategy $\sigma_j^\star$. This action is feasible for the centroid $\bar{\theta}_j$ (i.e., $\mathbf{a}_k \in \mathcal{A}(\bar{\theta}_j)$). The projection $\operatorname{Prj}(\mathbf{a}_k; \theta_n)$ maps it to the closest point in the true feasible set $\mathcal{A}(\theta_n)$. Since the feasible set correspondence $\theta \rightrightarrows \mathcal{A}(\theta)$ is Lipschitz continuous in the Hausdorff metric, the distance between the sets is bounded by $L_2 \delta$. Thus:
$$\|\mathbf{a}_k - \operatorname{Prj}(\mathbf{a}_k; \theta_n)\| \le L_2 \delta.$$
Since the profit function $\pi_n(\cdot)$ is linear (and thus Lipschitz with constant $L_\pi$), the profit loss from projection is:
$$|\pi_n(\mathbf{a}_k) - \pi_n(\operatorname{Prj}(\mathbf{a}_k; \theta_n))| \le L_\pi L_2 \delta.$$
Combining these bounds:
\begin{align*}
\text{Regret}_n &= V(\theta_n) - \mathbb{E}_{\mathbf{a} \sim \sigma_j^\star} [\pi_n(\operatorname{Prj}(\mathbf{a}; \theta_n))] \\
&= \underbrace{V(\theta_n) - V(\bar{\theta}_j)}_{\le L_1 \delta}
+ \underbrace{V(\bar{\theta}_j) - \mathbb{E}_{\mathbf{a} \sim \sigma_j^\star}[\pi_n(\mathbf{a})]}_{= 0 \text{ (Optimality of } \sigma_j^\star \text{ for } \bar{\theta}_j)} 
\quad + \underbrace{\mathbb{E}_{\mathbf{a} \sim \sigma_j^\star}[\pi_n(\mathbf{a}) - \pi_n(\operatorname{Prj}(\mathbf{a}; \theta_n))]}_{\le L_\pi L_2 \delta}.
\end{align*}
The middle term is zero because $\sigma_j^\star$ is optimal for the representative type $\bar{\theta}_j$, so its expected payoff equals $V(\bar{\theta}_j)$. The regret is thus bounded by $(L_1 + L_\pi L_2) \delta$. Letting $\kappa = L_1 + L_\pi L_2$, we have $\text{Regret}_n \le \kappa \delta$. \qed 

\end{document}